\documentclass[11pt]{article} 
\usepackage{amsfonts}
\usepackage{amsmath, amsthm, amssymb}
\usepackage{multirow}
\usepackage{bm}
\usepackage{color}
\usepackage{natbib}
\usepackage{url}
\usepackage{booktabs}
\usepackage{makecell}
\usepackage{float}
\usepackage{setspace}
\usepackage{latexsym}
\usepackage[pdftex]{graphicx}
\usepackage{fullpage}
\usepackage{longtable}
\usepackage{authblk}

\usepackage{array}
\newcommand{\PreserveBackslash}[1]{\let\temp=\\#1\let\\=\temp}
\newcolumntype{C}[1]{>{\PreserveBackslash\centering}p{#1}}
\newcolumntype{R}[1]{>{\PreserveBackslash\raggedleft}p{#1}}
\newcolumntype{L}[1]{>{\PreserveBackslash\raggedright}p{#1}}

\bibpunct{(}{)}{;}{a}{}{,}
\usepackage{dcolumn}
\newcolumntype{.}{D{.}{.}{-1}}
\newcolumntype{d}[1]{D{.}{.}{#1}}
\usepackage{bbding}
\usepackage[top=1in, left=1in, right=1in,
bottom=1in]{geometry}
\usepackage[compact]{titlesec}
\usepackage{threeparttable}
\usepackage{tikz,pgfplots}
\pgfplotsset{compat=1.10}
\usetikzlibrary{intersections}
\usepackage{perpage}

\def\bSig\mathbf{\Sigma}

\def\var{\mbox{var}}

\def\calT{{\cal T}}
\def\calR{{\cal R}}
\def\mh{\text{MH}}

\newcommand{\black}{\color{black}}

\newtheorem{assum}{Assumption}
\newtheorem{theo}{Theorem}
\usepackage{caption}
\usepackage{subcaption}
\usepackage{microtype}
\expandafter\def\expandafter\normalsize\expandafter{\normalsize\setlength\abovedisplayskip{0pt}}
\expandafter\def\expandafter\normalsize\expandafter{\normalsize\setlength\belowdisplayskip{0pt}}
\expandafter\def\expandafter\normalsize\expandafter{\normalsize\setlength\abovedisplayshortskip{0pt}}
\expandafter\def\expandafter\normalsize\expandafter{\normalsize\setlength\abovedisplayshortskip{0pt}}
\MakePerPage[2]{footnote}

\makeatother

\begin{document}
\pagestyle{plain}
\title{Clarifying the Role of the Mantel-Haenszel Risk Difference Estimator in Randomized Clinical Trials}

\author{Xiaoyu Qiu$^{1*}$, Yuhan Qian$^{2*}$, Jaehwan Yi$^3$, Jinqiu Wang$^4$, Yu Du$^5$, Yanyao Yi$^5$, Ting Ye$^2$\\
$^1$Department of Statistics, University of Michigan \\
$^2$Department of Biostatistics, University of Washington\footnote{Correspond to Ting Ye (tingye1@uw.edu).}\\
$^3$Department of Statistics, Pennsylvania State University\\
$^4$Newark Academy\\
$^5$Global Statistical Sciences, Eli Lilly and Company  \\
$^*$Equal contribution\\
}
\maketitle
\thispagestyle{empty}
\abstract{
The Mantel-Haenszel (MH) risk difference estimator, commonly used in randomized clinical trials for binary outcomes, calculates a weighted average of stratum-specific risk difference estimators. Traditionally, this method requires the stringent assumption that risk differences are homogeneous across strata, also known as the common (constant) risk difference assumption. In our article, we relax this assumption and adopt a modern perspective, viewing the MH risk difference estimator as an approach for covariate adjustment in randomized clinical trials, distinguishing its use from that in meta-analysis and observational studies. {We demonstrate that, under reasonable restrictions on risk difference variability, the MH risk difference estimator consistently estimates the average treatment effect within a standard super-population framework,} which is often the primary interest in randomized clinical trials, in addition to estimating a weighted average of stratum-specific risk differences. We rigorously study its properties under the large-stratum and sparse-stratum asymptotic regimes, as well as under mixed-regime settings. 
Furthermore, for either estimand, we propose a unified robust variance estimator that 
improves over the popular variance estimators by \cite{Greenland1985} and \cite{Sato1989} and has provable consistency across these asymptotic regimes, regardless of assuming common risk differences. {Extensions of our theoretical results also provide new insights into the Mantel-Haenszel test, the post-stratification estimator, and settings with multiple treatments.} Our findings are thoroughly validated through simulations and a clinical trial example.

\vspace{2mm}
\noindent%

\newcommand{\n}{\noindent}

\clearpage

\section{Introduction}
\label{sec: intro}

 The Mantel-Haenszel (MH) methods are standard statistical tools for analyzing stratified 2 $\times$ 2 tables and are among the most highly cited statistical methods by the medical literature \citep{M-H1959}. These methods are popular due to their simplicity and versatility in handling both a small number of large strata (\emph{large-stratum asymptotics}) and a large number of small strata (\emph{sparse-stratum asymptotics}), while maintaining good efficiency \citep{Greenland1985}. Originally developed to adjust for confounding factors in observational studies, MH methods have since been widely adopted for adjusting prognostic factors to improve efficiency in randomized clinical trials and for synthesizing findings across multiple studies in meta-analyses \citep{agresti2000strategies}. A review of recent clinical trials using the MH methods is in the Supplement.

In this article, we focus on the use of MH risk difference in randomized clinical trials. Although this method is a classical statistical technique covered in textbooks \citep{catedata_analysis} and standard statistical software routines \citep{sas}, its interpretation remains somewhat elusive. Since its proposal by \cite{Greenland1985}, it has been viewed as estimating a common (constant) risk difference across all strata. This is why the recent guidance on covariate adjustment by the U.S. Food and Drug Administration \citep{fda:2019aa} describes the MH methods as ``estimating a conditional treatment effect, which is assumed to be constant across subgroups defined by a
covariate taking a discrete number of levels.''  
However, randomization does not justify the common risk difference assumption \citep{freedman2008randomization}, and it poses a significant limitation, as risk differences may vary meaningfully across strata in practice \citep{greenland1982interpretation, catedata_analysis}, and it is unclear what the MH risk difference estimator targets when a ``common risk difference'' does not exist.
Recently, \cite{Noma2016} investigated this issue and showed that the MH risk difference estimator generally converges to a weighted average of stratum-specific risk differences under a binomial model with fixed marginal treatment sums per stratum. While this provides insight into the estimand, it does not fully address the question, as the marginal treatment sums are usually not fixed for each stratum in randomized clinical trials.

The versatility of the MH risk difference estimator also presents challenges in variance estimation across various scenarios, leading to a multitude of variance estimators. This has made it difficult for practitioners to fully understand their differences and select the appropriate one for a given context. The most popular variance estimators are those by \cite{Greenland1985} and \cite{Sato1989}, both Wald-type variance estimators. Greenland and Robins' (GR) estimator was originally proposed for large-stratum asymptotics under the common risk difference assumption, while Sato's extension also accommodates sparse-stratum asymptotics and is now the default in SAS \citep{sas}. In addition, several score-based confidence interval methods have been developed. \cite{miettinen1985comparative} and \cite{Klingenberg2014} constructed confidence intervals by inverting tests that use variance estimators under the null hypothesis and the common risk difference assumption. \cite{YanSu2010} 
proposed a stratified Newcombe confidence interval. Notably, all of these methods, except the stratified Newcombe confidence interval, were developed under the common risk difference assumption.

\cite{Noma2016} is a notable exception in the literature, as they derived an asymptotic variance formula for the MH risk difference estimator without assuming a common risk difference. However, their study has key limitations. First, it only considers fixed marginal treatment sums. Second, it assumes a finite number of risk differences in sparse-stratum asymptotics, limiting heterogeneity as the number of strata increases. Third, it incorrectly states that Sato's variance estimator is consistent when the common risk difference assumption is violated and does not empirically assess variance estimator performance.


The purpose of this article is to clarify the estimand and variance estimation of the MH risk difference estimator, as a tool to adjust for discrete prognostic factors to improve the efficiency of treatment effect estimation for binary endpoints in randomized clinical trials. Specifically, we forgo the common risk difference assumption and rigorously study the properties of the MH risk difference estimator under the large-stratum and sparse-stratum asymptotic regimes, as well as under mixed-regime settings. {We show that the MH risk difference estimator consistently estimates a weighted average of stratum-specific risk differences under the common setup of fixed marginal treatment sums, and does so under weaker assumptions than those of \cite{Noma2016}.} {\color{black} More importantly, we prove that, under reasonable restrictions on risk difference variability, the MH risk difference estimator estimates the average treatment effect (ATE)} within a standard super-population framework \citep{fda:2019aa}, where the ATE does not depend on the construction or choice of strata and is often of primary interest in randomized clinical trials. To our knowledge, this has not been previously established in the literature. {These results clarify the common confusion over whether the MH estimator targets a common stratum-specific risk difference,
and what the appropriate estimand should be when that homogeneity does not hold. This clarity is important given the central role that clearly defined estimands play in clinical trials.}

For either estimand, we derive the asymptotic variance of the MH risk difference estimator and propose a unified robust variance estimator that is consistent across different asymptotic regimes without assuming common risk differences. Extensions of our theoretical results also provide new insights into the validity of the {MH test} under treatment effect heterogeneity and the properties of the post-stratification estimator under sparse-stratum asymptotics. {We also generalize our results to accommodate multiple treatment arms.} Lastly, we thoroughly evaluate our findings and other variance estimators in the literature through comprehensive simulations and a clinical trial example, and provide clear guidance to practitioners. 

The rest of the article is as follows. Section \ref{sec: review} describes the setup and reviews MH methods. Section \ref{sec: main cmh risk difference} establishes the asymptotic properties of the MH risk difference estimator and proposes a unified robust variance estimator for each estimand. {Section \ref{sec: extensions} discusses extensions.} Section \ref{sec: sim} and Section \ref{sec: real data} present simulations and a real-data application. Section \ref{sec: discussion} concludes with a discussion. The proposed methods can be computed using \textsf{RobinCar R} package.

\section{Setup and Review}
\label{sec: review}

\subsection{Setup}

Unlike existing MH literature, which assumes binomial or hypergeometric models, we begin with a typical data-generating process of randomized clinical trials.
Consider a trial with 2 treatments, where treatment $a=0,1$ is assigned with probability $\pi_a>0$, and $\pi_0+\pi_1=1$. Let $Y^{(a)}$ be the binary potential outcome under treatment $a$, and $Z$ a vector of discrete baseline covariates (e.g., site, disease stage) used to stratify patients into $K$ strata defined by joint levels $z_1,\dots, z_K$. Let $A$ denote the treatment assignment, with $P(A=a)= \pi_a$, and the observed outcome $Y=Y^{(a)}$ for those assigned to treatment $a$.
A random sample of $n$ patients is obtained. For the $i$th patient, let $Y_i^{(0)}, Y_i^{(1)}, Z_i, A_i$ be the realizations of $Y^{(0)}, Y^{(1)}, Z, A$. We assume that $(Y_i^{(0)}, Y_i^{(1)}, Z_i, A_i), i=1, \dots, n$ are independent and identically distributed.  The observed data are $(Z_i, A_i, Y_i), i=1,\dots, n$, which can be summarized into a series of $2\times 2$ contingency tables, one for each stratum, as in Table \ref{Table 1}. Let ${\cal T}_k = \{n_{.1k},n_{.0k} \}$ denote the treatment marginal totals in stratum $k$, with ${\cal T} = \{{\cal T}_1,\dots, {\cal T}_K\}$. Similarly, Let ${\cal R}_k = \{n_{1.k},n_{0.k} \}$ denote the response marginal totals, and ${\cal R} = \{{\cal R}_1,\dots, {\cal R}_K\}$.    

\begin{table}[ht]
    \centering
        \caption{Summary of $2\times 2$ table for the $k$th stratum}
    \label{Table 1}
    \begin{tabular}{lccc}
    \hline
         & Treated & Control &    Total   \\
         \hline 
    Responder & $n_{11k}$ & $n_{10k}$ & $n_{1.k}$ \\
    Non-responder & $n_{01k}$ & $n_{00k}$ & $n_{0.k}$ \\
     Total    & $n_{.1k}$  & $n_{.0k}$  & $n_{..k}$  \\
         \hline
    \multicolumn{4}{l}{\footnotesize Treated: $A_i=1$; Control: $A_i=0$}\\
        \multicolumn{4}{l}{\footnotesize Responder: $Y_i=1$; Non-responder: $Y_i=0$}\\
    \end{tabular}
\end{table}

\subsection{Review: {MH} test}

The {MH test} was proposed by \cite{M-H1959} to test the null hypothesis of conditional independence within each of several $2\times 2$ tables. In randomized clinical trials, the {MH} test is used to test the null hypothesis of no treatment effect across all strata.

In \cite{M-H1959}, both the treatment marginal totals and the response marginal totals are treated as fixed, i.e. the analysis is conditional on  $\calT$ and $\cal R$. Consequently, the cell count $n_{11k}$ follows a hypergeometric distribution, and all other cell counts are determined once $n_{11k}$ and the marginal totals are known. Under the null hypothesis and conditional on all marginal totals, the hypergeometric mean and variance of $n_{11k}$ are: $ E(n_{11k} \mid  \calT_k, \calR_k )  = n_{1.k}n_{.1k}/n_{..k}$ and $ \var (n_{11k} \mid   \calT_k, \calR_k )  = n_{1.k}n_{0.k} n_{.1k} n_{.0k} / \{ n_{..k}^2 (n_{..k} - 1)\} $. 
Since strata are independent, the MH test statistic is formulated by aggregating across strata:
\begin{align}
\text{MH test statistic} = \frac{\{ \sum_{k=1}^K n_{11k} -    E(n_{11k}\mid  \calT_k, \calR_k) \}^2}{\sum_{k=1}^K  \var (n_{11k} \mid\calT_k, \calR_k)}.  \label{eq: cmh}
\end{align}
This statistic follows a large-sample $\chi^2$ distribution with 1 degree of freedom under the null hypothesis.  In practice, for tests at the 0.05 level, this approximation is considered adequate if the absolute values of $n_{11k} -    E(n_{11k}\mid  \calT_k, \calR_k)$ exceed 5 for every stratum $k$ \citep{mantel1980minimum}. Alternatively, an exact p-value can be derived from the null distribution of $\sum_{k=1}^K n_{11k}$ based on the convolution of the hypergeometric distributions \citep{catedata_analysis}. A key strength of {the MH test} is its robustness under both large-stratum and sparse-stratum asymptotics under the null hypothesis of no treatment effect across all strata.

\cite{Cochran1954} proposed a similar test statistic that 
considers only the treatment marginal sums as fixed (i.e., inference is conditional on $\calT$) and models the columns as two independent binomials rather than a hypergeometric. Therefore, Cochran's statistic resembles the MH statistic in form but uses a different denominator. {However, as noted by an anonymous expert reviewer, the Cochran and MH statistics differ fundamentally in their respective likelihood frameworks (unconditional versus conditional likelihood) and behavior under sparse-stratum asymptotics. These distinctions underscore the importance of clearly differentiating the two statistics. Additional details are provided in Section S1 of the Supplement.}




\subsection{Review:  MH risk difference estimator}
\label{subsec: review cmh RD}

In practice, it is important to estimate the treatment effect, not just test its existence. If the treatment effect in the chosen contrast (e.g., risk difference, risk ratio, or odds ratio) is homogeneous across all strata, we can combine the $K$ stratum-specific treatment effects into a single measure. In this section, we focus on reviewing the MH risk difference.

Consider $2K$ mutually independent binomial variables given the treatment marginal sums: $n_{11k}\mid \calT_k \sim {\rm Binom}(n_{.1k}, p_{1k})$, $n_{10k} \mid \calT_k \sim {\rm Binom}(n_{.0k}, p_{0k})$ for $k=1,\dots, K$, where $p_{1k}$ and $p_{0k}$ are the probabilities of response in the treated and control groups. The risk difference $\delta_k$ in stratum $k$ is defined as $
\delta_k := p_{1k} - p_{0k}$, and can be estimated by $\hat{\delta}_k = n_{11k}/n_{.1k} - n_{10k}/n_{.0k}$. The MH risk difference estimator $\hat{\delta}$ is given by \citep{Greenland1985}:
    \begin{align}
    \label{eq:hat_delta}
&\hat{\delta} = \frac{\sum_{k=1}^K w_{k}\hat{\delta}_k}{\sum_{k=1}^K w_{k}} = \frac{\sum_{k=1}^K (n_{.0k}n_{11k}-n_{.1k}n_{10k})/n_{..k} }{\sum_{k=1}^K w_{k}},
    \end{align}
where $ w_{k} = n_{.1k}n_{.0k}/n_{..k}$ is the MH weight for stratum $k$.

Typically, the literature assumes a \emph{common risk difference} assumption, meaning $\delta_1=\delta_2=\cdots=\delta_K := \delta.$  Under this assumption, the MH risk difference estimator is consistent for $\delta$ under both large- and sparse-stratum asymptotics. In Section \ref{sec: main cmh risk difference}, we relax this assumption.

As noted in Section \ref{sec: intro},  \cite{Greenland1985} and \cite{Sato1989} proposed two widely used variance estimators under the common risk difference assumption. Specifically, Greeland and Robins (GR) derived $\var(\hat{\delta} \mid \calT)$, denoted $\sigma_n^2$, as follows:
\begin{align}
\label{eq:cmh var}
\sigma_n^2=  \frac{\sum_{k=1}^K \var(  w_{k} (\hat\delta_k - \delta)\mid \calT_k )}{(\sum_{k=1}^K w_{k})^2} = \frac{\sum_{k=1}^K w_{k}^2 \{p_{1k}(1-p_{1k}) /n_{.1k}+ p_{0k}(1-p_{0k})/n_{.0k} \}}{(\sum_{k=1}^K w_{k})^2} ,
\end{align} 
using the identity $\var( w_{k} (\hat\delta_k - \delta)\mid {\calT_k} )= w^2_{k}\{\var(n_{11k}\mid {\calT_k})/n_{.1k}^2+\var(n_{10k}\mid {\calT_k})/n_{.0k}^2\}$, with ${\var}(n_{11k}\mid {\calT_k}) = n_{.1k}p_{1k}(1-p_{1k}) $ and ${\var}(n_{10k}\mid {\calT_k}) = n_{.0k}p_{0k}(1-p_{0k}) $. Using $n_{11k}/n_{.1k}$ and $n_{10k}/n_{.0k}$ to estimate $p_{1k}$ and  $p_{0k}$, respectively, gives the estimator:
 $\widehat{\var}(w_{k} (\hat\delta_k - \delta) \mid \calT_k) = \{n_{11k}(n_{.1k} - n_{11k}) n_{.0k}^3 + n_{10k}(n_{.0k} - n_{10k})n_{.1k}^3\}/\{n_{.1k}n_{.0k}n_{..k}^2\}.$
Replacing $\var(  w_{k} (\hat\delta_k - \delta) \mid {\calT})$ with $ \widehat{\var}(w_{k} (\hat\delta_k - \delta_k)  \mid {\calT} )$ in \eqref{eq:cmh var} gives the GR variance estimator for $\hat\delta$, which was originally proposed for the large-stratum asymptotics under the common risk difference assumption.


However, the GR variance estimator is not consistent under the sparse-stratum asymptotics. To address this issue, \cite{Sato1989} proposed a different way to estimate $\var( w_{k} (\hat\delta_k - \delta)\mid {\calT_k} )$ in \eqref{eq:cmh var}. Specifically, under the common risk difference assumption, Sato used the relationship $p_{1k} = \delta + p_{0k}$ to express $\operatorname{Var}(w_k(\hat\delta_k - \delta) \mid \mathcal{T}_k)$ in two different forms, each leading to a separate variance estimator. The final estimate of $\operatorname{Var}(w_k(\hat\delta_k - \delta) \mid \mathcal{T}_k)$ is then obtained by averaging these two estimates. The detailed derivations are included in the Supplement. Sato's variance estimator is applicable to both the large-stratum and sparse-stratum asymptotics under the common risk difference assumption.

\section{Main results: MH risk difference}
\label{sec: main cmh risk difference}

\subsection{Two asymptotic regimes}\vspace{-2mm}

To rigorously establish the statistical properties of the MH risk difference estimator, we first provide precise definitions of the large-stratum and sparse-stratum asymptotics, as well as the mixed-stratum asymptotics. Let $p_{ak} = E(Y^{(a)}\mid Z= z_k)$ for $a=0,1$ and $k=1,\dots, K$.

\begin{assum}[Large-stratum asymptotics]\label{as1_assumption} 
(i) The number of strata $K<\infty$ is fixed, and $P(Z= z_k)>0$ for all $k$; (ii) $\epsilon <p_{1k}, p_{0k}<1-\epsilon$ for all $k$ and a fixed positive constant $\epsilon$. 
\end{assum} 

\begin{assum}[Sparse-stratum asymptotics]\label{sparse_assumption}
(i) The number of strata $K \rightarrow \infty$ as $n\to \infty$, with each stratum size bounded by $1\leq n_{..k}\leq c$ for all $k$ and a fixed constant $c$; (ii) $\epsilon <p_{1k}, p_{0k}<1-\epsilon$ for all $k$ and a fixed positive constant  $\epsilon$; (iii)  Define $K_1$ as the number of strata where $n_{.1k} = 0$ or $n_{.0k} = 0$. Then,
    $\lim\limits_{K\rightarrow \infty} \frac{K_1}{K}\le C$ for some $ C \in [0,1)$, almost surely.
\end{assum}

\begin{assum}[Mixed-stratum asymptotics]\label{mix_assumption}
    The set of strata $\{1, \dots, K\}$ can be partitioned into two disjoint subsets $\mathcal{K}_l$ and $\mathcal{K}_s$, where the strata in $\mathcal K_l$ satisfy the Assumption \ref{as1_assumption} (large-stratum) and those in $\mathcal K_s$ satisfy Assumption \ref{sparse_assumption} (sparse-stratum).
\end{assum}

Assumption \ref{as1_assumption} defines the large-stratum asymptotic regime, where a few strata grow in size proportionally with the total sample size. 
 Assumption \ref{sparse_assumption} describes the sparse-stratum asymptotic regime, characterized by many small strata of bounded size. In this regime,  
Assumption \ref{sparse_assumption}(iii) is a mild condition requiring that a non-negligible proportion of strata include participants from both arms, which are the strata receiving non-zero weights in the MH risk difference estimator. This condition is needed for the variance of the MH risk difference estimator to decrease as the sample size increases. Assumption \ref{mix_assumption} describes the mixed-stratum asymptotic regime, combining a few large strata and many small strata.



\subsection{Asymptotic theory without assuming a common risk difference}

We next study the properties of the MH risk difference estimator under the large-stratum, sparse-stratum, and mixed-stratum asymptotic regimes, all without assuming a common risk difference. Theorem \ref{theo:1} demonstrates that the MH risk difference estimator consistently estimates a weighted average of stratum-specific risk differences. Theorem \ref{theo: ate} shows that it is also a consistent estimator of the ATE $\delta_{\rm ATE } = E(Y^{(1)}) - E(Y^{(0)})$, an  estimand that is often of primary interest in clinical trials. Depending on the estimand, the MH estimator  has different asymptotic variances and therefore requires different variance estimators.

The idea that a single estimator can estimate different parameters and the related variance implications is not unfamiliar to the clinical trials community. For instance, the difference in means estimator can estimate the sample ATE under a randomization inference framework \citep{Neyman:1923a}, or the population ATE under a super-population framework \citep{Tsiatis:2008aa}. Similar distinctions are discussed in causal inference \citep{imbens2004nonparametric}.

{\begin{theo}\label{theo:1}
    Suppose that one of the following holds: Assumption \ref{as1_assumption} (large-stratum), Assumption \ref{sparse_assumption} (sparse-stratum), or Assumption \ref{mix_assumption} (mixed). Then, as 
     $n\rightarrow \infty,$ 
\begin{align*}
   & \frac{\hat\delta-\delta_{\rm MH}}{\sigma_{n}}\xrightarrow{d}N(0,1 ), \quad \text{with} \quad 
     \delta_{\rm MH} = \frac{\sum_{k=1}^K w_{k}\delta_k}{\sum_{k=1}^K w_{k}}, 
\end{align*}
where $ w_{k} = n_{.1k}n_{.0k}/n_{..k}$ is the MH weight for stratum $k$, $\delta_k = p_{1k} - p_{0k}$, $\sigma_{n}^2$ is defined in \eqref{eq:cmh var}, and $\xrightarrow{d}$ denotes convergence in distribution. 
\end{theo}}

All technical proofs are in the Supplement. Theorem \ref{theo:1} shows that $\hat\delta$ is asymptotically normal with the same asymptotic variance formula $\sigma_{n}^2$, under both asymptotic regimes or mixed-regime settings. We also show that $\sigma_{n}^2 $ converges at rate $n^{-1}$ in the Supplement, implying that $\hat\delta$ consistently estimates $\delta_\mh$ as a weighted average of stratum-specified risk differences. However, this estimand may be difficult to interpret clinically, especially regarding why the stratum-specific risk differences should be weighted by this particular set of weights.



\begin{theo}\label{theo: ate} 
{\color{black}Suppose that one of the following holds: (i) Assumption \ref{as1_assumption} (large-stratum);  (ii) Assumption \ref{sparse_assumption} (sparse-stratum), and conditions \(\sum_{k=1}^K(\delta_k-\delta_{\rm ATE})=o(\sqrt{n})\)  and $\max_{k=1,\cdots,K}(\delta_k-\delta_{\rm ATE})^2/\sum_{k=1}^K(\delta_k-\delta_{\rm ATE})^2=o(1)$, where \(\delta_k = p_{1k} - p_{0k}\); or (iii) Assumption \ref{mix_assumption} (mixed), with the strata in $\mathcal{K}_s$ (the sparse strata) satisfying the conditions in (ii). Then, as $n\to\infty$,}
\begin{align*}
   & \frac{\hat\delta-\delta_{\rm ATE}}{\sqrt{\sigma^2_{n}+ \nu^2_{n} 
 }}\xrightarrow{d}N(0,1 ), \quad \text{with} \quad  \delta_{\rm ATE} = E(Y^{(1)}) -E(Y^{(0)}), 
\end{align*}
where 
$ \sigma_{n}^2$ is defined in \eqref{eq:cmh var},  $
     \nu_{n}^2 = n^{-1}\sum_{k=1}^K \big\{ (\delta_k-\delta_{\rm ATE})^2\pi_1\pi_0\frac{n_{..k}-1}{n_{..k}}\frac{n_{..k}-1-(4n_{..k}-6)\pi_1\pi_0}{n} + \pi_1^2\pi_0^2\rho_k(\delta_k^2-\delta_{\rm ATE}^2)  \big\} / (n^{-1}\sum_{k=1}^K w_{k})^2$, 
    $w_{k}= n_{.1k} n_{.0k}/n_{..k}$, and $\rho_k = P(Z= z_k) $. 
\end{theo} 
In this theorem, for the sparse strata,  we make two additional assumptions on $\delta_k$'s. {\color{black} The first assumption $\sum_{k=1}^K(\delta_k-\delta_{\rm ATE})=o(\sqrt{n}) $  limits the variability of $\delta_k$  to what we may reasonably expect in applications}. For example, if $\delta_k$ can be viewed as randomly sampled from some distribution $F$ and $\var(\delta_k)= o(1)$, then $\sum_{k=1}^K(\delta_k-\delta_{\rm ATE})= \sum_{k=1}^K\{\delta_k-E_F(\delta_k)\} (1- K \rho_k)$ is $o(\sqrt{n})$ with probability approaching 1. {\color{black} In our simulations of sparse-stratum in Section \ref{sec: sim}, the values of
var($\delta_k$) are 0, 0.003, and 0.11 under Factors (4a)-(4c), respectively. Our results indicate that values of var($\delta_k$) as large as 0.11 can be reasonably viewed as o(1).} The second assumption $\max_k(\delta_k-\delta_{\rm ATE})^2/\sum_{k=1}^K(\delta_k-\delta_{\rm ATE})^2=o(1)$ says that the distribution of $\delta_k$ cannot be too uneven and is used to verify Lindeberg's condition for the central limit theorem.
Theorem \ref{theo: ate} shows that when the MH risk difference estimator is used to estimate the ATE, its asymptotic variance is larger than targeting \(\delta_{\mh}\) without  common effect assumption, due to the need to additionally account for the variability in the size and marginal treatment sums for each stratum, reflecting the nature of randomized clinical trials that these are often not fixed.


\subsection{Modified GR variance estimator}
\label{subsec: variance estimator} 

For either $\delta_{\mh}$ or $\delta_{\rm ATE}$, we provide a variance estimator that is applicable across different asymptotic regimes and does not rely on the assumption of common risk differences.


For the estimand $\delta_{\mh}$, the variance $\sigma_{n}^2$ in  \eqref{eq:cmh var} uses $\var (n_{11k}\mid \calT_k)= n_{.1k}p_{1k}(1-p_{1k}),$  which the GR estimator estimates by $n_{11k}n_{01k}/n_{.1k}$ under large-stratum asymptotics. However, $n_{11k}n_{01k}/n_{.1k}$ is not an unbiased estimator of $\var (n_{11k}\mid \calT_k)$, making the GR variance estimator anti-conservative under sparse-stratum asymptotics. This observation motivates us to use $n_{11k}n_{01k}/(n_{.1k}-1)$, which is an unbiased estimator of $\var (n_{11k}\mid \calT_k)$ under both regimens (applicable when $n_{.1k}$ is larger than 1). We apply the same correction procedure to estimate  $\var (n_{10k}\mid \calT_k)$.  Then, we obtain a modified GR variance estimator for $\hat{\delta}$:
\begin{align*} 
    \hat\sigma^2  =\frac{1}{(\sum_{k=1}^K w_{k})^2}  \sum_{k=1}^K w_{k}^2 \left\{  \frac{n_{11k}n_{01k}}{n_{.1k}^3} \left( \frac{n_{.1k}}{n_{.1k}-1}\right)^{ I(n_{.1k}>1)}+  \frac{n_{10k}n_{00k}}{n_{.0k}^3} \left( \frac{n_{.0k}}{n_{.0k}-1}\right)^{ I(n_{.0k}>1)} \right\}.
\end{align*}
We refer to this as the $\text{mGR}_{\mh}$ variance estimator. Crucially,  $\hat\sigma^2 $ is a unified and consistent variance estimator for $\hat\delta$ across large-stratum, sparse-stratum, and mixed asymptotic regimes.

When focusing on the ATE, we additionally need a consistent estimator of $\nu^2_{n}$ (defined in Theorem \ref{theo: ate}). Our proposed estimator of $\nu^2_{n}$ is  
\begin{align*}
     \hat\nu^2 = \frac{ n^{-1}\sum_{k=1}^K I(n_{.1k}n_{.0k}\neq 0) \left\{(\hat{\delta}_k^2-2\hat\delta_k\hat\delta+\hat\delta^2) \hat\pi_1\hat\pi_0 \frac{n_{..k}-1}{n_{..k}}\frac{n_{..k}-1-(4n_{..k}-6)\hat\pi_1\hat\pi_0}{n}+\hat\pi_1^2\hat\pi_0^2\hat\rho_k(\widehat{\delta_k^2}-\hat\delta^2)\right\}}{(n^{-1} \sum_{k=1}^K w_{k})^2},
\end{align*}
where $\hat{\delta}_k = \hat p_{1k} - \hat p_{0k}$ is an unbiased estimator of $\delta_k$ and $\widehat{\delta_k^2}=\hat p_{1k}^2-S_{1k}^2I(n_{.1k}>1)/n_{.1k} + \hat p_{0k}^2 - S_{0k}^2I(n_{.0k}>1)/n_{.0k} -2 \hat p_{1k}\hat p_{0k}$ is an unbiased estimator of $\delta_k^2$ for strata with $n_{.1k}, n_{.0k}>1$, $\hat p_{1k}=n_{11k}/n_{.1k}$, $\hat p_{0k}=n_{10k}/n_{.0k}$, 
$S_{ak}^2$ is the sample variance of the $Y_i$'s in stratum $k$ under treatment $a=0,1$, $ \hat{\rho}_k ={n_{..k}}/{n}$, $\hat\pi_1= \sum_{k=1}^K n_{.1k}/ n $, and $\hat\pi_0=\sum_{k=1}^K n_{.0k}/ n $. Hence, when the estimand is the ATE,
$ \hat \sigma^2 + \hat{\nu}^2 $ is a consistent variance estimator for $\hat\delta$ across large-stratum, sparse-stratum, and mixed asymptotic regimes.
 We call this the $\text{mGR}_{\rm ATE}$ variance estimator.

For both variance estimators, the proof of consistency is in the Supplement, under the mild condition that the number of strata with $n_{.1k} = 1$ or $n_{.0k} = 1$ is negligible. Table \ref{tb: variance estimators} provides a summary of the applicability of variance estimators for the MH risk difference estimator.

\begin{table}[ht]
\caption{A summary of applicability of variance estimation and interval estimation methods for the MH risk difference estimator. \label{tb: variance estimators}}
\resizebox{\linewidth}{!}{
\begin{tabular}{llccccccc}
\hline
&& \multicolumn{5}{c}{Estimand: $\delta_{\mh}$} &&  \multicolumn{1}{c}{Estimand: $\delta_{\rm ATE}$} \\  \cline{3-7} \cline{9-9} 
&& \multicolumn{3}{c}{Wald-type variance estimator} & \multicolumn{2}{c}{Score-type confidence interval}&& Wald-type\\ 
Asymptotic    regime     & Effect & GR & Sato & $\text{mGR}_{\mh}$ &Klingenberg & Newcombe && $\text{mGR}_{\rm ATE}$
\\
\hline
Large-stratum  & Common &  \checkmark  &  \checkmark    & \checkmark                &\checkmark &\checkmark  && \checkmark  \\
Sparse-stratum  & Common &   $\times$        &   \checkmark   & \checkmark   &   \checkmark   &\checkmark &&  \checkmark           \\
Large-stratum  & {Varying} &   \checkmark$^*$      &  $\times$    &  \checkmark &       $\times$      &$\times^\dagger$  &&  \checkmark      \\
Sparse-stratum   & {Varying} &  $\times$       &  $\times$    &  \checkmark     & $\times$        & $\times^\dagger$  && \checkmark     \\
\hline 
\end{tabular}}
{\footnotesize  \singlespacing $^*$ The GR variance estimator was originally developed under the common risk difference assumption. Our analysis shows it also works for {varying} effects, but only under the large-stratum regime.\\
\footnotesize $^\dagger$ The stratified Newcombe estimator can be conservative with {varying} effects; see our simulations in Table \ref{tb: MH1a2a}.}
\end{table}

\section{Extensions}
\label{sec: extensions}
\subsection{{MH test}}

Since the  MH risk difference estimator and {MH test statistic} are closely related, the results above offer new insights into the MH test. The {MH test statistic} in \eqref{eq: cmh} can be written as:
\begin{align*}
\frac{\big( n^{-1/2}\sum_{k=1}^K n_{11k} - n_{.1k}n_{1.k}/n_{..k}\big)^2}{n^{-1}\sum_{k=1}^K \frac{ n_{1.k} n_{0.k}n_{.1k} n_{.0k} }{(n_{..k}- 1)n_{..k}^2}} .
\end{align*}
The numerator is exactly equal to $(n^{-1/2}\sum_{k=1}^K w_{k}\hat\delta_k)^2$ and the denominator converges to $\frac{1}{n}\sum_{k=1}^K\frac{n_{.1k}n_{.0k}}{n_{..k}}p_{k}(1-p_k)$, where $p_k = \pi_1 p_{1k} + \pi_0 p_{0k}$. In contrast, the asymptotic variance for $n^{-1/2} \sum_{k=1}^K w_{k} \hat{\delta}_k$, as derived in Theorems \ref{theo:1}, converges to 
$\frac{1}{n}\sum_{k=1}^K\frac{n_{.1k}n_{.0k}^2 }{n_{..k}^2}p_{1k}(1-p_{1k}) + \frac{n_{.0k}n_{.1k}^2}{n_{..k}^2}p_{0k}(1-p_{0k})$.
Thus,  the MH test statistic's denominator is generally not a consistent variance estimator for the numerator's square root, except under the sharp null hypothesis of no treatment effect in any stratum ($p_{1k} = p_{0k}$ for all $k$). Consequently, {the MH test} does not control the type-I-error rate when testing weaker null hypotheses such as $\delta_{\mh} = 0$ or $\delta_{\rm ATE} = 0$. Valid tests for these null hypotheses under different asymptotic regimes can be obtained using the variance estimators developed in Section \ref{subsec: variance estimator}.


 
\subsection{{Post-stratification (PS) estimator}}

The PS estimator is another estimator of the ATE and is similar to the MH risk difference estimator. It is defined as
$
 \hat{\delta}_{\rm PS} =  \sum_{k=1}^K w_{k, {\rm PS}}  I(n_{.0k} n_{.1k}\neq 0) \left( \bar Y_{1k} -  \bar Y_{0k}  \right),$ 
where $w_{k, {\rm PS}}= n_{..k}/n$, and $\bar Y_{ak}$ is the sample mean of $Y_i$ in stratum $k$ under treatment $a=0,1$. Section S1 of the Supplement discusses its asymptotic properties. In a nutshell, the PS estimator is consistent and asymptotically normal if $\frac{K_1}{\sqrt{K}} \to 0$ almost surely as $n \to \infty$, where $K_1$ (defined in Assumption \ref{sparse_assumption}) denotes the number of strata with $n_{.1k} = 0$ or $n_{.0k} = 0$. However, the usual variance estimators developed for large-stratum asymptotics (e.g., \cite{Ye2020}) are invalid under sparse- or mixed-stratum asymptotics. To address this, we propose a new consistent variance estimator in Section S1 that remains valid across different asymptotic regimes, provided the number of strata with $n_{.1k}=1$ or $n_{.0k}=1$ is negligible.

\section{Simulations}
\label{sec: sim}

{We conduct two simulation studies to evaluate various variance estimators for the MH risk difference estimator, targeting either $\delta_{\mh}$ or the ATE $\delta_{\rm ATE}$. In both studies, the simulation settings follow a factorial design that specifies a total of 54 scenarios, each with 1,000 runs.}\\
\textbf{Factor 1:} Sample size. (1a) $n=500$; (1b) $n=300$; (1c) $n=200$.\\
\textbf{Factor 2:} Treatment assignment ratio. (2a) $\pi_0=1/3$ and $\pi_1=2/3$; (2b) $\pi_0=\pi_1=0.5$.\\
\textbf{Factor 3:} Asymptotic regime. (3a) (large-stratum) $Z$ follows a multinomial distribution with $K=3$ and probabilities  $(0.2, 0.3, 0.5)$;  (3b) (sparse-stratum) $Z$ is multinomial with $K=30$ for $n=500$, $K=18$ for $n=300$, and $K=15$ for $n=200$; probabilities are drawn once from ${\rm Uniform}(0.2, 0.5)$ and normalized to sum to 1.    (3c) (mixed-stratum)  The three large strata have fixed probabilities $(0.1, 0.15, 0.25)$, and the remaining small strata (15 for $n=500$, 9 for $n=300$, 12 for $n=200$) have probabilities drawn once from ${\rm Uniform}(0.2,0.5)$ and normalized to sum to 0.5. Note that for $n=200$, both sparse and mixed settings have 15 strata to evaluate performance under smaller sample sizes and sparser data. \\
\textbf{Factor 4:} Common or {varying} risk difference. For the large-stratum,  (4a) (common)  $(p_{0k},k=1,2,3) = (0.5, 0.2, 0.6)$  and  $\delta_k = - 0.1$ for all $k$; (4b) ({varying}) $(p_{0k},k=1,2,3) = (0.1, 0.1 ,0.7)$  and  $(\delta_{k},k=1,2,3) = (0,0, 0.2)$. (4c) ({varying}) 
$(p_{0k},k=1,2,3) = (0.8, 0.9, 0.5)$  and  $(\delta_{k},k=1,2,3) = (-0.5, -0.3, 0.2)$. 
For the sparse-stratum, 
(4a) (common) $p_{0k}$'s are generated once from $ {\rm Uniform}(0.4, 0.7)$ and $\delta_k=- 0.1$  for all $k$; (4b) ({varying})  For half the strata, $p_{0k}$'s generated once from $ {\rm Uniform}(0.1,0.2)$ and $\delta_k$'s from $ {\rm TruncNorm}(0,0.1,0.05,0.05)$; for the other half, $p_{0k}$'s generated once from $ {\rm Uniform}(0.7,0.8)$ and $\delta_k$'s from ${\rm TruncNorm}(0.1,0.2,0.15,0.05)$, where {\rm TruncNorm}($a, b, \mu, \sigma$) denotes a truncated normal distribution on $[a,b]$ with mean $\mu$ and standard deviation $\sigma$; (4c) ({varying}) two-thirds of the strata have $p_{0k}$'s generated once from $ {\rm Uniform}(0.8,0.9)$ and $\delta_k$'s from ${\rm Uniform}(-0.6,-0.5)$; for the remaining strata, $p_{0k}$'s generated once from ${\rm Uniform}(0.4,0.5)$ and $\delta_k$'s from ${\rm Uniform}(0.1,0.2)$. {\color{black} Note: 
var($\delta_k$) of sparse stratum under Factors (4a)-(4c) are 0, 0.003, 0.110,  based on the sample variance of $10^6$ simulated values.}


\subsection{MH risk difference estimator as an estimator for $\delta_{\mh}$}
\label{subsec: sim: CMH RD}

We first consider the estimand $\delta_{\mh}$, a random quantity that varies across simulation runs. We evaluate six variance estimators:  the GR estimator, Sato's estimator, our proposed $\text{mGR}_{\mh}$ estimator (Section \ref{subsec: variance estimator}), and two score-based confidence interval methods, \cite{Klingenberg2014} interval and the stratified Newcombe interval \citep{YanSu2010}. We also implement the nonparametric bootstrap (200 iterations) to estimate the variance of $\hat\delta - \delta_{\mathrm{MH}}$, although this is infeasible in practice since $\delta_{\mathrm{MH}}$ is unknown. Table \ref{tb: MH1a2a} presents results for $n=500$ (Factor 1a) and unbalanced allocation (Factor 2a), including average bias ($\hat\delta - \delta_{\mh}$), standard deviation (SD), average standard error (SE), empirical 
power for testing $\delta_{\mh} = 0$, and the probability of the 95\% confidence interval covering $ \delta_{\mh}$ (CP). More results are in the Supplement.

\begin{table}[ht]  
\centering  
\caption{Simulation results for estimating $\delta_{\mh}$ under unbalanced allocation (Factor 2a), based on $n=500$ (Factor 1a) and 1,000 simulation runs. Entries have all been multiplied by 100. Since $\delta_{\mh}$ varies across simulation runs, we report the average true value of $\delta_{\mh}$ (denoted as Ave. Truth).}
\label{tb: MH1a2a} 
\renewcommand{\arraystretch}{0.9}
\resizebox{0.9\textwidth}{!}{
\begin{tabular}{@{}cccccccccc@{}}
\toprule
Effect & Regime  & {Ave. Truth}   &    & GR   & Sato & mGR  & Boot & Newc & Klin  \\ \midrule
\multirow{15}{*}{\makecell{(4a) \\ Common}} & \multirow{5}{*}{\makecell{(3a) \\ Large}} & \multirow{5}{*}{-10.0} & Bias  & \multicolumn{6}{c}{0.21}           \\
                    &                    &                        & SD    & \multicolumn{6}{c}{4.39}          \\
                    &                    &                        & SE    & 4.34 & 4.37 & 4.38 & 4.37 & 4.64 & 4.37  \\
                    &                    &                        & CP    & 94.4 & 94.4 & 94.4 & 94.3 & 96.0 & 94.7  \\
                    &                    &                        & Power & 61.5 & 61.2 & 61.0 & 61.2 & 56.2 & 61.0  \\ \cmidrule(l){2-10} 
                    & \multirow{5}{*}{\makecell{(3b) \\ Sparse}} & \multirow{5}{*}{-10.0} & Bias  & \multicolumn{6}{c}{0.01}           \\
                    &                    &                        & SD    & \multicolumn{6}{c}{4.86}           \\
                    &                    &                        & SE    & 4.37 & 4.77 & 4.76 & 4.80 & 4.82 & 4.77  \\
                    &                    &                        & CP    & 92.2 & 94.4 & 94.2 & 94.6 & 94.9 & 94.7  \\
                    &                    &                        & Power & 61.5 & 54.9 & 54.7 & 53.7 & 53.9 & 53.8  \\ \cmidrule(l){2-10} 
                    & \multirow{5}{*}{\makecell{(3c) \\ Mixed}} & \multirow{5}{*}{-10.0} & Bias  & \multicolumn{6}{c}{0.00}        \\ 
                    &                    &                        & SD    & \multicolumn{6}{c}{4.52}         \\
  &                                            &                        & SE    & 4.34 & 4.56 & 4.57 & 4.57 & 4.76 & 4.56 \\
&                                            &                        & CP    & 94.9 & 96.0 & 96.1 & 95.9 & 96.9 & 96.0 \\
&                                            &                        & Power & 62.3 & 58.7 & 59.0 & 58.9 & 56.3 & 58.2 \\ \midrule
\multirow{15}{*}{\makecell{(4b) \\ Varying}} & \multirow{5}{*}{\makecell{(3a) \\ Large}} & \multirow{5}{*}{10.0}  & Bias  & \multicolumn{6}{c}{0.20}          \\
                    &                    &                        & SD    & \multicolumn{6}{c}{3.45}         \\
                    &                    &                        & SE    & 3.40 & 3.33 & 3.43 & 3.42 & 4.66 & 3.33  \\
                    &                    &                        & CP    & 94.6 & 94.0 & 94.9 & 94.9 & 99.0 & 94.4  \\
                    &                    &                        & Power & 84.8 & 85.9 & 84.6 & 84.3 & 61.9 & 85.0  \\ \cmidrule(l){2-10} 
                    & \multirow{5}{*}{\makecell{(3b) \\ Sparse}} & \multirow{5}{*}{9.84}  & Bias  & \multicolumn{6}{c}{-0.15}         \\
                    &                    &                        & SD    & \multicolumn{6}{c}{3.75}          \\
                    &                    &                        & SE    & 3.34 & 3.61 & 3.66 & 3.69 & 4.83 & 3.61  \\
                    &                    &                        & CP    & 91.5 & 94.0 & 94.2 & 94.1 & 98.8 & 94.4  \\
                    &                    &                        & Power & 80.4 & 76.7 & 75.7 & 75.1 & 52.1 & 76.5  \\ \cmidrule(l){2-10} 
                    & \multirow{5}{*}{\makecell{(3c) \\ Mixed}} & \multirow{5}{*}{10.1}  & Bias  & \multicolumn{6}{c}{-0.14}        \\ 
                    &                    &                        & SD    & \multicolumn{6}{c}{3.69}     \\
& & & SE  & 3.41  & 3.53  & 3.60  & 3.60  & 4.76  & 3.53  \\
& & & CP  & 92.3  & 92.9  & 93.6  & 93.3  & 99.0  & 93.5  \\
& & & Power & 81.1  & 79.7  & 78.9  & 79.0  & 55.4  & 79.4  \\ \midrule
\multirow{15}{*}{\makecell{(4c) \\ Varying}} & \multirow{5}{*}{\makecell{(3a) \\ Large}} & \multirow{5}{*}{-8.99} & Bias  & \multicolumn{6}{c}{0.01}         \\
                    &                    &                        & SD    & \multicolumn{6}{c}{4.15}         \\
                    &                    &                        & SE    & 4.18 & 4.67 & 4.21 & 4.22 & 4.49 & 4.67  \\
                    &                    &                        & CP    & 94.7 & 97.5 & 94.8 & 95.5 & 97.1 & 97.9  \\
                    &                    &                        & Power & 55.8 & 47.4 & 55.4 & 54.7 & 51.5 & 45.9  \\ \cmidrule(l){2-10} 
                    & \multirow{5}{*}{\makecell{(3b) \\ Sparse}} & \multirow{5}{*}{-32.3} & Bias  & \multicolumn{6}{c}{-0.12}       \\
                    &                    &                        & SD    & \multicolumn{6}{c}{4.21}          \\
                    &                    &                        & SE    & 3.83 & 4.78 & 4.17 & 4.23 & 4.51 & 4.78  \\
                    &                    &                        & CP    & 93.2 & 97.6 & 95.4 & 95.3 & 96.8 & 97.8  \\
                    &                    &                        & Power & 100  & 100  & 100  & 100  & 100  & 100   \\ \cmidrule(l){2-10} 
                    & \multirow{5}{*}{\makecell{(3c) \\ Mixed}} & \multirow{5}{*}{-20.4} & Bias  & \multicolumn{6}{c}{0.15}        \\
                    &                    &                        & SD    & \multicolumn{6}{c}{4.20}          \\
&                           &                        & SE    & 4.01 & 4.77 & 4.20 & 4.23 & 4.53 & 4.77 \\
&                                            &                        & CP    & 93.8 & 97.4 & 95.0 & 95.5 & 96.8 & 97.6 \\
&                                            &                        & Power & 99.6 & 98.7 & 99.5 & 99.4 & 99.0 & 98.6 \\ \bottomrule 
  \end{tabular}
}                  
                    
                    \begin{flushleft} 
  \footnotesize
  Note: For the two score‐based methods (Newc and Klin), 
  the SEs are calculated by dividing the average interval length 
  by \(2\times1.96=3.92\).
\end{flushleft} 

\end{table}  

Here is a summary of the results in Table \ref{tb: MH1a2a} (results from the other tables are similar): 
\begin{itemize}
    \item The MH risk difference estimator has negligible biases for estimating $\delta_{\mh}$. 
    \item Both the $\text{mGR}_{\mh}$ and (infeasible) bootstrap variance estimator give SEs that are close to the SDs in all cases, including large-, sparse-, and mixed-stratum, and under both common and {varying} risk differences. As a result, the CPs are all close to the nominal 95\%.  When risk differences are homogeneous (Factor 4a), 
  both methods achieve power comparable to Sato’s estimator, indicating no noticeable loss of power even under homogeneity.
    \item The GR variance estimator performs well under large-stratum asymptotics but tends to underestimate under sparse- or mixed-stratum asymptotics.
    \item {The Sato and Klingenberg variance estimators perform similarly. They perform well under Factor (4a), where the common risk difference assumption holds, but become conservative under Factor (4c), leading to reduced power. For instance, under Factor (4c), both yield CPs around 97.5\% and up to 8\% lower power compared to $\text{mGR}_{\mh}$.
    Although generally conservative when the common risk difference assumption is violated, they can be slightly anti-conservative in some cases (e.g., under Factor 4b).
    \item The stratified Newcombe variance estimator performs well under Factor (4a) when the common risk difference assumption holds. However, under Factors (4b) and (4c), it becomes conservative, with SEs 1.07–1.35 times the SD and CPs ranging from 96.8\% to 99.0\%. This leads to up to 23.6\% lower power compared to $\text{mGR}_{\mh}$.}
\end{itemize}

\subsection{MH risk difference estimator as an estimator for $\delta_{\rm ATE}$} 
\label{subsec: cmh ate sim}

We next consider the estimand $\delta_{\rm ATE}$. Unlike $\delta_{\mh}$, the ATE remains fixed across runs. We evaluate our proposed $\text{mGR}_{\rm ATE}$ variance estimator alongside other variance estimators of $\hat \delta$,   although only $\text{mGR}_{\rm ATE}$ and the bootstrap are applicable to the ATE. The bootstrap variance estimator (200 iterations) is implemented to estimate the variance of $\hat\delta$, which is practically feasible. For comparison, we also include standard ATE estimators: the unadjusted difference in means, the G-computation estimator \citep{bannick2025general} with logistic regression (from the \textsf{RobinCar R} package), and the PS estimator in Section 4.2. Table \ref{tb: ATE1a2a} presents results for $n=500$ (Factor 1a) and unbalanced allocation (Factor 2a), including average bias ($\hat\delta - \delta_{\rm ATE}$), SD of $\hat\delta$, average SE, empirical power for testing $\delta_{\rm ATE} = 0$, and the probability of the 95\% confidence interval covering $ \delta_{\rm ATE}$. More results are in the Supplement.

\begin{table}[ht]   
\centering  
\caption{Simulation results for estimating the ATE $\delta_{\rm ATE}$ under unbalanced allocation (Factor 2a), based on $n=500$ (Factor 1a) and 1,000 simulation runs. Entries have all been multiplied by 100. \label{tb: ATE1a2a}}   
\centering
\resizebox{\textwidth}{!}{

\begin{tabular}{@{}ccccccccccccc@{}}
\toprule
Effect & Regime & Truth &  & GR & Sato & mGR & Boot & Newc & Klin & Gcomp & Unadj & PS \\
\midrule
\multirow{15}{*}{\makecell{(4a) \\ Common}} & \multirow{5}{*}{\makecell{(3a) \\ Large}} & \multirow{5}{*}{-10.0} & Bias & \multicolumn{6}{c}{0.21}  & 0.20 & 0.29 & 0.20 \\
 &  &  & SD & \multicolumn{6}{c}{4.39}  & 4.40 & 4.60 & 4.40 \\
 &  &  & SE & 4.34 & 4.37 & 4.37 & 4.37 & 4.64 & 4.37 & 4.36 & 4.68 & 4.38 \\
 &  &  & CP & 94.4 & 94.4 & 94.4 & 94.3 & 96.0 & 94.7 & 94.4 & 95.3 & 94.4 \\
 &  &  & Power & 61.5 & 61.2 & 60.8 & 61.2 & 56.2 & 61.0 & 61.1 & 54.8 & 60.8 \\ \cmidrule(l){2-13}
 & \multirow{5}{*}{\makecell{(3b) \\ Sparse}} & \multirow{5}{*}{-10.0} & Bias & \multicolumn{6}{c}{0.01} & 0.02 & -0.05 & 0.07 \\
 &  &  & SD & \multicolumn{6}{c}{4.86}  & 4.98 & 4.74 & 4.95 \\
 &  &  & SE & 4.37 & 4.77 & 4.77 & 4.80 & 4.82 & 4.77 & 4.47 & 4.72 & 4.79 \\
 &  &  & CP & 92.2 & 94.4 & 94.4 & 94.6 & 94.9 & 94.7 & 91.9 & 95.1 & 93.6 \\
 &  &  & Power & 61.5 & 54.9 & 54.6 & 53.7 & 53.9 & 53.8 & 59.0 & 54.0 & 54.1 \\ \cmidrule(l){2-13}
 & \multirow{5}{*}{\makecell{(3c) \\ Mixed}} & \multirow{5}{*}{-10.0} & Bias & \multicolumn{6}{c}{0.00}  & -0.01 & -0.02& 0.01 \\
 &  &  & SD & \multicolumn{6}{c}{4.52}  & 4.60 & 4.69 & 4.59 \\
 &  &  & SE    & 4.34 & 4.56 & 4.57 & 4.57 & 4.76 & 4.56 & 4.39  & 4.73  & 4.59 \\
&  &  & CP    & 94.9 & 96.0 & 96.1 & 95.9 & 96.9 & 96.0 & 94.9  & 94.9  & 95.6 \\
&  &  & Power & 62.3 & 58.7 & 58.6 & 58.9 & 56.3 & 58.2 & 60.6  & 56.2  & 56.9 \\ \midrule
\multirow{15}{*}{\makecell{(4b) \\ Varying}} & \multirow{5}{*}{\makecell{(3a) \\ Large}} & \multirow{5}{*}{10.0} & Bias & \multicolumn{6}{c}{0.20} & 0.19 & 0.28 & 0.19 \\
 &  &  & SD & \multicolumn{6}{c}{3.49}  & 3.47 & 4.51 & 3.47 \\
 &  &  & SE & 3.40 & 3.33 & 3.46 & 3.46 & 4.66 & 3.33 & 3.44 & 4.68 & 3.45 \\
 &  &  & CP & 93.5 & 92.9 & 93.8 & 94.2 & 98.6 & 93.3 & 94.6 & 94.5 & 94.4 \\
 &  &  & Power & 84.8 & 85.9 & 83.9 & 83.9 & 61.9 & 85.0 & 84.9 & 58.5 & 84.5 \\ \cmidrule(l){2-13}
 & \multirow{5}{*}{\makecell{(3b) \\ Sparse}} & \multirow{5}{*}{10.5} & Bias & \multicolumn{6}{c}{-0.12}  & -0.10 & -0.08 & -0.13 \\
 &  &  & SD & \multicolumn{6}{c}{3.77} & 3.88 & 4.79 & 3.87 \\
 &  &  & SE & 3.34 & 3.61 & 3.69 & 3.70 & 4.83 & 3.61 & 3.44 & 4.73 & 3.70 \\
 &  &  & CP & 91.7 & 94.2 & 94.7 & 93.8 & 98.7 & 94.4 & 92.2 & 94.6 & 94.0 \\
 &  &  & Power & 80.4 & 76.7 & 75.6 & 74.9 & 52.1 & 76.5 & 78.1 & 54.8 & 73.5 \\ \cmidrule(l){2-13}
 & \multirow{5}{*}{\makecell{(3c) \\ Mixed}} & \multirow{5}{*}{10.1} & Bias & \multicolumn{6}{c}{-0.10} & -0.16 & -0.22 & -0.16 \\
 &  &  & SD & \multicolumn{6}{c}{3.74} & 3.83 & 4.81 & 3.82 \\
 &  &  & SE    & 3.41  & 3.53  & 3.63  & 3.63  & 4.76  & 3.53  & 3.48  & 4.71  & 3.64 \\
&   &   & CP    & 91.7  & 93.0  & 93.5  & 93.7  & 98.9  & 93.1  & 91.7  & 94.4  & 92.6 \\
&  & & Power & 81.1  & 79.7  & 78.2  & 78.1  & 55.4  & 79.4  & 79.6  & 54.8  & 77.2 \\ \midrule
\multirow{15}{*}{\makecell{(4c) \\ Varying}} & \multirow{5}{*}{\makecell{(3a) \\ Large}} & \multirow{5}{*}{-9.00} & Bias & \multicolumn{6}{c}{0.03}  & -0.02 & -0.06 & -0.02 \\
 &  &  & SD & \multicolumn{6}{c}{4.43}  & 4.37 & 4.43 & 4.37 \\
 &  &  & SE & 4.18 & 4.67 & 4.51 & 4.54 & 4.49 & 4.67 & 4.39 & 4.51 & 4.41 \\
 &  &  & CP & 94.0 & 96.1 & 95.5 & 96.0 & 95.6 & 96.4 & 96.0 & 95.6 & 95.8 \\
 &  &  & Power & 55.8 & 47.4 & 50.4 & 49.6 & 51.5 & 45.9 & 53.1 & 49.5 & 53.2 \\ \cmidrule(l){2-13}
 & \multirow{5}{*}{\makecell{(3b) \\ Sparse}} & \multirow{5}{*}{-32.4} & Bias & \multicolumn{6}{c}{-0.12}  & -0.18 & -0.07 & 0.05 \\
 &  &  & SD & \multicolumn{6}{c}{4.62}  & 4.52 & 4.36 & 4.50 \\
 &  &  & SE & 3.83 & 4.78 & 4.60 & 4.70 & 4.51 & 4.78 & 4.17 & 4.39 & 4.41 \\
 &  &  & CP & 89.7 & 95.8 & 94.8 & 94.3 & 94.9 & 96.4 & 92.8 & 95.6 & 94.6 \\
 &  &  & Power & 100 & 100 & 100 & 100 & 100 & 100 & 100 & 100 & 100 \\ \cmidrule(l){2-13}
 & \multirow{5}{*}{\makecell{(3c) \\ Mixed}} & \multirow{5}{*}{-20.8} & Bias & \multicolumn{6}{c}{0.48}  & 0.13 & 0.07 & 0.22 \\
 &  &  & SD & \multicolumn{6}{c}{4.64}  & 4.57 & 4.53 & 4.58 \\
&  &  & SE    & 4.01 & 4.77 & 4.61 & 4.67 & 4.53 & 4.77 & 4.30 & 4.47 & 4.45 \\
& & & CP    & 89.2 & 95.3 & 94.3 & 94.3 & 94.3 & 95.7 & 93.1 & 94.4 & 94.0 \\
&  &  & Power & 99.6 & 98.7 & 98.9 & 99.1 & 99.0 & 98.6 & 99.6 & 99.6 & 99.4 \\  \bottomrule 

\end{tabular}}
\begin{flushleft}      
  \footnotesize
  Note: For the two score‐based methods (Newc and Klin), 
  the SEs are calculated by dividing the average interval length 
  by \(2\times1.96=3.92\).
\end{flushleft}

\end{table}

The conclusions for the first six variance estimators are similar to earlier findings. The proposed $\text{mGR}_{\rm ATE}$ and  the bootstrap variance estimator perform similarly, achieving CPs near the nominal level across all scenarios. This demonstrates their adaptability to different asymptotic regimes and robustness under both common and varying risk differences. Compared to the bootstrap, $\text{mGR}_{\rm ATE}$ offers similar performance with lower computational cost.  As expected, the GR estimator underestimates variance even under large-stratum asymptotics, as it ignores the $\nu_n^2$ term (see Theorem \ref{theo: ate}). Under Factor (4c), Sato and Klingenberg overestimate $\sigma_n^2$ but also omit $\nu_n^2$, which is why the CPs are not too large. Their undercoverage is more evident under Factors (4b) and (3a), with CPs of 92.9\% and 93.3\%, respectively. The stratified Newcombe estimator remains overly conservative under Factor (4b).

The unadjusted estimator, G-computation, and PS estimator all show negligible bias when estimating the ATE. G-computation and PS yield nearly identical point estimates, biases, and SDs, as both adjust for stratum indicators using saturated models. 

Under large-stratum asymptotics, covariate-adjusted estimators (MH, G-computation, and PS estimator) have smaller SDs than the unadjusted estimator. G-computation and PS show comparable SDs to MH under Factors 4a–4b and slightly smaller SDs under 4c. In sparse or mixed-stratum settings, covariate-adjusted estimators may have either smaller (Factor 4b) or larger SDs (Factors 4a and 4c) than the unadjusted estimator, suggesting that adjusting for many strata can sometimes reduce efficiency. G-computation and PS estimators have larger SDs than MH under Factors 4a–4b but smaller under 4c. Thus, in sparse or mixed settings, no method consistently outperforms the others in efficiency.

The CPs of the unadjusted estimator are close to the nominal 95\% across all scenarios. However, the SE for the G-computation estimator tends to underestimate the true SD under sparse or mixed-stratum settings, leading to CPs as low as 91.9\%. The proposed variance estimator for the PS estimator generally performs well, but it also underestimates the SD in certain cases, such as under Factor 4b and 3c. 

We also repeat the simulations 100 times to generate boxplots of the coverage probabilities, shown in Figure \ref{fig:cp_comp}. The results are consistent with those presented in the table.

%

\begin{figure}[ht]
  \centering

  \begin{subfigure}[b]{0.49\textwidth}
    \includegraphics[width=\linewidth]{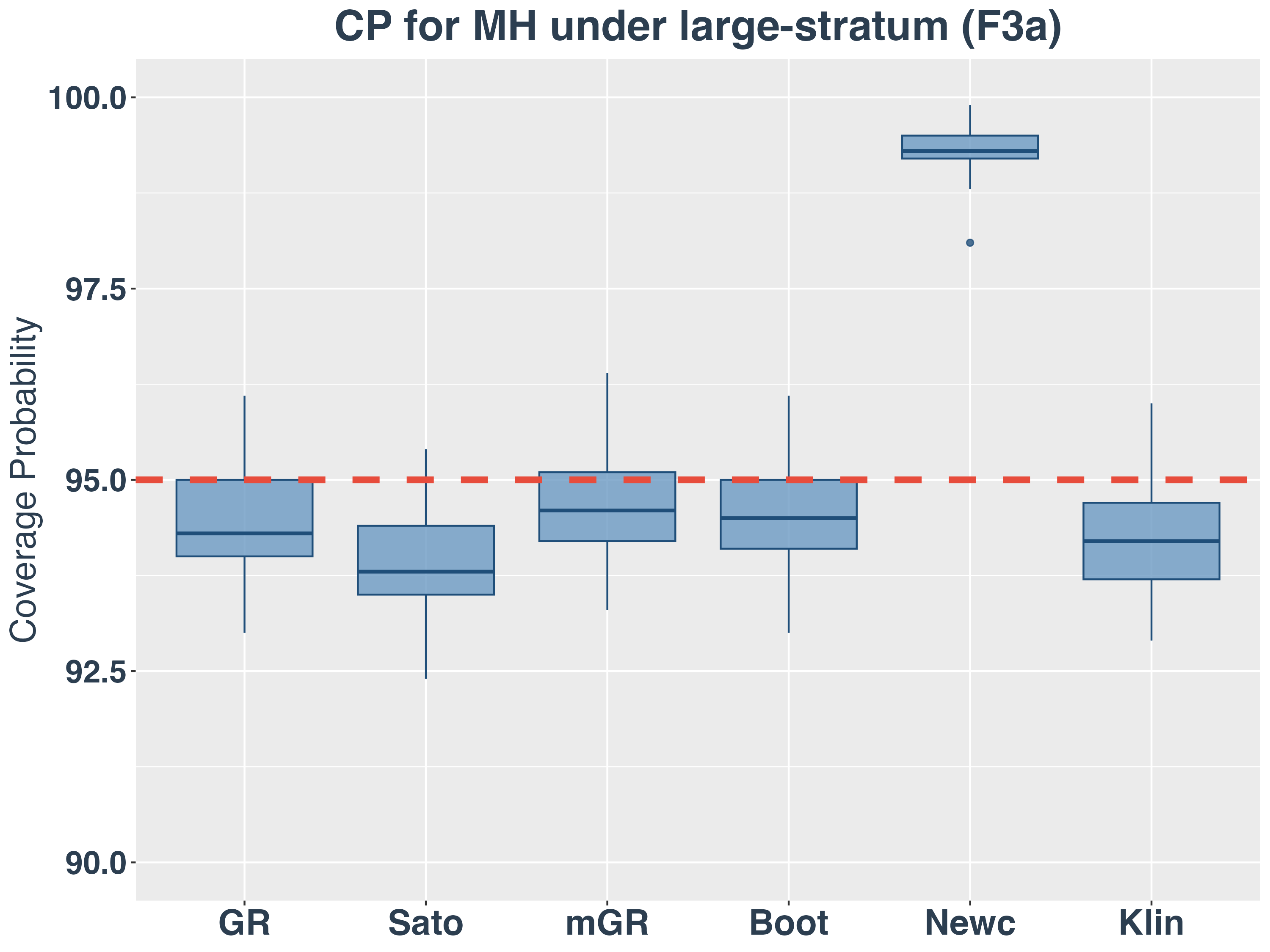}
    \label{fig:sub1}
  \end{subfigure}\hfill
  \begin{subfigure}[b]{0.49\textwidth}
    \includegraphics[width=\linewidth]{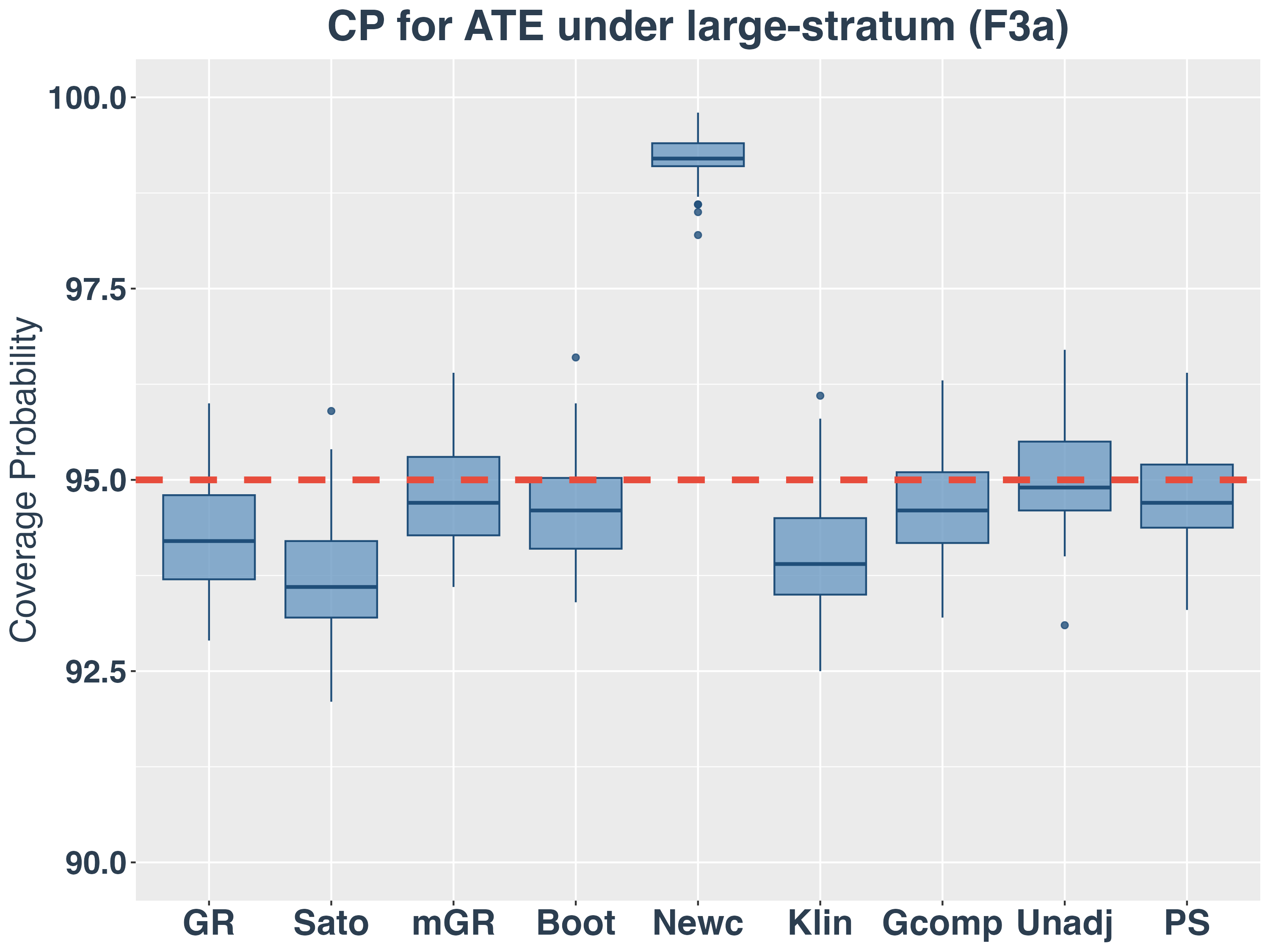}
    \label{fig:sub2}
  \end{subfigure}

  \vspace{1em} 
  \begin{subfigure}[b]{0.49\textwidth}
    \includegraphics[width=\linewidth]{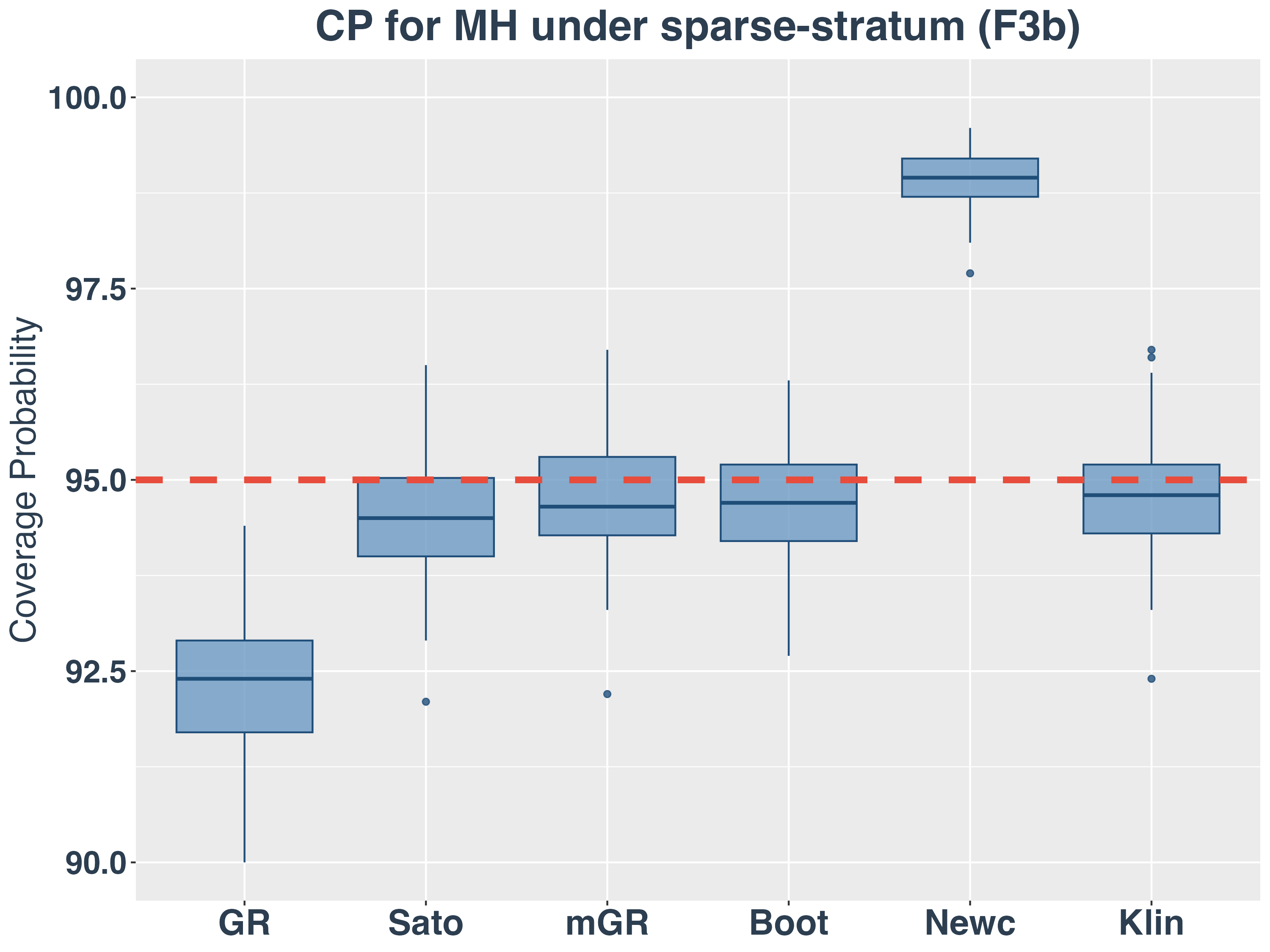}
    \label{fig:sub3}
  \end{subfigure}\hfill
  \begin{subfigure}[b]{0.49\textwidth}
    \includegraphics[width=\linewidth]{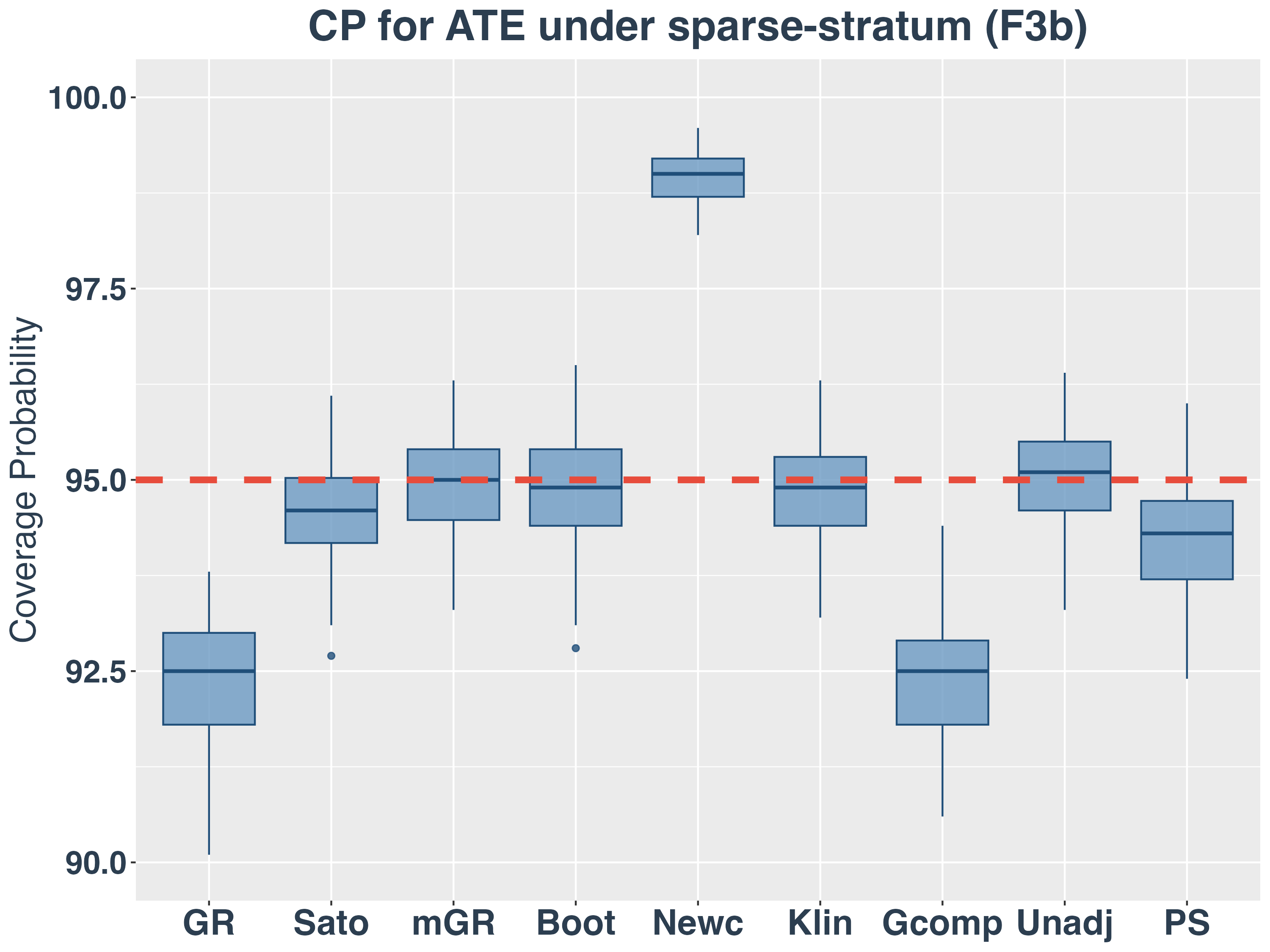}
    \label{fig:sub4}
  \end{subfigure}

  \vspace{1em}
  \begin{subfigure}[b]{0.49\textwidth}
    \includegraphics[width=\linewidth]{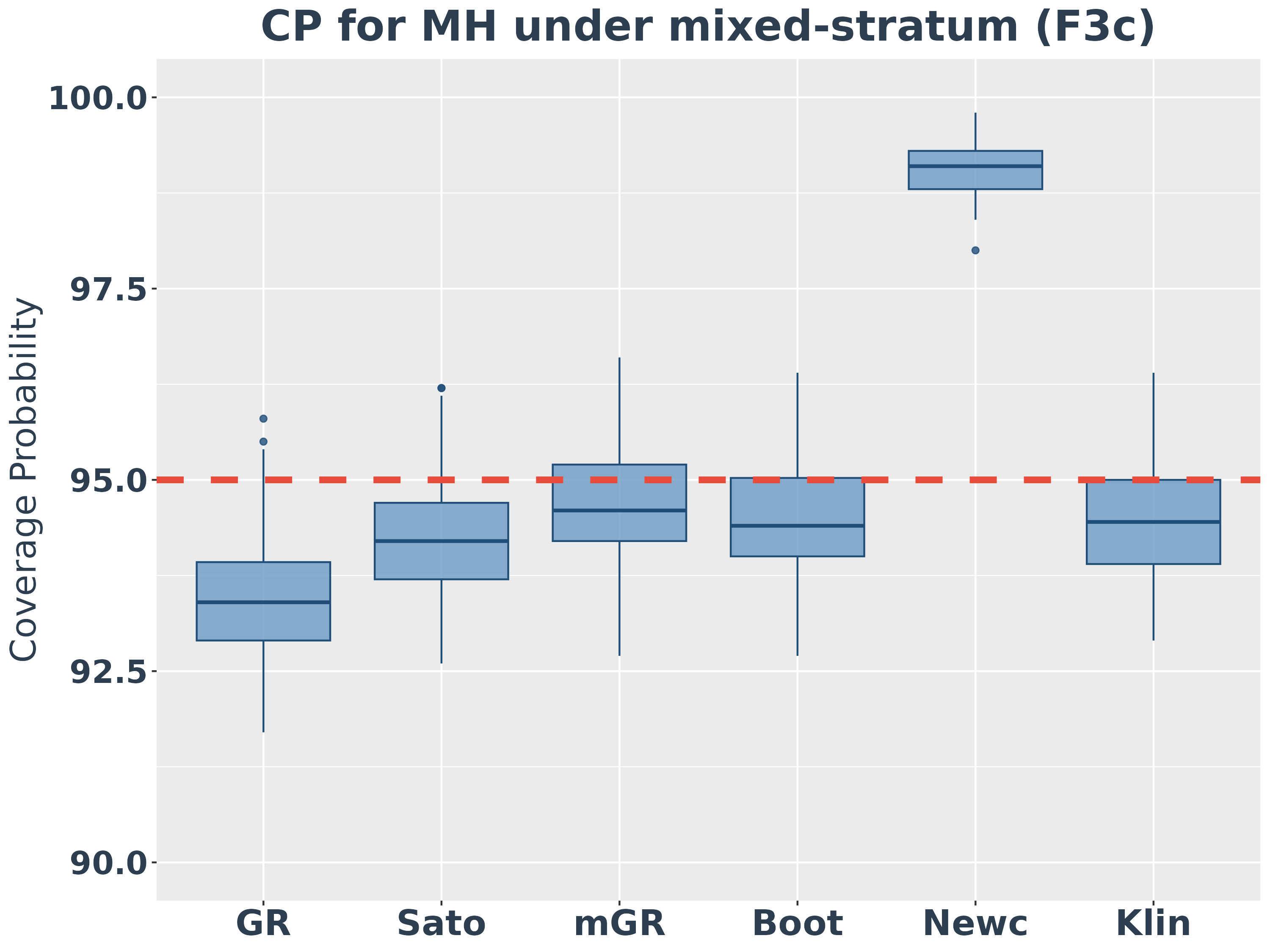}
    \label{fig:sub5}
  \end{subfigure}\hfill
  \begin{subfigure}[b]{0.49\textwidth}
    \includegraphics[width=\linewidth]{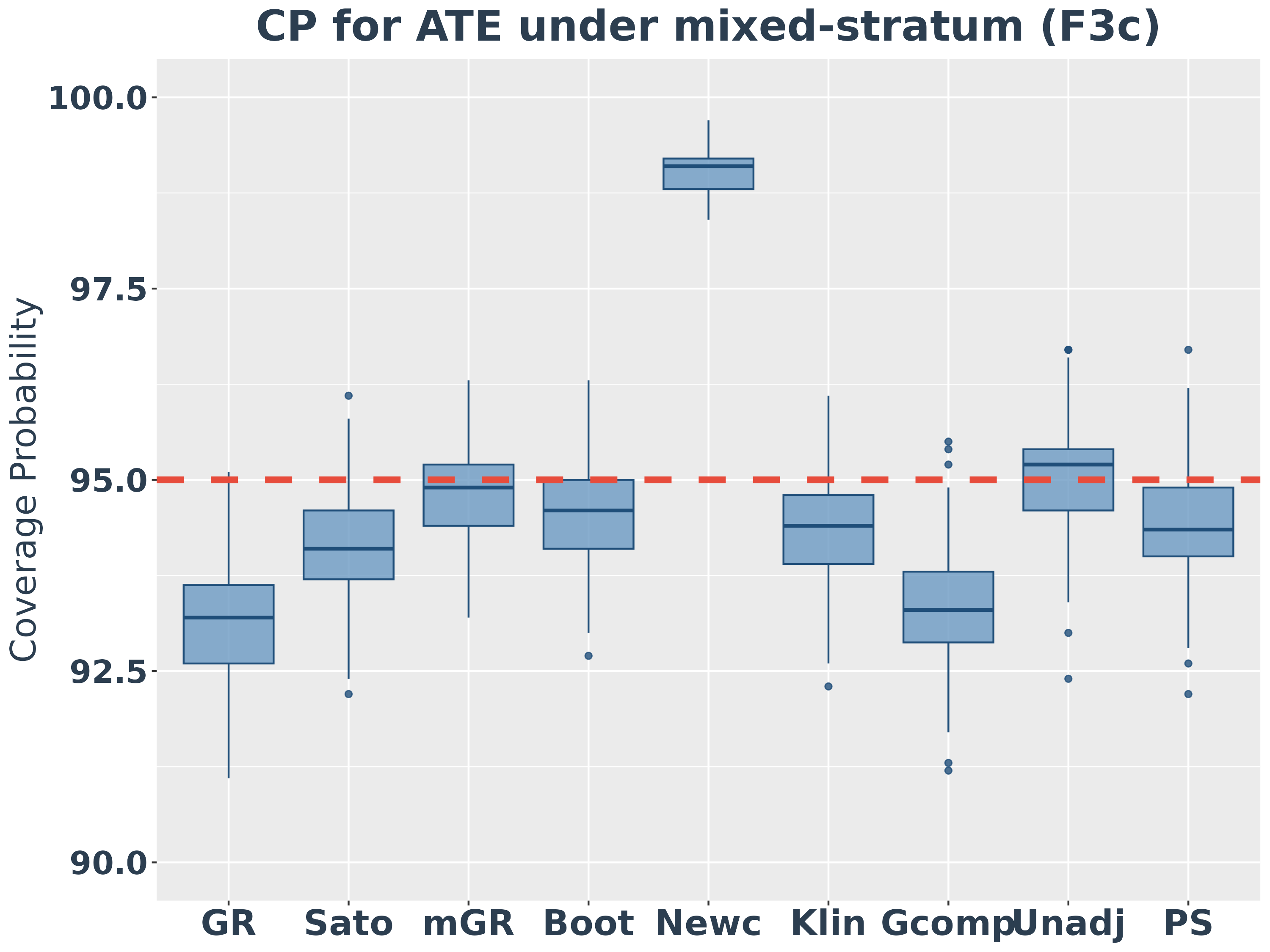}
    \label{fig:sub6}
  \end{subfigure}

  \caption{Coverage probability under different asymptotic regimes with sample size $n=500$ (Factor 1a), unbalanced treatment allocation (Factor 2a), and varying treatment effect (Factor 4b). Each boxplot represents the distribution of coverage probabilities computed from 100 repetitions, where each coverage probability is estimated based on 1,000 simulation runs.}
  \label{fig:cp_comp}
\end{figure}

\section{A clinical trial example}
\label{sec: real data}

We apply our proposed variance estimators to a Cancer and Leukemia Group B (CALGB) 
randomized clinical trial described in  \cite{Klingenberg2014}. This trial enrolled $n=156$ patients across $K=21$ institutions and compared two chemotherapy treatments using a binary survival outcome (lived/died by the end of the study) in patients with multiple myeloma. On average, each institution had 7.4 patients. The smallest stratum has 4 patients, in which two patients were randomized to each treatment.  The data are available in the Appendix of \cite{Klingenberg2014} and also included in the Supplement.


\begin{table}[ht]   
\centering  
\caption{Analysis results in the CALGB trial, all the entries multiplied by 100. ``–'' indicates that the method is not applicable.
\label{tb: real data results} }  
\centering
\resizebox{\textwidth}{!}{
\begin{tabular}{@{}ccccccccccc@{}}
\toprule
          &          & GR            & Sato         & mGR          & Newc         & Klin         & Boot         & Gcomp        & Unadj        & PS           \\ \midrule
\multirow{3}{*}{$\delta_{\mh}$}  & Est & 5.72 & 5.72& 5.72& 5.72&5.72    & -&-&-&-                                                            \\
                     & SE       & 6.32          & 7.99         & 7.30         & 7.98         & 7.99         &    -&-&-&-                                          \\
                     & 95\% CI       & (-6.67,18.10) & (-9.94,21.4) & (-8.60,20.0) & (-10.1,21.2) & (-10.2,21.1) &    -&-&-&-                             \\ \midrule
\multirow{3}{*}{$\delta_{\rm ATE}$} & Est  & -&-      & 5.72 &-&-&5.72                                                           & 5.69         & 1.79         & 5.69         \\
                     & SE       & -          & -        & 7.74         & -          & -          & 7.74         & 7.18         & 8.06         & 7.66         \\
                     & 95\% CI       & - & - & (-9.46,20.9) & - & - & (-9.46,20.9) & (-8.38,19.8) & (-14.0,17.6) & (-9.33,20.7) \\\bottomrule 
\end{tabular}
}
\end{table}

{\black
Table \ref{tb: real data results} presents the MH risk difference and other estimators for the ATE $\delta_{\rm ATE}$, including point estimates, standard errors, and 95\% confidence intervals. In this example, small stratum sizes invalidate the GR variance estimator, likely explaining its smaller values relative to others. The high variability in $\hat\delta_k$ also renders the Sato, stratified Newcombe, and Klingenberg estimators overly conservative. In contrast, our proposed mGR$_{\mh}$ estimator performs robustly across asymptotic regimes, regardless of risk difference heterogeneity.

Given the substantial heterogeneity in risk differences across institutions, $\delta_{\mh}$ represents a weighted average of institution-specific effects. However, the clinical relevance of these weights is unclear, making $\delta_{\mh}$ less interpretable. In contrast, the ATE offers a more interpretable estimand, reflecting the population risk difference. For the ATE estimand,  the mGR$_{\rm ATE}$ and the bootstrap variance estimators are nearly identical and slightly larger than mGR$_{\mh}$, as expected since they account for additional variability (see Theorem \ref{theo: ate}). We also report results from the G-computation, unadjusted, and PS estimators along with their variances. G-computation and PS yield identical point estimates (0.0569), similar to that of the MH estimator.  However, G-computation’s variance estimator is smaller—likely underestimating the true variance under sparse-stratum settings—while the PS variance is more aligned with mGR$_{\rm ATE}$ and bootstrap. The unadjusted estimator yields a lower point estimate (0.018) and a larger standard error, due to the influence of chance imbalance.}

\section{Discussion}
\label{sec: discussion}

We have extensively investigated the estimand and variance estimation for the MH risk difference estimator without assuming a common risk difference, under large-stratum, sparse-stratum, and mixed asymptotic regimes. This work yields several novel insights that enhance the understanding and applicability of the MH estimator in randomized clinical trials. Our main findings, all derived without assuming a common risk difference, are summarized below:
\begin{description}
\item 1. The MH risk difference estimator consistently estimates either $\delta_{\mh} $ or the ATE $\delta_{\rm ATE}$. Without assuming a common risk difference, the ATE—representing the population risk difference in the trial population—is generally more interpretable and widely accepted, as it does not depend on the choice or construction of strata. 
\item 2. The asymptotic variance of the MH risk difference estimator depends on the target estimand. When targeting the ATE, the variance is generally larger than when targeting $\delta_{\mh}$, as it accounts for variability in stratum sizes and marginal sums, reflecting the nature of randomized clinical trials where these numbers are often not fixed. 
\item 3. For either estimand, we propose a unified robust variance estimator applicable across large-, sparse-, and mixed-stratum regimes without assuming a common risk difference. This simplifies implementation by removing the need to choose an asymptotic regime or variance estimator. Simulations show that our mGR$_{\mh}$ and mGR$_{\rm ATE}$ estimators outperform commonly used alternatives, including the GR and Sato estimators and score-based methods by Klingenberg and stratified Newcombe, across a range of realistic scenarios.
\item 4. Unlike methods such as G-computation, which typically require a small number of adjustment variables for valid inference,
the MH risk difference estimator paired with our mGR$_{\rm ATE}$ variance estimator remains valid in sparse or mixed-stratum settings.
\item 5. We show that the MH test may be invalid under treatment effect heterogeneity. We also propose a unified robust variance estimator for the PS estimator that is valid across asymptotic regimes. We also discuss an extension to multiple arms in the Supplement.
\end{description}  
Our results offer two practical guidelines for covariate adjustment using the MH risk difference estimator. First, in randomized trials, stratification is primarily used to improve efficiency rather than adjust for confounding, as in observational studies or meta-analyses. Since variance estimation requires at least two participants per arm in each stratum, overly fine stratification should be avoided—supporting the common practice of pooling small strata in analysis.
 {Second, as reviewed in the Supplement, many trials use the MH test for hypothesis testing and separately report treatment effect estimates. However, under treatment effect heterogeneity, the test and estimator may target different estimands and rely on different assumptions, potentially leading to inconsistencies between p-values and confidence intervals. To address this, we recommend a unified approach: use the MH risk difference estimator with the mGR$_{\rm ATE}$ variance estimator for both testing and interval estimation to ensure consistency and interpretability.}

Several important topics related to the MH methods are beyond the scope of this paper. These include variance estimation for matched pair studies \citep{Sato1989}, rare outcomes \citep{hansen2023exact}, MH risk ratio and odds ratio estimators \citep{breslow1982variance, Greenland1985,  robins_breslow1986},  categorical outcomes \citep{generalization_MH, catedata_analysis}, adjustment for multiple covariates 
\citep{Liang_generalization}, and general estimating equation frameworks \citep{Davis1985, FUJII_generalize20051}. {
Another appealing property of MH methods is their consistency under certain models with dependent outcomes \citep{liang1985, Liang_generalization}. Whether this extends to the MH risk difference, and how to modify the variance formula to account for such dependencies, remains an open question.}

\section*{Data Availability}
The dataset of the Cancer and Leukemia Group B (CALGB) is available at \url{https://sites.williams.edu/bklingen/files/2013/06/myel.txt}.

\bibliographystyle{apalike} 
\bibliography{export}

\end{document}


\title{Supplementary Materials for ``Clarifying Mantel-Haenszel Risk Difference Estimator in Randomized Clinical Trials'' by}
\author{Xiaoyu Qiu$^{1*}$, Yuhan Qian$^{2*}$, Jaehwan Yi$^3$, Jinqiu Wang$^4$, Yu Du$^5$, Yanyao Yi$^5$, Ting Ye$^2$}

\affil{
$^1$Department of Statistics, University of Michigan \\
$^2$Department of Biostatistics, University of Washington\\
$^3$Department of Statistics, Pennsylvania State University\\
$^4$Newark Academy\\
$^5$Global Statistical Sciences, Eli Lilly and Company \\
$^*$Equal contribution
}

\maketitle
\thispagestyle{empty}

\setcounter{equation}{0}
\setcounter{table}{0}
\setcounter{section}{0}
\renewcommand{\theequation}{S\arabic{equation}}
\renewcommand{\thetable}{S\arabic{table}}
\renewcommand{\thefigure}{S\arabic{figure}}
\renewcommand{\thesection}{S\arabic{section}}
\allowdisplaybreaks

\begin{abstract}
Section S1 discusses additional results on methods, Section S2 presents additional simulation and real data analysis results, and Section S3 provides technical proofs.
\end{abstract}

\section{Additional results on methods}
\subsection{A review of MH methods in practice}

{\red The MH methods are commonly used in clinical trials with binary endpoints. A search of the New England Journal of Medicine on April 15, 2025 (using the keyword “Mantel-Haenszel”) identified 103 publications over the past five years. We review nine papers published in the New England Journal of Medicine over the past six months that use the MH test: \cite{dimopoulos2024daratumumab, humbert2025sotatercept, blockmans2025phase, furie2025efficacy, fiorella2025embolization, brown2025fixed, liu2024middle, hou2024xalnesiran, facon2024isatuximab}.  Many of these papers used the MH test for formal hypothesis testing and reported treatment effects as risk differences, odds ratios, or risk ratios. However, only two papers explicitly stated in the main text how the treatment effect estimates and confidence intervals were calculated: \cite{liu2024middle} used the MH risk difference with confidence intervals computed using the Miettinen–Nurminen method, and \cite{furie2025efficacy} used the MH risk difference with confidence intervals computed using the stratified Newcombe method. This emphasizes the need for clearer understanding of both the estimand and variance estimation.}

\subsection{Discussion of Cochran's test and MH test}

{\red 
   An important distinction between the Cochran and MH statistics lies in their respective likelihood frameworks and behavior under sparse-stratum asymptotics.  The Cochran statistic is a Wald-type statistic derived from the unconditional binomial likelihood, with stratum-specific intercepts treated as nuisance parameters and estimated via maximum likelihood \citep{Cochran1954}. However, as the number of strata increases, this approach suffers from the classical Neyman–Scott problem: the number of nuisance parameters grows with the sample size, leading to inconsistent maximum likelihood estimators. Consequently, unconditional binomial likelihood-based methods are inconsistent in sparse-stratum settings \citep{Greenland1985}. In contrast, the MH statistic, as shown by \citet{day1979testing}, is the score statistic from the conditional logistic likelihood, which eliminates the stratum-specific nuisance parameters through conditioning. While this conditioning introduces a small finite-sample efficiency loss, the loss is small and vanishes under standard large-stratum asymptotics and in sparse-stratum limits when binomial parameters approach zero, as discussed by \citet[Section 5]{Sprott1975Marginal} and \citet{Greenland1985}. Crucially, under both large- and sparse-stratum asymptotics, the MH (score), likelihood ratio, and Wald tests derived from the conditional likelihood remain consistent for testing any fixed odds ratio, more than compensating for the slight efficiency loss in sparse-stratum settings. 

These distinctions underscore the importance of clearly differentiating the MH and Cochran statistics. While their formulas may appear similar, they have fundamentally different inferential properties in finite samples and under sparse-stratum conditions.
}

\subsection{Extension of MH risk difference to multiple treatments}
\citet{Mickey_generalization} and \citet{generalization_MH} discussed extensions of the Mantel-Haenszel methods to settings with multiple treatment arms. In a similar spirit, our results can also be extended to accommodate multiple treatments ($J$ treatment arms).
\begin{table}[H]
    \centering
        \caption{Summary of $2\times J$ table for the $k^\text{th}$ stratum}
    \label{Table: multiple trt}
    \begin{tabular}{lcccc}
    \hline
      Treatment   &  A &B & $\cdots$ &    Total   \\
         \hline 
    Responder & $n_{11k}$ & $n_{12k}$ && $n_{1.k}$ \\
    Non-responder & $n_{01k}$ & $n_{02k}$ & $\cdots$ &$n_{0.k}$ \\
     Total    & $n_{.1k}$  & $n_{.2k}$  & &$n_{..k}$  \\
         \hline
        \multicolumn{4}{l}{\footnotesize Responder: $Y_i=1$; Non-responder: $Y_i=0$}\\
    \end{tabular}
\end{table}

Denote the stratum-specific risk difference between treatments \( j \) and \( l \) by \( \delta_{k}^{jl} = p_{1jk} - p_{1lk} \), and the corresponding stratum weight by \( w_{k}^{jl} = n_{.jk} n_{.lk} / n_{..k} \). The MH risk difference between treatments \( j \) and \( l \) is then defined as  $\delta^{jl}_{\mh} = (\sum_{k=1}^K \delta_{k}^{jl}w_{k}^{jl})/(\sum_{k=1}^K w_k^{jl})$ and the ATE is  
$\delta^{jl}_{\rm ATE} = E(Y^{(j)}) -E(Y^{(l)})$. The MH risk difference estimator is given by
\begin{align*}
    \hat{\delta}^{jl} = \frac{\sum_{k=1}^K w_k^{jl}\hat{\delta_k}^{jl}}{\sum_{k=1}^K w_k^{jl}} = \frac{\sum_{k=1}^K (n_{1jk}n_{.lk} - n_{1lk}n_{.jk})/n_{..k}}{\sum_{k=1}^K w_{k}^{jl}},
\end{align*}
where $\hat\delta_k^{jl} = n_{1jk}/n_{.jk} - n_{1lk}/n_{.lk}$.
The results established in Theorems \ref{theo:1} and \ref{theo: ate} for binary treatment settings still hold in the multiple-treatment setting, because the proofs of Theorems \ref{theo:1} and \ref{theo: ate} do not rely on the constraint $n_{.1k} + n_{.0k} = n_{..k}$, and thus remain valid when more than two treatments are present.

{\red \subsection{Post-stratification estimator} }

Recall that the PS estimator is defined as:
$$
 \hat{\delta}_{\rm PS} =  \sum_{k=1}^K w_{k, {\rm PS}}  I(n_{.0k} n_{.1k}\neq 0) \left( \bar Y_{1k} -  \bar Y_{0k}  \right),
$$ 
where $w_{k, {\rm PS}}= n_{..k}/n$ for $ k=1,\dots, K$, and $\bar Y_{ak}$ denotes the sample mean of the $Y_i$'s in stratum $k$ under treatment $a=0,1$. The weight $I(n_{.0k} n_{.1k}\neq 0)$ is added to explicitly indicate that strata with either $n_{.0k} =0$ or $n_{.1k}= 0$ are dropped since in these strata either $ \bar Y_{1k} $ or $  \bar Y_{0k} $ is undefined  \citep{miratrix2013adjusting}.  The PS estimator is well-known in sample surveys \citep{fuller2009} and causal inference \citep{rosenbaum1984reducing}.

The PS estimator is not limited to binary outcomes and can be applied to any continuous or discrete outcomes. When the outcome is binary, the PS estimator can be written as $ \sum_{k=1}^K w_{k, {\rm PS}}  I(n_{.0k} n_{.1k}\neq 0) 
 \hat\delta_k$, making it a weighted average of stratum-specific risk difference estimators. Unlike the MH weights, the PS weights are proportional to stratum size.

In randomized clinical trials under large-stratum asymptotics, the properties of the PS estimator have been studied in \cite{Ye2020}. In the following theorem, we establish the properties of the PS estimator under {\red both sparse- and mixed-stratum asymptotics} for either continuous or discrete outcomes. 
\begin{theoremS}\label{gen_th}
   {\red Suppose that one of the following holds: (i) Assumption \ref{sparse_assumption} (sparse-stratum), with $\lim_{K\to \infty} \frac{K_1}{\sqrt{K}}= 0$ almost surely; or (ii) Assumption \ref{mix_assumption} (mixed), with the subset of strata $\mathcal{K}_s$ (the sparse strata) satisfying the condition in (i). Then, as $n\to\infty$, 
  $$
  \frac{\hat{\delta}_{\rm PS} - \delta_{\rm ATE}}{\sqrt{\sigma_{n, \rm PS}^2 + \nu_{n,\rm PS}^2} } \xrightarrow{d} N(0,1),
  $$
where $\sigma_{n, \rm PS}^2= \sum_{k=1}^K I(n_{.1k}n_{.0k}\neq 0) w_{k,\rm PS}^2 \left\{\frac{{\rm var} (Y^{(1)} \mid Z=z_k)}{n_{.1k}} + \frac{{\rm var} (Y^{(0)} \mid Z=z_k)}{n_{.0k}} \right\}
$ and $\nu_{n, \rm PS}^2= n^{-1} {\rm var} \{E(Y^{(1)}-Y^{(0)}|Z)\}$.
}
\end{theoremS}
 The condition $\lim_{K\to \infty} \frac{K_1}{\sqrt{K}}= 0$ in Theorem \ref{gen_th} offers practical guidance for covariate adjustment in randomized clinical trials using the PS estimator. Since the choice of strata impacts efficiency but not consistency, and the PS estimator drops the $K_1$ strata with data only in one arm,  this condition suggests that stratification should not be too fine. In other words, the number of strata with participants only in one arm should remain small compared to $\sqrt{K}$, ensuring that dropping such strata does not introduce asymptotic bias. This requirement aligns with the common practice of pooling small strata before applying PS method in clinical trials.

Although Theorem \ref{gen_th} focuses on the sparse-stratum regime, we see that $n \sigma_{n, \rm PS}^2 $ converges in probability to $E\{\pi_1^{-1} \var (Y^{(1)}\mid Z) + \pi_0^{-1} \var (Y^{(0)}\mid Z) \}$  under the large-stratum regime. Therefore, the result in Theorem \ref{gen_th} aligns with the large-stratum results in \cite{Ye2020}.

For robust variance estimation, let $\bar Y_{ak}$ and $S_{ak}^2$ respectively denote the sample mean and sample variance of the $Y_i$'s in stratum $k$ under treatment $a=0,1$. We propose the following estimators of $\sigma_{n, \rm PS}^2 $ and $\nu_{n, \rm PS}^2$:
\begin{align*}
    &\hat\sigma_{\rm PS}^2 = \sum_{k=1}^KI(n_{.1k}n_{.0k}\neq 0)\frac{n_{..k}^2}{n^2}\left\{\frac{S_{1k}^2}{n_{.1k}}I(n_{.1k}>1)+\frac{S_{0k}^2}{n_{.0k}}I(n_{.0k}>1)\right\}\\
    &\hat\nu_{\rm PS}^2=\frac{1}{n}\left[\sum_{k=1}^KI(n_{.1k}n_{.0k}\neq 0)\frac{n_{..k}}{n}\left\{\bar{Y}_{1k}^2-\frac{S_{1k}^2}{n_{.1k}}I(n_{.1k}>1)+\bar{Y}_{0k}^2-\frac{S_{0k}^2}{n_{.0k}}I(n_{.0k}>1)-2\Bar{Y}_{1k}\Bar{Y}_{0k}\right\}-\hat\delta_{\rm PS}^2\right],
\end{align*}
As shown in Section S3, $\hat\sigma_{\rm PS}^2+ \hat\nu_{\rm PS}^2$ is a consistent variance estimator under both asymptotic regimes or a mix of the two regimes, under the mild condition that the number of strata with $n_{.1k} = 1$ or $n_{.0k} = 1$ is negligible.

\section{Additional simulation and real data analysis results}

\subsection{MH risk difference estimator as an estimator for $\delta_{\mh}$}
\begin{table}[H] 

\centering  
\caption{
Simulation results for estimating $\delta_{\mh}$ under unbalanced allocation (Factor 2a), based on $n=300$ (Factor 1b) and 1,000 simulation runs. Entries have all been multiplied by 100. Since $\delta_{\mh}$ varies across simulation runs, we report the average true value of $\delta_{\mh}$ (denoted as Ave. Truth). \label{tb: MH 1b2a}} 
\centering
\renewcommand{\arraystretch}{0.9}
\resizebox{0.75\textwidth}{!}{
\begin{tabular}{@{}clllllllll@{}}
\toprule
Effect & Regime & Ave.Truth &       & GR    & Sato  & mGR   & Boot  & Newc  & Klin  \\ \midrule
\multirow{15}{*}{\makecell{(4a) \\ Common}} & \multirow{5}{*}{\makecell{(3a) \\ Large}} & \multirow{5}{*}{-10.0} & Bias  & \multicolumn{6}{c}{-0.04} \\
                    &                    &                        & SD    & \multicolumn{6}{c}{5.66} \\
                    &                    &                        & SE    & 5.59  & 5.64  & 5.67  & 5.66  & 5.96  & 5.64  \\
                    &                    &                        & CP    & 94.4  & 94.6  & 94.8  & 94.7  & 95.9  & 94.8  \\
                    &                    &                        & Power & 43.1  & 42.8  & 42.8  & 43.2  & 38.7  & 42.4  \\ \cmidrule(l){2-10} 
                    & \multirow{5}{*}{\makecell{(3b) \\ Sparse}} & \multirow{5}{*}{-10.0} & Bias  & \multicolumn{6}{c}{0.05} \\
                    &                    &                        & SD    & \multicolumn{6}{c}{5.82} \\
\bottomrule

\end{tabular}
}
\end{table}

\begin{table}[H]  
\centering  
\caption{
Simulation results for estimating $\delta_{\mh}$ under unbalanced allocation (Factor 2a), based on $n=200$ (Factor 1c) and 1,000 simulation runs. Entries have all been multiplied by 100. Since $\delta_{\mh}$ varies across simulation runs, we report the average true value of $\delta_{\mh}$ (denoted as Ave. Truth). \label{tb: MH 1c2a}}   
\centering
\renewcommand{\arraystretch}{0.95}
\resizebox{0.75\textwidth}{!}{
\begin{tabular}{@{}clllllllll@{}}
\toprule
Effect & Regime & Ave.Truth &       & GR    & Sato  & mGR   & Boot  & Newc  & Klin  \\ \midrule
\multirow{15}{*}{\makecell{(4a) \\ Common}} & \multirow{5}{*}{\makecell{(3a) \\ Large}} & \multirow{5}{*}{-10.0} & Bias  & \multicolumn{6}{c}{-0.16} \\
                    &                    &                        & SD    & \multicolumn{6}{c}{7.14} \\
                    &                    &                        & SE    & 6.81  & 6.91  & 6.95  & 6.93  & 7.26  & 6.91  \\
                    &                    &                        & CP    & 94.5  & 94.6  & 94.8  & 94.5  & 96.4  & 94.9  \\
                    &                    &                        & Power & 33.7  & 32.7  & 32.7  & 33.4  & 29.4  & 32.1  \\ \cmidrule(l){2-10} 
                    & \multirow{5}{*}{\makecell{(3b) \\ Sparse}} & \multirow{5}{*}{-10.0} & Bias  & \multicolumn{6}{c}{0.09} \\
                    &                    &                        & SD    & \multicolumn{6}{c}{7.46} \\
\bottomrule

\end{tabular}
}
\end{table}

\begin{table}[H]  
\centering  
\caption{
Simulation results for estimating $\delta_{\mh}$ under balanced allocation (Factor 2b), based on $n=500$ (Factor 1a) and 1,000 simulation runs. Entries have all been multiplied by 100. Since $\delta_{\mh}$ varies across simulation runs, we report the average true value of $\delta_{\mh}$ (denoted as Ave. Truth). \label{tb: MH 1a2b}}   
\centering
\resizebox{0.75\textwidth}{!}{
\begin{tabular}{@{}clllllllll@{}}
\toprule
Effect & Regime & Ave.Truth &       & GR    & Sato  & mGR   & Boot  & Newc  & Klin  \\ \midrule
\multirow{15}{*}{\makecell{(4a) \\ Common}} & \multirow{5}{*}{\makecell{(3a) \\ Large}} & \multirow{5}{*}{-10.0} & Bias  & \multicolumn{6}{c}{0.15} \\
                    &                    &                        & SD    & \multicolumn{6}{c}{4.04} \\
                    &                    &                        & SE    & 4.07  & 4.09  & 4.10  & 4.09  & 4.36  & 4.09  \\
                    &                    &                        & CP    & 94.7  & 94.8  & 94.8  & 94.6  & 96.5  & 94.7  \\
                    &                    &                        & Power & 68.2  & 68.1  & 68.1  & 69.1  & 63.1  & 67.7  \\ \cmidrule(l){2-10} 
                    & \multirow{5}{*}{\makecell{(3b) \\ Sparse}} & \multirow{5}{*}{-10.0} & Bias  & \multicolumn{6}{c}{-0.02} \\
                    &                    &                        & SD    & \multicolumn{6}{c}{4.58} \\
\bottomrule
\end{tabular}
}
\end{table}

\begin{table}[H]  
\centering  
\caption{
Simulation results for estimating $\delta_{\mh}$ under balanced allocation (Factor 2b), based on $n=300$ (Factor 1b) and 1,000 simulation runs. Entries have all been multiplied by 100. Since $\delta_{\mh}$ varies across simulation runs, we report the average true value of $\delta_{\mh}$ (denoted as Ave. Truth). \label{tb: MH 1b2b}}   
\centering
\resizebox{0.75\textwidth}{!}{
\begin{tabular}{@{}clllllllll@{}}
\toprule
Effect & Regime & Ave.Truth &       & GR    & Sato  & mGR   & Boot  & Newc  & Klin  \\ \midrule
\multirow{15}{*}{\makecell{(4a) \\ Common}} & \multirow{5}{*}{\makecell{(3a) \\ Large}} & \multirow{5}{*}{-10.0} & Bias  & \multicolumn{6}{c}{0.13} \\
                    &                    &                        & SD    & \multicolumn{6}{c}{5.28} \\
                    &                    &                        & SE    & 5.25  & 5.29  & 5.31  & 5.30  & 5.61  & 5.29  \\
                    &                    &                        & CP    & 94.6  & 94.8  & 95.2  & 94.6  & 97.1  & 95.4  \\
                    &                    &                        & Power & 45.6  & 45.3  & 45.0  & 44.7  & 41.0  & 44.5  \\ \cmidrule(l){2-10} 
                    & \multirow{5}{*}{\makecell{(3b) \\ Sparse}} & \multirow{5}{*}{-10.0} & Bias  & \multicolumn{6}{c}{-0.27} \\
                    &                    &                        & SD    & \multicolumn{6}{c}{5.89} \\
\bottomrule

\end{tabular}
}
\end{table}

\begin{table}[H]  
\centering  
\caption{
Simulation results for estimating $\delta_{\mh}$ under balanced allocation (Factor 2b), based on $n=200$ (Factor 1c) and 1,000 simulation runs. Entries have all been multiplied by 100. Since $\delta_{\mh}$ varies across simulation runs, we report the average true value of $\delta_{\mh}$ (denoted as Ave. Truth). \label{tb: MH 1c2b}}  
\renewcommand{\arraystretch}{0.95}
\centering
\resizebox{0.75\textwidth}{!}{
\begin{tabular}{@{}clllllllll@{}}
\toprule
Effect & Regime & Ave.Truth &       & GR    & Sato  & mGR   & Boot  & Newc  & Klin  \\ \midrule
\multirow{15}{*}{\makecell{(4a) \\ Common}} & \multirow{5}{*}{\makecell{(3a) \\ Large}} & \multirow{5}{*}{-10.0} & Bias  & \multicolumn{6}{c}{-0.38} \\
                    &                    &                        & SD    & \multicolumn{6}{c}{6.38} \\
                    &                    &                        & SE    & 6.41  & 6.48  & 6.51  & 6.49  & 6.84  & 6.48  \\
                    &                    &                        & CP    & 95.1  & 95.5  & 95.6  & 95.4  & 96.7  & 96.0  \\
                    &                    &                        & Power & 37.5  & 36.6  & 36.0  & 36.3  & 32.3  & 34.9  \\ \cmidrule(l){2-10} 
                    & \multirow{5}{*}{\makecell{(3b) \\ Sparse}} & \multirow{5}{*}{-10.0} & Bias  & \multicolumn{6}{c}{0.14} \\
                    &                    &                        & SD    & \multicolumn{6}{c}{6.90} \\
\bottomrule
\end{tabular}
}
\end{table}

\subsection{MH risk difference estimator as an estimator for the ATE $\delta_{\rm ATE}$}
\begin{table}[H]  
\centering  
\caption{Simulation results for estimating the ATE $\delta_{\rm ATE}$ under unbalanced allocation (Factor 2a), based on $n=300$ (Factor 1b) and 1,000 simulation runs. Entries have all been multiplied by 100. \label{tb: ATE 1b2a}}   
\centering
\renewcommand{\arraystretch}{0.95}
\resizebox{\textwidth}{!}{
\begin{tabular}{@{}cllllllllllll@{}}
\toprule
Effect & Regime & Truth &       & GR    & Sato  & mGR   & Boot  & Newc  & Klin  & Gcomp & Unadj & PS   \\ \midrule
\multirow{15}{*}{\makecell{(4a) \\ Common}} & \multirow{5}{*}{\makecell{(3a) \\ Large}} & \multirow{5}{*}{-10.0} & Bias  & \multicolumn{6}{c}{-0.04} & -0.04 & -0.07 & -0.04 \\
                    &                    &                          & SD    & \multicolumn{6}{c}{5.66} & 5.69 & 6.01 & 5.69 \\
                    &                    &                          & SE    & 5.59  & 5.64  & 5.65  & 5.66  & 5.96  & 5.64  & 5.62  & 6.04    & 5.67  \\
                    &                    &                          & CP    & 94.4  & 94.6  & 94.7  & 94.7  & 95.9  & 94.8  & 94.4  & 94.7    & 94.6  \\
                    &                    &                          & Power & 43.1  & 42.8  & 42.8  & 43.2  & 38.7  & 42.4  & 42.6  & 38.8    & 42.2  \\ \cmidrule(l){2-13} 
                    & \multirow{5}{*}{\makecell{(3b) \\ Sparse}} & \multirow{5}{*}{-10.0} & Bias  & \multicolumn{6}{c}{0.05} & 0.01 & 0.07 & 0.06 \\
                    &                    &                          & SD    & \multicolumn{6}{c}{5.82} & 5.95 & 5.69 & 5.92 \\
\bottomrule
\end{tabular}
}
\end{table}

\begin{table}[H] 

\centering  
\caption{Simulation results for estimating the ATE $\delta_{\rm ATE}$ under unbalanced allocation (Factor 2a), based on $n=200$ (Factor 1c) and 1,000 simulation runs. Entries have all been multiplied by 100. \label{tb: ATE 1c2a}}   
\centering
\resizebox{\textwidth}{!}{
\begin{tabular}{@{}cllllllllllll@{}}
\toprule
Effect & Regime & Truth &       & GR    & Sato  & mGR   & Boot  & Newc  & Klin  & Gcomp & Unadj & PS   \\ \midrule
\multirow{15}{*}{\makecell{(4a) \\ Common}} & \multirow{5}{*}{\makecell{(3a) \\ Large}} & \multirow{5}{*}{-10.0} & Bias  & \multicolumn{6}{c}{-0.16} & -0.12 & -0.28 & -0.12 \\
                    &                    &                          & SD    & \multicolumn{6}{c}{7.14} & 7.17 & 7.56 & 7.17 \\
                    &                    &                          & SE    & 6.81  & 6.91  & 6.92  & 6.93  & 7.26  & 6.91  & 6.86  & 7.40    & 6.95  \\
                    &                    &                          & CP    & 94.5  & 94.6  & 94.7  & 94.5  & 96.4  & 94.9  & 94.3  & 94.3    & 94.4  \\
                    &                    &                          & Power & 33.7  & 32.7  & 32.7  & 33.4  & 29.4  & 32.1  & 33.9  & 29.7    & 32.8  \\ \cmidrule(l){2-13} 
                    & \multirow{5}{*}{\makecell{(3b) \\ Sparse}} & \multirow{5}{*}{-10.0}   & Bias  & \multicolumn{6}{c}{0.09} & 0.08 & -0.03 & 0.19 \\
                    &                    &                          & SD    & \multicolumn{6}{c}{7.46} & 7.66 & 7.20 & 7.59 \\
\bottomrule
\end{tabular}
}
\end{table}

\begin{table}[H] 

\centering  
\caption{Simulation results for estimating the ATE $\delta_{\rm ATE}$ under balanced allocation (Factor 2b), based on $n=500$ (Factor 1a) and 1,000 simulation runs. Entries have all been multiplied by 100. \label{tb: ATE 1a2b}}   
\centering
\resizebox{\textwidth}{!}{
\begin{tabular}{@{}cllllllllllll@{}}
\toprule
Effect & Regime & Truth &       & GR    & Sato  & mGR   & Boot  & Newc  & Klin  & Gcomp & Unadj & PS   \\ \midrule
\multirow{15}{*}{\makecell{(4a) \\ Common}} & \multirow{5}{*}{\makecell{(3a) \\ Large}} & \multirow{5}{*}{-10.0} & Bias  & \multicolumn{6}{c}{0.15} & 0.15 & 0.03 & 0.15 \\
                    &                    &                          & SD    & \multicolumn{6}{c}{4.04} & 4.04 & 4.26 & 4.04 \\
                    &                    &                          & SE    & 4.07  & 4.09  & 4.09  & 4.09  & 4.36  & 4.09  & 4.08  & 4.38    & 4.09  \\
                    &                    &                          & CP    & 94.7  & 94.8  & 94.8  & 94.6  & 96.5  & 94.7  & 94.8  & 95.9    & 94.8  \\
                    &                    &                          & Power & 68.2  & 68.1  & 68.1  & 69.1  & 63.1  & 67.7  & 68.3  & 63.7    & 68.1  \\ \cmidrule(l){2-13} 
                    & \multirow{5}{*}{\makecell{(3b) \\ Sparse}} & \multirow{5}{*}{-10.0}   & Bias  & \multicolumn{6}{c}{-0.02} & -0.03 & -0.03 & -0.02 \\
                    &                    &                          & SD    & \multicolumn{6}{c}{4.58} & 4.59 & 4.54 & 4.58 \\
\bottomrule
\end{tabular}
}
\end{table}

\begin{table}[H] 

\centering  
\caption{Simulation results for estimating the ATE $\delta_{\rm ATE}$ under balanced allocation (Factor 2b), based on $n=300$ (Factor 1b) and 1,000 simulation runs. Entries have all been multiplied by 100. \label{tb: ATE 1b2b}}   
\centering
\resizebox{\textwidth}{!}{
\begin{tabular}{@{}cllllllllllll@{}}
\toprule
Effect & Regime & Truth &       & GR    & Sato  & mGR   & Boot  & Newc  & Klin  & Gcomp & Unadj & PS   \\ \midrule
\multirow{15}{*}{\makecell{(4a) \\ Common}} & \multirow{5}{*}{\makecell{(3a) \\ Large}} & \multirow{5}{*}{-10.0} & Bias  & \multicolumn{6}{c}{0.13} & 0.13 & 0.15 & 0.13 \\
                    &                    &                          & SD    & \multicolumn{6}{c}{5.28} & 5.28 & 5.73 & 5.28 \\
                    &                    &                          & SE    & 5.25  & 5.29  & 5.30  & 5.30  & 5.61  & 5.29  & 5.27  & 5.66    & 5.30  \\
                    &                    &                          & CP    & 94.6  & 94.8  & 95.1  & 94.6  & 97.1  & 95.4  & 95.0  & 94.6    & 95.2  \\
                    &                    &                          & Power & 45.6  & 45.3  & 45.1  & 44.7  & 41.0  & 44.5  & 45.4  & 41.3    & 45.1  \\ \cmidrule(l){2-13} 
                    & \multirow{5}{*}{\makecell{(3b) \\ Sparse}} & \multirow{5}{*}{-10.0} & Bias  & \multicolumn{6}{c}{-0.27} & -0.30 & -0.24 & -0.29 \\
                    &                    &                          & SD    & \multicolumn{6}{c}{5.89} & 5.92 & 5.83 & 5.92 \\
\bottomrule
\end{tabular}
}
\end{table}

\begin{table}[H] 

\centering  
\caption{Simulation results for estimating the ATE $\delta_{\rm ATE}$ under balanced allocation (Factor 2b), based on $n=200$ (Factor 1c) and 1,000 simulation runs. Entries have all been multiplied by 100. \label{tb: ATE 1c2b}}   
\centering
\resizebox{\textwidth}{!}{
\begin{tabular}{@{}cllllllllllll@{}}
\toprule
Effect & Regime & Truth &       & GR    & Sato  & mGR   & Boot  & Newc  & Klin  & Gcomp & Unadj & PS   \\ \midrule
\multirow{15}{*}{\makecell{(4a) \\ Common}} & \multirow{5}{*}{\makecell{(3a) \\ Large}} & \multirow{5}{*}{-10.0} & Bias  & \multicolumn{6}{c}{-0.38} & -0.38 & -0.13 & -0.38 \\
                    &                    &                          & SD    & \multicolumn{6}{c}{6.38} & 6.38 & 6.93 & 6.38 \\
                    &                    &                          & SE    & 6.41  & 6.48  & 6.49  & 6.49  & 6.84  & 6.48  & 6.45  & 6.93    & 6.50  \\
                    &                    &                          & CP    & 95.1  & 95.5  & 95.6  & 95.4  & 96.7  & 96.0  & 95.1  & 95.7    & 95.5  \\
                    &                    &                          & Power & 37.5  & 36.6  & 36.4  & 36.3  & 32.3  & 34.9  & 36.6  & 32.3    & 36.2  \\ \cmidrule(l){2-13} 
                    & \multirow{5}{*}{\makecell{(3b) \\ Sparse}} & \multirow{5}{*}{-10.0}   & Bias  & \multicolumn{6}{c}{0.14} & 0.11 & 0.10 & 0.15 \\
                    &                    &                          & SD    & \multicolumn{6}{c}{6.90} & 7.02 & 6.72 & 6.98 \\
\bottomrule
\end{tabular}
}
\end{table}

{ \red \subsection{Simulation with individual risk difference}

Next, we evaluate the MH risk difference estimator on alternative data-generating mechanisms. In this simulation with individual risk difference, suppose that in stratum \( k \), there are three sets of possible values of potential outcomes:
\begin{itemize}
    \item A proportion \( \lambda_{k,-1} \) of patients with individual-level RD = –1, meaning their potential outcomes are \( (Y^{(0)}, Y^{(1)}) = (1, 0) \)
    \item A proportion \( \lambda_{k,1} \) with RD = 1, i.e., \( (Y^{(0)}, Y^{(1)}) = (0, 1) \)
    \item A proportion \( \lambda_{k,0} \) with RD = 0, split between those with \( (0, 0) \) and \( (1, 1) \).
\end{itemize}

To fully determine the data-generating process, we additionally specify \( \lambda_{k,0}^{(0)} \), the proportion of patients with \( (Y^{(0)}, Y^{(1)}) = (0, 0) \). The full breakdown is shown in the table below (where $\lambda_{k,0}^{(1)}= \lambda_{k,0}-\lambda_{k,0}^{(0)} $): 

\[
\begin{array}{llll}
\hline
\textbf{Proportion} & Y^{(0)} & Y^{(1)} & \textbf{RD} \\
\hline
\lambda_{k, -1} & 1 & 0 & -1 \\
\lambda_{k, 1} & 0 & 1 & 1 \\
\lambda_{k, 0}^{(0)} & 0 & 0 & 0 \\
\lambda_{k, 0}^{(1)} & 1 & 1 & 0 \\
\hline
\end{array}
\]

Under this setup, the stratum-specific risk difference is given by:
\[
\delta_k = E(Y^{(1)} \mid Z = z_k) - E(Y^{(0)} \mid Z = z_k) = \lambda_{k,1} - \lambda_{k,-1}.
\]
That is, even when treatment has a deterministic effect at the individual level, each stratum will typically consist of a mix of patients who are helped or harmed by the treatment. We implemented simulations using this data-generating process, and the results are shown in Tables \ref{tb:individual_mh}-\ref{tb:individual_ate}}. We find that the results are nearly identical to those from our original approach (Table \ref{tb: MH1a2a} and Table \ref{tb: ATE1a2a}) provided that \( p_{0k} \) and \( \delta_k \) match those derived from the table above (i.e., \( p_{0k} = \lambda_{k,-1} + \lambda_{k,0}^{(1)} \) and \( \delta_k = \lambda_{k,1} - \lambda_{k,-1} \)).

To stress test our estimators under an extreme scenario where \( \delta_k \) takes values only from \{-1, 0, 1\}—that is, the treatment either has no effect, causes the outcome in all patients within a stratum, or prevents the outcome in all patients—we conducted simulations under the following settings:
\begin{itemize}
    \item Large-stratum setting:
  \( (p_{0k}, k=1,2,3) = (0, 1, 0) \) and \( (\delta_k, k=1,2,3) = (1, -1, 0) \)
  \item Sparse-stratum setting:  
  Each \( \delta_k \) is drawn independently from a multinomial distribution over \{-1, 0, 1\} with equal probability. 
  \begin{itemize}
      \item If \( \delta_k = -1 \), then \( p_{0k} = 1 \)  
      \item If \( \delta_k = 1 \), then \( p_{0k} = 0 \)  
      \item If \( \delta_k = 0 \), then \( p_{0k} \) is set to 0 or 1 with equal probability
  \end{itemize}
\end{itemize}
Under these setting, $\var(\hat{\delta}-\delta_{\mh}) = \sigma_n^2=0,$ since $\hat{\delta}-\delta_{\mh}$ is always 0 when all $p_{0k}, p_{1k}$ are either 0 or 1. In contrast, $\hat{\delta} - \delta_{\rm ATE}$ still exhibits variability due to randomness from treatment allocation, captured by $\nu_n^2.$  Simulation results for estimating the ATE $\delta_{\rm ATE}$ under Factor (1a) and (2a) are shown in Table \ref{tb:individual_extreame_ate}. The GR and Newcombe estimators fail to provide estimates under this extreme scenario, while Sato’s and Klingenberg’s estimators become overly conservative. In contrast, our proposed mGR variance estimator, and PS, Bootstrap, G-computation, and Unadjusted estimators perform well, with standard errors close to the empirical standard deviations and coverage probabilities near the nominal 95\% level. Notably, in this simulation, the G-computation and PS methods achieve higher efficiency with smaller standard deviations. However, it is important to note that the validity of G-computation inference is not guaranteed under sparse- or mixed-stratum regimes.

\begin{table}[H]   
\centering  
\caption{Simulation results with individual risk difference generation for estimating $\delta_{\mh}$ under unbalanced allocation (Factor 2a), based on $n=500$ (Factor 1a) and 1,000 simulation runs. Entries have all been multiplied by 100.\label{tb:individual_mh} }   
\renewcommand{\arraystretch}{0.8}
\resizebox{0.75\textwidth}{!}{
\begin{tabular}{@{}clllllllll@{}}
\toprule
Effect & Regime & Ave.Truth &       & GR    & Sato  & mGR   & Boot  & Newc  & Klin  \\
\midrule
\multirow{15}{*}{\makecell{(4a) \\ Common}}
  & \multirow{5}{*}{\makecell{(3a) \\ Large}}
  & \multirow{5}{*}{-10.0}
    & Bias  & \multicolumn{6}{c}{-0.11}   \\
  &  &  & SD    & \multicolumn{6}{c}{4.62}    \\
  &  &  & SE    & 4.35  & 4.37  & 4.38  & 4.38  & 4.64  & 4.37  \\
  &  &  & CP    & 93.6  & 93.6  & 93.7  & 93.7  & 95.7  & 93.7  \\
  &  &  & Power & 62.6  & 62.6  & 62.2  & 62.9  & 58.4  & 62.3  \\ \cmidrule(l){2-10}

  & \multirow{5}{*}{\makecell{(3b) \\ Sparse}}
  & \multirow{5}{*}{-10.0}
    & Bias  & \multicolumn{6}{c}{-0.05}   \\
  &  &  & SD    & \multicolumn{6}{c}{4.83}    \\
  &  &  & SE    & 4.37  & 4.77  & 4.76  & 4.81  & 4.82  & 4.77  \\
  &  &  & CP    & 92.1  & 94.6  & 94.6  & 94.8  & 95.1  & 94.8  \\
  &  &  & Power & 61.5  & 56.2  & 56.7  & 56.1  & 55.7  & 55.3  \\ \cmidrule(l){2-10}

  & \multirow{5}{*}{\makecell{(3c) \\ Mixed}}
  & \multirow{5}{*}{-10.0}
    & Bias  & \multicolumn{6}{c}{-0.03}   \\
  &  &  & SD    & \multicolumn{6}{c}{4.57}    \\
  &  &  & SE    & 4.34  & 4.56  & 4.56  & 4.58  & 4.76  & 4.56  \\
  &  &  & CP    & 94.4  & 95.0  & 95.4  & 95.4  & 96.1  & 95.3  \\
  &  &  & Power & 61.1  & 57.7  & 57.6  & 57.9  & 54.7  & 57.4  \\ \midrule

\multirow{15}{*}{\makecell{(4b) \\ Varying}}
  & \multirow{5}{*}{\makecell{(3a) \\ Large}}
  & \multirow{5}{*}{10.0}
    & Bias  & \multicolumn{6}{c}{0.01}    \\
  &  &  & SD    & \multicolumn{6}{c}{3.39}    \\
  &  &  & SE    & 3.41  & 3.33  & 3.43  & 3.43  & 4.67  & 3.34  \\
  &  &  & CP    & 95.7  & 95.5  & 95.7  & 95.6  & 99.4  & 95.5  \\
  &  &  & Power & 83.1  & 84.6  & 83.0  & 82.5  & 58.6  & 83.6  \\ \cmidrule(l){2-10}

  & \multirow{5}{*}{\makecell{(3b) \\ Sparse}}
  & \multirow{5}{*}{9.88}
    & Bias  & \multicolumn{6}{c}{0.03}    \\
  &  &  & SD    & \multicolumn{6}{c}{3.74}    \\
  &  &  & SE    & 3.34  & 3.62  & 3.66  & 3.68  & 4.83  & 3.62  \\
  &  &  & CP    & 91.6  & 93.4  & 93.7  & 94.1  & 98.6  & 93.7  \\
  &  &  & Power & 81.4  & 77.6  & 76.9  & 77.3  & 54.9  & 77.0  \\ \cmidrule(l){2-10}

  & \multirow{5}{*}{\makecell{(3c) \\ Mixed}}
  & \multirow{5}{*}{10.1}
    & Bias  & \multicolumn{6}{c}{-0.17}   \\
  &  &  & SD    & \multicolumn{6}{c}{3.72}    \\
  &  &  & SE    & 3.41  & 3.52  & 3.59  & 3.59  & 4.75  & 3.52  \\
  &  &  & CP    & 93.3  & 93.6 & 94.0 & 94.1  & 98.3 & 93.9  \\
  &  &  & Power &81.3 &79.3 & 78.3 &78.3&56.3&78.8  \\ \midrule

\multirow{15}{*}{\makecell{(4c) \\ Varying}}
  & \multirow{5}{*}{\makecell{(3a) \\ Large}}
  & \multirow{5}{*}{-8.96}
    & Bias  & \multicolumn{6}{c}{0.05}    \\
  &  &  & SD    & \multicolumn{6}{c}{4.37}    \\
  &  &  & SE    & 4.18  & 4.67  & 4.21  & 4.22  & 4.49  & 4.67  \\
  &  &  & CP    & 94.2  & 96.0  & 94.4  & 94.5  & 95.8  & 96.3  \\
  &  &  & Power & 56.9  & 48.7  & 56.2  & 56.2  & 51.2  & 47.3  \\ \cmidrule(l){2-10}

  & \multirow{5}{*}{\makecell{(3b) \\ Sparse}}
  & \multirow{5}{*}{-32.5}
    & Bias  & \multicolumn{6}{c}{0.18}    \\
  &  &  & SD    & \multicolumn{6}{c}{4.23}    \\
  &  &  & SE    & 3.85  & 4.78  & 4.18  & 4.23  & 4.52  & 4.79  \\
  &  &  & CP    & 93.2  & 97.4  & 95.3  & 95.8  & 97.0  & 97.7  \\
  &  &  & Power & 100   & 100   & 100   & 100   & 100   & 100   \\ \cmidrule(l){2-10}

  & \multirow{5}{*}{\makecell{(3c) \\ Mixed}}
  & \multirow{5}{*}{-20.5}
    & Bias  & \multicolumn{6}{c}{-0.21}   \\
  &  &  & SD    & \multicolumn{6}{c}{4.27}    \\
  &  &  & SE    & 4.00  & 4.77  & 4.19  & 4.21  & 4.53  & 4.77  \\
  &  &  & CP    & 93.4  & 97.4  & 94.9  & 94.9  & 96.9  & 97.8  \\
  &  &  & Power & 99.6  & 99.0  & 99.6  & 99.5  & 99.4  & 98.8  \\ 
\bottomrule
\end{tabular}

}
\end{table}

\begin{table}[H]   
\centering  
\caption{Simulation results with individual risk difference generation for estimating the  $\delta_{ATE}$ under unbalanced allocation (Factor 2a), based on $n=500$ (Factor 1a) and 1,000 simulation runs. Entries have all been multiplied by 100.\label{tb:individual_ate} }   

\renewcommand{\arraystretch}{0.95}
\resizebox{\textwidth}{!}{
\begin{tabular}{@{}cllllllllllll@{}}
\toprule
Effect & Regime & Truth & & GR & Sato & mGR & Boot & Newc & Klin & Gcomp & Unadj & PS \\
\midrule
 \multirow{15}{*}{\makecell{(4a) \\ Common}} &  \multirow{5}{*}{\makecell{(3a) \\ Large}} & \multirow{5}{*}{-10.0} & Bias & \multicolumn{6}{c}{-0.11} & -0.08 & -0.07 & -0.11 \\
& & & SD & \multicolumn{6}{c}{4.62} & 4.64 & 4.94 & 4.64 \\
& & & SE & 4.35 & 4.37 & 4.38 & 4.38 & 4.64 & 4.37 & 4.36 & 4.68 & 4.38 \\
& & & CP & 93.6 & 93.6 & 93.7 & 93.7 & 95.7 & 93.7 & 93.7 & 94.0 & 93.7 \\
& & & Power & 62.6 & 62.6 & 62.4 & 62.9 & 58.4 & 62.3 & 62.9 & 57.0 & 62.5 \\ \cmidrule(lr){2-13}

&  \multirow{5}{*}{\makecell{(3b) \\ Sparse}} & \multirow{5}{*}{-10.0} & Bias & \multicolumn{6}{c}{-0.05} & 0.04 & -0.09 & -0.00 \\
& & & SD & \multicolumn{6}{c}{4.83} & 4.78 & 4.75 & 4.91 \\
& & & SE & 4.37 & 4.77 & 4.78 & 4.81 & 4.82 & 4.77 & 4.47 & 4.72 & 4.79 \\
& & & CP & 92.1 & 94.6 & 94.6 & 94.8 & 95.1 & 94.8 & 92.7 & 94.8 & 94.3 \\
& & & Power & 61.5 & 56.2 & 56.1 & 56.1 & 55.7 & 55.3 & 60.3 & 57.3 & 55.1 \\ \cmidrule(lr){2-13}


& \multirow{5}{*}{\makecell{(3c) \\ Mixed}} & \multirow{5}{*}{-10.0} & Bias  &\multicolumn{6}{c}{-0.03}& -0.03 & -0.10 & -0.01 \\
&                                            &                        & SD    & \multicolumn{6}{c}{4.57}& 4.66  & 4.85  & 4.65  \\
&                                            &                        & SE    & 4.34  & 4.56  & 4.56  & 4.58  & 4.76  & 4.56  & 4.39  & 4.73  & 4.59  \\
&                                            &                        & Power & 61.1  & 57.7  & 57.7  & 57.9  & 54.7  & 57.4  & 61.5  & 55.7  & 56.3  \\
&                                            &                        & CP    & 94.4  & 95.0  & 95.2  & 95.4  & 96.1  & 95.3  & 94.4  & 94.2  & 95.1  \\ \midrule

 \multirow{15}{*}{\makecell{(4b) \\ Varying}} &  \multirow{5}{*}{\makecell{(3a) \\ Large}} & \multirow{5}{*}{10.0} & Bias & \multicolumn{6}{c}{0.01} & -0.10 & 0.08 & 0.00 \\
& & & SD & \multicolumn{6}{c}{3.44} & 3.40 & 4.62 & 3.42 \\
& & & SE & 3.41 & 3.33 & 3.47 & 3.47 & 4.67 & 3.34 & 3.44 & 4.69 & 3.46 \\
& & & CP & 95.5 & 95.2 & 95.8 & 95.4 & 99.4 & 95.3 & 95.9 & 96.0 & 95.9 \\
& & & Power & 83.1 & 84.6 & 82.6 & 82.2 & 58.6 & 83.6 & 83.2 & 58.3 & 82.7 \\ \cmidrule(lr){2-13}

&  \multirow{5}{*}{\makecell{(3b) \\ Sparse}} & \multirow{5}{*}{10.0} & Bias & \multicolumn{6}{c}{0.07} & -0.02 & -0.09 & 0.02 \\
& & & SD & \multicolumn{6}{c}{3.74} & 3.70 & 4.72 & 3.88 \\
& & & SE & 3.34 & 3.62 & 3.68 & 3.69 & 4.83 & 3.62 & 3.44 & 4.73 & 3.70 \\
& & & CP & 91.9 & 93.7 & 94.1 & 94.2 & 98.8 & 93.9 & 91.5 & 95.9 & 93.4 \\
& & & Power & 81.4 & 77.6 & 76.3 & 77.1 & 54.9 & 77.0 & 79.9 & 53.0 & 75.3 \\ \cmidrule(lr){2-13}


& \multirow{5}{*}{\makecell{(3c) \\ Mixed}} & \multirow{5}{*}{10.1} & Bias  & \multicolumn{6}{c}{-0.14} & -0.16 & -0.35 & -0.16 \\
&                                            &                        & SD    & \multicolumn{6}{c}{3.74}  & 3.82  & 4.87  & 3.81  \\
&                                            &                        & SE    & 3.41  & 3.52  & 3.62  & 3.62  & 4.75  & 3.52  & 3.48  & 4.71  & 3.63  \\
&                                            &                        & CP    & 93.2  & 93.8  & 94.3  & 94.5  & 98.1  & 94.0  & 93.1  & 94.3  & 94.2  \\
&                                            &                        & Power & 81.3  & 79.3  & 77.8  & 77.5  & 56.3  & 78.8  & 79.3  & 53.8  & 77.7  \\ \midrule

 \multirow{15}{*}{\makecell{(4c) \\ Varying}} & \multirow{5}{*}{\makecell{(3a) \\ Large}} & \multirow{5}{*}{-9.00} & Bias & \multicolumn{6}{c}{0.09} & 0.25 & 0.03 & 0.02 \\
& & & SD & \multicolumn{6}{c}{4.64} & 4.60 & 4.60 & 4.51 \\
& & & SE & 4.18 & 4.67 & 4.52 & 4.53 & 4.49 & 4.67 & 4.40 & 4.51 & 4.41 \\
& & & CP & 92.7 & 95.6 & 94.9 & 95.1 & 95.1 & 96.0 & 94.6 & 94.6 & 95.0 \\
& & & Power & 56.9 & 48.7 & 50.8 & 50.3 & 51.2 & 47.3 & 54.2 & 52.2 & 54.1 \\ \cmidrule(lr){2-13}

&  \multirow{5}{*}{\makecell{(3b) \\ Sparse}} & \multirow{5}{*}{-32.3} & Bias & \multicolumn{6}{c}{0.19} & 0.52 & 0.21 & 0.37 \\
& & & SD & \multicolumn{6}{c}{4.66} & 4.61 & 4.45 & 4.59 \\
& & & SE & 3.85 & 4.78 & 4.61 & 4.70 & 4.52 & 4.79 & 4.18 & 4.40 & 4.42 \\
& & & CP & 90.1 & 95.6 & 95.0 & 95.3 & 95.0 & 95.9 & 92.7 & 95.0 & 94.2 \\
& & & Power & 100 & 100 & 100 & 100 & 100 & 100 & 100 & 100 & 100 \\ \cmidrule(lr){2-13}

& \multirow{5}{*}{\makecell{(3c) \\ Mixed}} & \multirow{5}{*}{-20.8} & Bias  & \multicolumn{6}{c}{0.10} & -0.24 & -0.27 & -0.14 \\
&                                            &                        & SD    & \multicolumn{6}{c}{4.73}  & 4.56  & 4.58  & 4.55  \\
&                                            &                        & SE    & 4.00  & 4.77  & 4.61  & 4.65  & 4.53  & 4.77  & 4.29  & 4.47  & 4.44  \\
&                                            &                        & CP    & 90.5  & 95.2  & 94.7  & 94.1  & 94.7  & 95.5  & 93.3  & 94.0  & 94.5  \\
&                                            &                        & Power & 99.6  & 99.0  & 99.3  & 99.2  & 99.4  & 98.8  & 99.7  & 99.7  & 99.5  \\ \bottomrule

\end{tabular}
}
\end{table}

\begin{table}[H]   
\centering  
\caption{Simulation results with individual risk difference generation for estimating the ATE $\delta_{\rm ATE}$ under extreme scenario where \( \delta_k \) takes values only from \{-1, 0, 1\} and unbalanced allocation (Factor 2a), based on $n=500$ (Factor 1a) and 1,000 simulation runs. Entries have all been multiplied by 100. \label{tb:individual_extreame_ate}}   
\centering
\resizebox{\textwidth}{!}{
\begin{tabular}{@{}clllllllllll@{}}
\toprule
 Regime & Truth &  & Gr & Sato & mGR & Boot & Newc & Klin &  Gcomp &  Unadj &  PS \\
\midrule
 \multirow{5}{*}{\makecell{(3a) \\ Large}} & \multirow{5}{*}{-10.0} & Bias & \multicolumn{6}{c}{-0.21}  & -0.08 & -0.02 & -0.08  \\
   &  & SD & \multicolumn{6}{c}{3.86}  & 3.23 & 4.09 & 3.23  \\
   &  & SE & 0.00 & 4.78 & 3.87 & 3.92 & NA & 4.78 & 3.13 & 4.18 & 3.13  \\
   &  & CP & 0.00 & 98.5 & 95.5 & 95.0 & NA & 98.7 & 94.2 & 95.6 & 94.1  \\
   &  & Power & 100 & 59.2 & 74.7 & 73.8 & NA & 60.1 & 88.6 & 65.3 & 88.4  \\ \midrule
 \multirow{5}{*}{\makecell{(3b) \\ Sparse}} & \multirow{5}{*}{2.50} & Bias & \multicolumn{6}{c}{0.07} & 0.00 & 0.18 & -0.00  \\
   &  & SD & \multicolumn{6}{c}{3.95}  & 3.10 & 4.63 & 3.10  \\
   &  & SE & 0.00 & 4.57 & 3.91 & 4.12 & NA & 4.57 & 2.95 & 4.74 & 2.95  \\
   &  & CP & 0.00 & 97.8 & 94.1 & 93.8 & NA & 97.8 & 93.9 & 96.2 & 93.8  \\
  &  & Power & 100 & 5.90 & 10.1 & 8.60 & NA & 5.40 & 15.0 & 7.30 & 14.9  \\ \midrule
 \multirow{5}{*}{\makecell{(3c) \\Mixed}} & \multirow{5}{*}{-10.4} & Bias & \multicolumn{6}{c}{-0.18}  & 0.19 & -0.08 & 0.02  \\
   &  & SD & \multicolumn{6}{c}{4.67} & 3.67 & 4.83 & 3.61  \\
   &  & SE & 0.00 & 5.34 & 4.49 & 4.64 & NA & 5.34 & 3.50 & 4.68 & 3.49  \\
   &  & CP & 0.00 & 97.7 & 93.9 & 93.9 & NA & 97.9 & 93.8 & 94.8 & 93.9  \\
   &  & Power & 100 & 51.5 & 65.5 & 62.3 & NA & 51.2 & 80.8 & 60.2 & 82.7  \\ \bottomrule
\end{tabular}

}

\end{table}

\subsection{Data from the  CALGB trial}

\begin{table}[H]
\centering  
\caption{Data from the  CALGB trial.}  
\label{real data}  
\begin{tabular}{crrrrrr} \toprule
Institution & $n_{.1k}$ & $\hat p_{1k}$   & $n_{.0k}$ & $\hat p_{0k}$   & $\hat\delta_k$ & $w_{k}$    \\\hline 
1           & 4  & 0.75 & 3  & 0.33 & 0.42  & 1.71 \\
2           & 4  & 0.75 & 11 & 0.73 & 0.02  & 2.93 \\
3           & 2  & 1.00 & 3  & 0.67 & 0.33  & 1.20 \\
4           & 2  & 1.00 & 2  & 1.00 & 0.00  & 1.00 \\
5           & 2  & 1.00 & 3  & 0.00 & 1.00  & 1.20 \\
6           & 3  & 0.33 & 3  & 0.67 & -0.33 & 1.50 \\
7           & 2  & 1.00 & 3  & 0.67 & 0.33  & 1.20 \\
8           & 5  & 0.20 & 4  & 1.00 & -0.80 & 2.22 \\
9           & 2  & 1.00 & 3  & 0.67 & 0.33  & 1.20 \\
10          & 2  & 0.00 & 3  & 0.67 & -0.67 & 1.20 \\
11          & 3  & 1.00 & 3  & 1.00 & 0.00  & 1.50 \\
12          & 2  & 1.00 & 2  & 0.00 & 1.00  & 1.00 \\
13          & 4  & 0.25 & 5  & 0.20 & 0.05  & 2.22 \\
14          & 3  & 0.67 & 4  & 0.50 & 0.17  & 1.71 \\
15          & 4  & 0.50 & 6  & 0.67 & -0.17 & 2.40 \\
16          & 12 & 0.33 & 9  & 0.33 & 0.00  & 5.14 \\
17          & 2  & 0.50 & 3  & 0.67 & -0.17 & 1.20 \\
18          & 3  & 1.00 & 4  & 0.25 & 0.75  & 1.71 \\
19          & 4  & 0.25 & 3  & 0.67 & -0.42 & 1.71 \\
20          & 3  & 0.00 & 2  & 0.00 & 0.00  & 1.20 \\
21          & 4  & 0.50 & 5  & 0.20 & 0.30  & 2.22 \\\bottomrule
\end{tabular}
\end{table}

\section{Technical proofs}

In the following proof, we will use the weak law of large numbers repeatedly, which can be proved directly from Chebyshev's inequality \citep[Chapter 1.5]{shao2003mathematical}:
\begin{lemma}\label{lemma1}
    Let $X_1, X_2, \dots, X_K$ be independent random variables with finite expectations $\mu_1,\dots, \mu_K$. If 
\[
\lim_{K\to\infty}
\frac{1}{K^2} \sum_{k=1}^K \var (X_k) = 0,
\]
then 
\[
\frac{1}{K}\sum_{k=1}^K (X_k - \mu_k) \xrightarrow{p} 0 .
\]
\end{lemma}

We will use the following order notations repeatedly: Let $\{X_n\}_{n=1}^\infty$ and $\{R_n\}_{n=1}^\infty$ be sequences of $\mathbb{R}-$valued random variables, all defined on the same probability space.

{\bf Big-O in probability notation:} We say that $X_n=O_p(R_n)$ if, for all $\epsilon>0$, there exists some constant $M>0$ such that $\liminf_{n\rightarrow\infty} P(|X_n|\leq M|R_n|)>1-\epsilon$.

{\bf Little-o in probability notation:} We say that $X_n=o_p(R_n)$ if, for all constant $M>0$, $\lim_{n\rightarrow\infty}P(|X_n|\leq M|R_n|)=1$. 
\subsection{Sato's variance estimator in Eq. \eqref{eq: sato}}
\label{sec: sato var}
In this section, we show the derivations of Sato's variance estimator and that Sato's variance estimator is usually conservative.

As mentioned in the main text, Sato used the relationship $p_{1k} = \delta + p_{0k}$ to express $\operatorname{Var}(w_k(\hat\delta_k - \delta) \mid \mathcal{T}_k)$ in the following two ways:
\begin{align*}
   & \var(w_{k} (\hat\delta_k-\delta)\mid \calT_k)=  \frac{1}{n_{..k}^2}E[\delta(n_{.0k}^2n_{01k}-n_{.1k}^2n_{00k})+n_{.0k}n_{10k}n_{01k}+n_{.1k}n_{11k}n_{00k}\mid\calT_k] := A_k , \\
   &\var(w_{k}  (\hat \delta_k  - \delta)\mid\calT_k ) = \frac{1}{n_{..k}^2}E[\delta(n_{.1k}^2n_{10k}-n_{.0k}^2n_{11k})+n_{.0k}n_{11k}n_{00k}+n_{.1k}n_{10k}n_{01k}\mid\calT_k] := B_k.
\end{align*}
 Hence, $\var(w_{k} (\hat\delta_k-\delta)\mid \calT_k)$ can be estimated in two ways: $\hat A_k=  \{\hat\delta(n_{.0k}^2n_{01k}-n_{.1k}^2n_{00k})+n_{.0k}n_{10k}n_{01k}+n_{.1k}n_{11k}n_{00k}\}/n_{..k}^2$ and $\hat B_k= \{\hat\delta(n_{.1k}^2n_{10k}-n_{.0k}^2n_{11k})+n_{.0k}n_{11k}n_{00k}+n_{.1k}n_{10k}n_{01k}\}/n_{..k}^2$.
Finally, $\var(w_{k}  (\hat \delta_k  - \delta)\mid\calT_k )$ can be estimated by the average of $\hat{A}_k$ and $\hat{B}_k$. Replacing $\var(  w_{k} (\hat\delta_k - \delta) \mid {\calT_k})$ with $ (\hat{A}_k+\hat{B}_k)/2$ in \eqref{eq:cmh var} thus gives Sato's variance estimator for $\hat\delta$:
\begin{equation}\label{eq: sato}
   \frac{\hat{\delta} (\sum_{k=1}^K P_k) + (\sum_{k=1}^K Q_k)}{(\sum_{k=1}^K w_{k})^2}
\end{equation}
where $P_k = \{n_{.1k}^2n_{10k} - n_{.0k}^2 n_{11k} + 2^{-1} n_{.1k}n_{.0k}(n_{.0k} - n_{.1k}) \}/n_{..k}^2$ and $Q_k = \{n_{11k}(n_{.0k} - n_{10k}) + n_{10k}(n_{.1k} - n_{11k})\}/(2n_{..k})$. Details of the derivations are included below.

(a) {\bf Derivations of Sato's variance estimator.} First, note that the variance in Eq. \eqref{eq:cmh var}: 
\begin{align}
    &\var(w_{k}  (\hat \delta_k  - \delta)\mid\calT ) = w_{k}^2 \{\frac{p_{1k}(1-p_{1k})}{n_{.1k}}+\frac{p_{0k}(1-p_{0k})}{n_{.0k}}\}. \label{eq: sato variance}
\end{align}
Replacing $p_{1k}$ and $p_{0k}$ respectively by $\hat{p}_{1k}=\frac{n_{11k}}{n_{.1k}}$ and $\hat{p}_{0k}=\frac{n_{10k}}{n_{.0k}}$ gives GR variance estimator in Eq. \eqref{eq:cmh var}, which is not applicable to sparse-stratum asymptotics, as discussed in the main article and shown in the later section.  

Next, we derive Sato's variance estimator in Eq. \eqref{eq: sato}. Assuming the common risk difference across strata, $p_{1k}$ and $p_{0k}$ are related as below:
 \begin{align*}
     &p_{1k} = \delta +p_{0k},\\
     &p_{0k} = p_{1k} - \delta.
 \end{align*}
Replacing 
$\frac{p_{1k}(1-p_{1k})}{n_{.1k}}+\frac{p_{0k}(1-p_{0k})}{n_{.0k}} $ with $ \frac{(\delta+p_{0k})(1-p_{1k})}{n_{.1k}}+\frac{(p_{1k}-\delta)(1-p_{0k})}{n_{.0k}}$ in Eq. \eqref{eq: sato variance}, and using the facts that $(n_{11k},n_{01k})$ and $(n_{10k},n_{00k})$ are independent conditional on $\calT_k$, and
$p_{1k} = E(n_{11k}\mid \calT_k)/n_{.1k}, p_{0k} = E(n_{10k}\mid \calT_k)/n_{.0k}$, we obtain that:
\begin{align*}
    &\var \{w_{k}  (\hat \delta_k  - \delta)\mid\calT \}= w_{k}^2 \{\frac{(\delta+p_{0k})(1-p_{1k})}{n_{.1k}}+\frac{(p_{1k}-\delta)(1-p_{0k})}{n_{.0k}}\}\\
    &=\frac{n_{. 1k}^2n_{. 0k}^2}{n_{.. k}^2}\left[\frac{\{\delta+E(n_{10k}\mid\calT_k)/n_{. 0k}\}\{1-E(n_{11k}\mid\calT_k)/n_{. 1k}\}}{n_{. 1k}} +\frac{\{E(n_{11k}\mid\calT_k)/n_{. 1k}-\delta\}\{1-E(n_{10k}\mid\calT_k)/n_{. 0k}\}}{n_{. 0k}}  \right]\\
    &=\frac{1}{n_{..k}^2}\delta[n_{.0k}^2\{n_{.1k}-E(n_{11k}\mid \calT_k)\}-n_{.1k}^2\{n_{.0k}-E(n_{10k}\mid\calT_k)\}] \\ \ 
    &\quad +\frac{1}{n_{..k}^2} [n_{.0k}E(n_{10k}\mid \calT_k)\{n_{.1k}-E(n_{11k}\mid\calT_k)\}+n_{.1k}E(n_{11k}\mid\calT_k)\{n_{.0k}-E(n_{10k}\mid\calT_k)\} ] \\
    &=\frac{1}{n_{..k}^2}\delta\{n_{.0k}^2E(n_{01k}\mid \calT_k)-n_{.1k}^2E(n_{00k}\mid \calT_k)\} \\ \ 
    &\quad +\frac{1}{n_{..k}^2} \{n_{.0k}E(n_{10k}\mid\calT_k)E(n_{01k}\mid\calT_k)+n_{.1k}E(n_{11k}\mid\calT_k)E(n_{00k}\mid\calT_k) \}\\
    &=\frac{1}{n_{..k}^2}E\{\delta(n_{.0k}^2n_{01k}-n_{.1k}^2n_{00k})+n_{.0k}n_{10k}n_{01k}+n_{.1k}n_{11k}n_{00k}\mid\calT_k\}\\
    &:= A_k.
\end{align*}
Similarly, by replacing 
$\frac{p_{1k}(1-p_{1k})}{n_{.1k}}+\frac{p_{0k}(1-p_{0k})}{n_{.0k}} $ with $ \frac{p_{1k}(1-\delta-p_{0k})}{n_{.1k}}+\frac{p_{0k}(1+\delta-p_{1k})}{n_{.0k}}$, we can derive another formula:
\begin{align*}
    &\var\{w_{k}  (\hat \delta_k  - \delta)\mid\calT \} = \frac{1}{n_{..k}^2}E\{\delta(n_{.1k}^2n_{10k}-n_{.0k}^2n_{11k})+n_{.0k}n_{11k}n_{00k}+n_{.1k}n_{10k}n_{01k}\mid\calT_k\} := B_k.
\end{align*}
Then replacing $\delta$ with $\hat{\delta}_{MH},$ and the expectation of $n_{11k},n_{10k}$ with $n_{11k},n_{10k}$, we derive: 
\begin{align*} 
     &\hat{A}_k =\frac{1}{n_{..k}^2}\left\{  \hat{\delta}(n_{.0k}^2n_{01k}-n_{.1k}^2n_{00k})+n_{.0k}n_{10k}n_{01k}+n_{.1k}n_{11k}n_{00k}\right\}, \\
     &\hat{B}_k = \frac{1}{n_{..k}^2}\left\{  
 \hat{\delta}(n_{.1k}^2n_{10k}-n_{.0k}^2n_{11k})+n_{.0k}n_{11k}n_{00k}+n_{.1k}n_{10k}n_{01k} \right\}.
\end{align*}
Finally, the theoretical variance $\var\{w_{k}  (\hat \delta_k  - \delta)\mid\calT \}$ can be estimated by the average of $\hat{A}_k$ and $\hat{B}_k$ and used to estimate $\var(\hat\delta\mid\calT)$:
\begin{align*}
    &\widehat{\var}\{w_{k}  (\hat \delta_k  - \delta)\mid\calT \} = \frac{\hat{A}_k+\hat{B}_k}{2},\\
    &\widehat{\var}(\hat \delta \mid\calT )
    = \frac{\sum_{k=1}^K \widehat\var\{w_{k}(\hat\delta_k-\delta)\mid\calT\}}{(\sum_{k=1}^Kw_{k})^2}= 
    \frac{\sum_{k=1}^K (\hat{A}_k+\hat{B}_k)/2 }{(\sum_{k=1}^Kw_{k})^2}=\frac{\hat{\delta} (\sum_{k=1}^K P_k) + (\sum_{k=1}^K Q_k)}{(\sum_{k=1}^K w_{k})^2},
    \end{align*}
    where 
    \begin{align*}
    &P_k = \frac{2n_{.1k}^2n_{10k} - 2n_{.0k}^2 n_{11k} + n_{.1k}n_{.0k}(n_{.0k} - n_{.1k})}{2n_{..k}^2},\\
    &Q_k = \frac{n_{11k}(n_{.0k} - n_{10k}) + n_{10k}(n_{.1k} - n_{11k})}{2n_{..k}},
\end{align*}
following from that
$n_{.0k}^2(n_{01k}-n_{11k})+ n_{.1k}^2(n_{10k}-n_{00k})=n_{.0k}^2(n_{.1k}-2n_{11k})+ n_{.1k}^2(2n_{10k}-n_{.0k})=  
2n_{.1k}^2n_{10k} - 2n_{.0k}^2 n_{11k} + n_{.1k}n_{.0k}(n_{.0k} - n_{.1k})$ and $n_{.1k}n_{11k}n_{00k}+n_{.0k}n_{10k}n_{01k}+n_{.0k}n_{11k}n_{00k}+n_{.1k}n_{10k}n_{01k}=n_{..k}(n_{11k}n_{00k}+n_{10k}n_{01k})$.

This approach allows for the estimation of the variance under both asymptotic scenarios.

(b) {\bf Sato's variance estimator can be conservative.} 
Note that the theoretical variance $\var(\hat \delta \mid\calT)$ can be written as
\begin{align}\label{eq: theoretical variance}
 \frac{ (\sum_{k=1}^K P_k \delta_k) + (\sum_{k=1}^K Q_k)}{(\sum_{k=1}^K w_{k})^2},
\end{align}
which, under common risk differences, equals to
\begin{align}\label{eq: variance under common}
\frac{ \delta_{\mh} (\sum_{k=1}^K P_k) + (\sum_{k=1}^K Q_k)}{(\sum_{k=1}^K w_{k})^2}.
\end{align}
Then, Sato's variance estimator uses $\hat{\delta}$ to estimate $\delta_{\mh}$. However, under varying risk differences, \eqref{eq: theoretical variance} still holds while \eqref{eq: variance under common} does not, and the difference results from $(\sum_{k=1}^K P_k \delta_k) $ in \eqref{eq: theoretical variance} and $\delta_{\mh} (\sum_{k=1}^K P_k)$ in \eqref{eq: variance under common}. In the following, we show that Sato's variance estimator usually overestimates the theoretical variance under varying risk differences and large-stratum asymptotics.

Under large-stratum asymptotics, we have that $n_{..k}/n=\rho_k+o_p(1)$, $n_{1ak}/n_{.ak}=p_{ak}+ o_p(1)$, and $n_{.ak}/n_{..k}=\pi_a+o_p(1)$ for $a=0$ and $1$. Therefore, it follows that $n^{-1}w_{k}=\frac{n_{.1k}n_{.0k}}{n_{..k}n}=\rho_k\pi_1\pi_0+o_p(1)$ and
\begin{align*}
    n^{-1}P_k&= \frac{2n_{.1k}^2n_{10k} - 2n_{.0k}^2 n_{11k} + n_{.1k}n_{.0k}(n_{.0k} - n_{.1k})}{2nn_{..k}^2}\\
    &\ =\frac{2n_{..k}\pi_1^2\pi_0p_{0k}- 2n_{..k}\pi_0^2\pi_1p_{1k} +n_{..k}\pi_0\pi_1(\pi_0-\pi_1)}{2n}+o_p(1)\\
    &\ =\rho_k\pi_1\pi_0\{\pi_0(\frac{1}{2}-p_{1k})+\pi_1(p_{0k}-\frac{1}{2})\}+o_p(1).
\end{align*}
    We first consider the case where $K=2$. We can see that
\begin{align*}
    \frac{\delta_{\mh} (\sum_{k=1}^2 P_k)-(\sum_{k=1}^2 P_k \delta_k)}{n}& = \frac{w_{1}\delta_1+w_{2}\delta_2}{w_{1}+w_{2}}\frac{(P_1+P_2)}{n}-\frac{(\delta_1P_1+\delta_2P_2)}{n}\\
    &= (\rho_1\delta_1+\rho_2\delta_2)\frac{P_1+P_2}{n}-\frac{\delta_1P_1+\delta_2P_2}{n} +o_p(1)\\
    &=-\rho_1\rho_2\pi_1\pi_0(\delta_1-\delta_2)(H_1-H_2) + o_p(1),
\end{align*}
where $H_k = \pi_0(1/2-p_{1k})+\pi_1(p_{0k}-1/2).$ Note that $\delta_k$ and $H_k$ exhibit opposite monotonicity with respect to $p_{0k}$ and $p_{1k}$. Hence, $H_1-H_2$ can have the opposite sign with $\delta_1-\delta_2$.  For example, when $\pi_1=\pi_0$, $(\delta_1-\delta_2)(H_1-H_2)$ is equal to $-(\delta_1-\delta_2)^2 \leq 0$, which implies that $n^{-1}\{\delta_{\mh} (\sum_{k=1}^2 P_k)-(\sum_{k=1}^2 P_k \delta_k)\} = \rho_1\rho_2\pi_1\pi_0(\delta_1-\delta_2)^2\geq 0$ with probability approaching 1. With varying risk differences, the difference is strictly larger than $0$, explaining the overestimation of Sato's variance estimator.

To give a more general result, we claim that for any $K\geq2$, it holds that
\begin{align}\label{eq: induction}
    \frac{\delta_{\mh} (\sum_{k=1}^K P_k)-(\sum_{k=1}^K P_k \delta_k)}{n}&=-\pi_1\pi_0\sum_{1\leq i<j\leq K}\rho_i\rho_j(\delta_i-\delta_j)(H_i-H_j) + o_p(1).
\end{align}

{\bf Proof of \eqref{eq: induction}.} Under large-stratum asymptotics, we have that
\begin{align*}
    \frac{\delta_{\mh} (\sum_{k=1}^K P_k)-(\sum_{k=1}^K P_k \delta_k)}{n}&=\sum_{k=1}^K\rho_k\delta_k\sum_{k=1}^K\pi_1\pi_0\rho_kH_k-\sum_{k=1}^K\pi_1\pi_0\rho_kH_k\delta_k+o_p(1).
\end{align*}
Notice that $2\sum_{1\leq i<j\leq K}\rho_i\rho_j(\delta_i-\delta_j)(H_i-H_j)=\sum_{1\leq i,j\leq K}\rho_i\rho_j(\delta_i-\delta_j)(H_i-H_j)$. Then, to prove \eqref{eq: induction}, it only needs to show that
\begin{align}\label{eq: tmp3}
    2\sum_{k=1}^K\rho_k\delta_k\sum_{k=1}^K\rho_kH_k-2\sum_{k=1}^K\rho_kH_k\delta_k=-\sum_{1\leq i,j\leq K}\rho_i\rho_j(\delta_i-\delta_j)(H_i-H_j).
\end{align}
Since $\sum_{j=1}^K\rho_j=1$, the right hand side of \eqref{eq: tmp3} equals to 
\begin{align*}
    \sum_{1\leq i,j\leq K}\rho_i\rho_j(\delta_iH_j-\delta_iH_i-\delta_jH_j+\delta_jH_i)&=\sum_{i=1}^K\sum_{j=1}^K\rho_i\delta_i\rho_jH_j-\rho_i\delta_iH_i\rho_j-\rho_j\delta_jH_j\rho_i+\rho_j\delta_j\rho_iH_i\\
    &=2\sum_{k=1}^K\rho_k\delta_k\sum_{k=1}^K\rho_kH_k-2\sum_{k=1}^K\rho_kH_k\delta_k,
\end{align*}
which is the left hand side of \eqref{eq: tmp3}.

{\bf Takeaway of \eqref{eq: induction}.} If $\pi_0=\pi_1$, then 
$$-\pi_1\pi_0\sum_{1\leq i<j\leq K}\rho_i\rho_j(\delta_i-\delta_j)(H_i-H_j)=\frac{1}{2}\pi_1\pi_0\sum_{1\leq i<j\leq K}\rho_i\rho_j(\delta_i-\delta_j)^2\geq 0,$$
and the inequality holds strictly with varying risk differences, indicating that Sato's variance estimator can be conservative with probability approaching 1. 

Besides, Sato's variance estimator can underestimate the true variance. For example, when $\pi_0<\pi_1$, i.e., $\pi_0<1/2$, as long as $\pi_1(p_{0i}-p_{0j})\geq \pi_0 (p_{1i}-p_{1j})\geq 0,$ $p_{1i}-p_{1j}\geq p_{0i}-p_{0j}\geq 0$ for all $i,j$ and some inequalities hold strictly for some $i,j$, then it follows that
\begin{align*}
    &\quad-\sum_{1\leq i<j\leq K}\rho_i\rho_j(\delta_i-\delta_j)(H_i-H_j)\\
    &=\sum_{1\leq i<j\leq K}\rho_i\rho_j\{(p_{1i}-p_{1j})-(p_{0i}-p_{0j})\}\{\pi_0(p_{1i}-p_{1j})-\pi_1(p_{0i}-p_{0j})\}\\
    &<0,
\end{align*}
which means that Sato's variance estimator can also under-estimate the true variance.

\subsection{Connections between the GR and Sato's variance estimators}

It is interesting to note that both the GR variance estimator and Sato's variance estimator can be viewed in a common form: 
\begin{equation}
\label{eq:common_var}
    \hat{\var}(\hat{\delta}) = \frac{\sum_k \lambda_k \hat{A}_k +(1-\lambda_k)\hat{B}_k}{(\sum_k w_{k})^2}
\end{equation}
with weights $ \lambda_k = \frac{1}{2}$ for Sato's estimator as shown in \ref{sec: sato var}  and  $$
 \lambda_k = \frac{n_{.0k}n_{11k}/n_{.1k}-n_{.1k}n_{10k}/n_{.0k}}{n_{.0k}-n_{.1k}}: =\lambda_{k,\rm GR}
 $$ for the GR estimator.

To show that we can obtain the GR variance estimator, first notice that $\lambda_{k,\rm GR}(n_{.0k}^2n_{01k}-n_{.1k}^2n_{00k})+(1-\lambda_{k,\rm GR})(n_{.1k}^2n_{10k}-n_{.0k}^2n_{11k}) =0$. 
Then, we see that
 \begin{align*}
     &\quad n_{..k}^2\{\lambda_{k,\rm GR}\hat{A}_k+(1-\lambda_{k,\rm GR})\hat{B}_k\}\\ &=\hat{\delta}\{\lambda_{k,\rm GR}(n_{.0k}^2n_{01k}-n_{.1k}^2n_{00k})+(1-\lambda_{k,\rm GR})(n_{.1k}^2n_{10k}-n_{.0k}^2n_{11k})\} \\ &\ 
     + \{ \lambda_{k,\rm GR}(n_{.0k}n_{10k}n_{01k}+n_{.1k}n_{11k}n_{00k})+(1-\lambda_{k,\rm GR})(n_{.0k}n_{11k}n_{00k}+n_{.1k}n_{10k}n_{01k})\} \\
     &=\lambda_{k,\rm GR}(n_{.0k}n_{10k}n_{01k}+n_{.1k}n_{11k}n_{00k})+(1-\lambda_{k,\rm GR})(n_{.0k}n_{11k}n_{00k}+n_{.1k}n_{10k}n_{01k})\\
     &=n_{.0k}n_{11k}n_{00k}+n_{.1k}n_{10k}n_{01k}+\lambda_{k,\rm GR}(n_{.0k}-n_{.1k})(n_{10k}n_{01k}-n_{11k}n_{00k})\\
     &=n_{.0k}n_{11k}n_{00k}+n_{.1k}n_{10k}n_{01k}+\frac{n_{.0k}^2n_{11k}-n_{.1k^2}n_{10k}}{n_{.1k}n_{.0k}}(n_{10k}n_{01k}-n_{11k}n_{00k})\\
     &=n^{-1}_{.1k}n_{.0k}^{-1}\left(n_{.0k}^3n_{11k}n_{01k}+n_{.1k}^3n_{10k}n_{00k}\right)\\
     &=n_{..k}^2\hat\var\{w_{k}(\hat\delta_k-\delta)\mid\calT_k\},
 \end{align*}
which is used in the GR variance estimator.

\subsection{Proof for Theorem \ref{theo:1}}\label{sec: proof thm1}
(a) 
We first establish the asymptotic distribution under Assumption \ref{as1_assumption} for large-stratum asymptotics. Let $\mathcal{D}=\{A_1,\cdots,A_n,Z_1,\cdots Z_n\}$.

{\bf Asymptotic distribution under large-stratum asymptotics.} First, we calculate the expectation of $\hat\delta$ conditional on $\calD$:
\begin{align*}
    E(\hat\delta_k\mid\calD)&=\frac{\sum_{i=1}^nI(A_i=1,Z_i=z_k)E(Y_i^{(1)}\mid \mathcal{D})}{\sum_{i=1}^nI(A_i=1,Z_i=z_k)}-\frac{\sum_{i=1}^nI(A_i=0,Z_i=z_k)E(Y_i^{(0)}\mid \mathcal{D})}{\sum_{i=1}^nI(A_i=0,Z_i=z_k)}\\
    &=\frac{\sum_{i=1}^nI(A_i=1,Z_i=z_k)E(Y_i^{(1)}\mid Z_i=z_k)}{\sum_{i=1}^nI(A_i=1,Z_i=z_k)}-\frac{\sum_{i=1}^nI(A_i=0,Z_i=z_k)E(Y_i^{(0)}\mid Z_i=z_k)}{\sum_{i=1}^nI(A_i=0,Z_i=z_k)}\\
    &=\frac{\sum_{i=1}^nI(A_i=1,Z_i=z_k)E(Y^{(1)}\mid Z=z_k)}{\sum_{i=1}^nI(A_i=1,Z_i=z_k)}-\frac{\sum_{i=1}^nI(A_i=0,Z_i=z_k)E(Y^{(0)}\mid Z=z_k)}{\sum_{i=1}^nI(A_i=0,Z_i=z_k)}\\
    &=p_{1k}-p_{0k}.
\end{align*}
Then we calculate its conditional variance. 
Denote $n_{1ak}/n_{.ak}$ by $\hat p_{ak}$ for $a=0,1$. We have
for $a=0,1$,
\begin{align*}
    \var\{\sqrt{n}(\hat p_{ak}-p_{ak})\mid\calD\}&=\frac{n\sum_{i=1}^nI(A_i=a,Z_i=z_k)\var(Y_i^{(a)}\mid Z_i=z_k)}{n_{.ak}^2}\\&=\frac{n\sum_{i=1}^nI(A_i=a,Z_i=z_k)\var(Y^{(a)}\mid Z=z_k)}{n_{.ak}^2}\\&=\frac{n}{n_{.ak}}\var(Y^{(a)}\mid Z=z_k)\\
    &=\frac{n}{n_{.ak}}p_{ak}(1-p_{ak}).
\end{align*}
Since $\hat p_{ak}$ is an average of independent terms and $\hat p_{1k}\perp \hat p_{0k}$ conditional on $\calD$, by the central limit theorem (CLT), we have that
\begin{align*}
    \frac{\sqrt{n}(\hat\delta_k-\delta_k)}{\sigma_k}\mid \mathcal{D}\xrightarrow{d}N(0,1),
\end{align*}
where $\sigma_k^2=\frac{n}{n_{.0k}}p_{0k}(1-p_{0k})+\frac{n}{n_{.1k}}p_{1k}(1-p_{1k})$. 

Since $\hat\delta_k$'s are mutually independent conditional on $\calD$, and $K$ is fixed and bounded, we have that $(\hat\delta_{1},\cdots,\hat\delta_{K})^T$ is jointly normal conditional on $\calD$. This implies 
\begin{align*}
    \frac{\sqrt{n}(\hat\delta-\delta_{\mh})}{\sqrt{\sum_{k=1}^K \frac{w_{k}^2\sigma_k^2}{(\sum_{k=1}^Kw_{k})^2}}}\mid\calD\xrightarrow{d}N(0,1),
\end{align*}
where 
\begin{align*}
    \sum_{k=1}^K \frac{w_{k}^2\sigma_k^2}{(\sum_{k=1}^Kw_{k})^2}&=n\sigma_{n}^2.
\end{align*}
Therefore,
\begin{align*}
    \frac{(\hat\delta-\delta_{\mh})}{\sigma_{n}}\mid\calD\xrightarrow{d}N(0,1).
\end{align*}
From the bounded convergence theorem, this result still holds unconditionally, i.e.,
\begin{align*}
    \frac{(\hat\delta-\delta_{\mh})}{\sigma_{n}}\xrightarrow{d}N(0,1).
\end{align*}

{\bf Proof of $\sigma_{n}^2=O_p(n^{-1})$.} 
Since $w_{k}/n\xrightarrow{p}\pi_1\pi_0\rho_k$, it follows that as $n\rightarrow \infty$,
\begin{align*}
    n\sigma^2_{n}&=\frac{\sum_{k=1}^K\pi_1^2\pi_0^2\rho_k\left\{\pi_1^{-1}\var(Y^{(1)}\mid Z=z_k)+\pi_0^{-1}\var(Y^{(0)}\mid Z=z_k)\right\}}{(\sum_{k=1}^K\pi_1\pi_0\rho_k)^2}+o_p(1)\\&=E\{\frac{\var(Y^{(1)}\mid Z)}{\pi_1}+\frac{\var(Y^{(0)}\mid Z)}{\pi_0}\}+o_p(1),
\end{align*}
which implies that the asymptotic variance of $n^{-1/2}\sum_{k=1}^Kw_{k}\hat\delta_k$ is $\pi_0\pi_1E\{\pi_0\var(Y^{(1)}\mid Z)+\pi_1\var(Y^{(0)}\mid Z)\}$ and generally not equal to $\pi_1\pi_0E\{\var(Y\mid Z)\}$, and $\sigma_{n}^2=O_p(n^{-1})$.




(b) 
{\bf 
Asymptotic distribution under sparse-stratum asymptotics.}
We establish the asymptotic distribution of $\hat\delta$ under Assumption \ref{sparse_assumption} for sparse-stratum asymptotics. We define $T_1$ and $T_2$ as follows:
\begin{align*}
    &\sqrt{n}(\hat{\delta}- \delta_{\mh}) = \frac{n^{-1/2}\sum_k w_{k}(\hat{\delta}_k-\delta_k)}{\sum_k n^{-1}w_{k}} :=\frac{n^{-1/2}T_1}{T_2}.
\end{align*}
To derive the asymptotic distribution of the numerator $n^{-1/2}T_1$, we decompose it as follows:
\begin{align*}
    n^{-1/2}T_1&=\sum_kn^{-1/2}\{w_{k}(\hat{\delta}_{k}- \delta_{k}) \}\\
    &= \sum_kn^{-1/2}\{\frac{n_{.1k}n_{.0k}}{n_{..k}}(\frac{n_{11k}}{n_{.1k}}-p_{1k})I(n_{.1k}\neq 0) -
    \frac{n_{.1k}n_{.0k}}{n_{..k}}(\frac{n_{10k}}{n_{.0k}}-p_{0k})I(n_{.0k}\neq 0)\} \\
    & := \sum_k Y_{1k} - Y_{2k} := \sum_k Y_k.
\end{align*}
Then, we can calculate that
\begin{align*}
    &\mu_{1,k}=E(Y_{1k}\mid\calD) = n^{-1/2}\{\frac{n_{.1k}n_{.0k}}{n_{..k}}(\frac{n_{.1k}p_{1k}}{n_{.1k}}-p_{1k})I(n_{.1k}\neq 0)\}=0, \\
    &\mu_{2,k}=E(Y_{2k}\mid\calD) = n^{-1/2}\{
    \frac{n_{.1k}n_{.0k}}{n_{..k}}(\frac{n_{.0k}p_{0k}}{n_{.0k}}-p_{0k})I(n_{.0k}\neq 0)\}=0, \mbox{ and}\\
    &\sigma_{y,k}^2=\var(Y_k\mid \calD) = \frac{1}{n}\{\frac{n_{.1k}n_{.0k}^2 }{n_{..k}^2}p_{1k}(1-p_{1k})I(n_{.1k}\neq 0) + \frac{n_{.0k}n_{.1k}^2}{n_{..k}^2}p_{0k}(1-p_{0k})I(n_{.0k}\neq 0)\}.
\end{align*}
Note that $\{Y_k\}$ satisfies Lindeberg-Feller condition, which is proven at the end. Since $C<1$ and $\epsilon<p_{0k},p_{1k}<1-\epsilon$ for any $k$, $S_K:=\sum_k \sigma_{y,k}^2$ is positive and finite. Hence, by Lindeberg's CLT, we obtain that
\begin{align*}
    \frac{\sum_k Y_k}{S_K}\mid\calD \xrightarrow{d} N(0,1).
\end{align*}
As $\calT$ and $n_{..k}$'s are fixed conditioning on $\calD$, we derive the asymptotic distribution of $\hat\delta$ when $n\rightarrow \infty:$ 
$$\frac{\sqrt{n}(\hat{\delta} -\delta_{\mh})}{\sqrt{\frac{\sum_k \sigma_{y,k}^2}{(\sum_k n^{-1}w_{k})^2}}}\mid\calD \xrightarrow{d} N(0,1).$$ \\
Note that the denominator is exactly $\sqrt{n\sigma_{n}^2}.$ By the bounded convergence theorem, we complete the proof.

{\bf Proof of $\sigma_{n}^2=O_p(n^{-1})$.} 
To show $\sigma_{n}^2$ converges at the rate of $n^{-1}$, using the fact that $p_{ak}(1-p_{ak})\leq (1/2)^2=1/4$, we can bound $\sigma_n^2$ as follows:
\begin{align*}
    \sigma_{n}^2&=\frac{\sum_{k=1}^K\frac{n_{.1k}n_{.0k}^2}{n_{..k}^2}p_{1k}(1-p_{1k})+\frac{n_{.0k}n_{.1k}^2}{n_{..k}^2}p_{0k}(1-p_{0k})}{\left(\sum_{k=1}^K\frac{n_{.1k}n_{.0k}}{n_{..k}}\right)^2}\\
    &\leq \frac{\sum_{k=1}^K\frac{n_{.1k}n_{.0k}^2}{4n_{..k}^2}+\frac{n_{.0k}n_{.1k}^2}{4n_{..k}^2}}{\left(\sum_{k=1}^K\frac{n_{.1k}n_{.0k}}{n_{..k}}\right)^2}\\
    &=\frac{1}{4}\frac{\sum_{k=1}^K\frac{n_{.1k}n_{.0k}}{n_{..k}}}{\left(\sum_{k=1}^K\frac{n_{.1k}n_{.0k}}{n_{..k}}\right)^2}\\
    &=\frac{1}{4}\left\{\sum_{k=1}^K\frac{n_{.1k}n_{.0k}}{n_{..k}}I(n_{.1k}n_{.0k}\neq 0)\right\}^{-1}\\
    &\leq \frac{1}{4}\left\{\sum_{k=1}^K\frac{n_{..k}-1}{n_{..k}}I(n_{.1k}n_{.0k}\neq 0)\right\}^{-1}\\
    &\leq \frac{1}{2(K-K_1)}.
\end{align*}
Since $\lim_{K\rightarrow\infty}K_1/K\leq C<1$ a.s. and $K/n=O(1)$, we conclude that $\sigma_{n}^2=O_p(n^{-1})$ as desired.
\\

{\bf Proof of $\{Y_k\}$ satisfying Lindeberg's condition.} 
We complete the proof by verifying  Lindeberg's condition for 
$\{Y_k\}$. For any $\epsilon>0,$
\begin{align*}
    &\sum_{k=1}^K E\left\{\frac{(Y_{1k}-Y_{2k})^2}{S_K^2} I(|Y_{1k}-Y_{2k}| \ge \epsilon S_K)\mid\calD \right\} \\
    &= \sum_{k=1}^K \frac{\sigma_{y,k}^2}{S_K^2} E\left[\frac{(Y_{1k}-Y_{2k})^2}{\sigma_{y,k}^2} I\{\frac{(Y_{1k}-Y_{2k})^2}{S_K^2} \ge \epsilon^2 \}\mid \calD\right] \\
    & \le \max_k E\left[\frac{(Y_{1k}-Y_{2k})^2}{\sigma_{y,k}^2} I\{\frac{(Y_{1k}-Y_{2k})^2}{\sigma_{y,k}^2} \ge \epsilon^2 \frac{S_K^2}{\sigma_{y,k}^2} \}\mid\calD\right] \\
    & =o(1),
\end{align*}
where the inequality results from $S_K^2 = \sum_k \sigma_{y,k}^2$ and then $\sigma^2_{y,k}/S_K^2 \le 1.$ 
The last line holds because $(Y_{1k}-Y_{2k})/\sigma_{y,k}$ has zero expectation and unit variance, and $\max_k \sigma_{y,k}^2/S_K^2=o(1)$ since $K\rightarrow\infty$, $\epsilon<p_{1k},p_{0k}<1-\epsilon$, and $\sigma_{y,k}^2$ is bounded for any $k.$

{\red 
(c){ \bf Asymptotic distribution under mixed asymptotics.}
We denote the weight of asymptotic $a$  by 
$w_a=\frac{\sum_{k\in\calK_a}w_k}{\sum_{k=1}^K w_k}$ and let $\hat\delta_a$, $\delta_{\rm MH,a}$, and $\sigma_{n,a}^2$ be $\hat\delta$, $\delta_{\rm MH}$, and $\sigma_n^2$ in $\calK_a$, for $a=l$ and $s$. Note that $\sigma_n^2=w_l^2\sigma_{n,l}^2+w_s^2\sigma^2_{n,s}$. Some calculations show that
    \begin{align*}
        \frac{\hat\delta-\delta_{\rm MH}}{\sigma_n}&=\sum_{a=l,s}\frac{w_a\sigma_{n,a}}{\sigma_n}\frac{\hat\delta_a-\delta_{\rm MH,a}}{\sigma_{n,a}}.
    \end{align*}
    Conditional on $\calD$, from parts (a) and (b), it yields that for $a=l$ and $s$,
    \begin{align*}
        \frac{\hat\delta_a-\delta_{\rm MH,a}}{\sigma_{n,a}}\mid \calD \xrightarrow{d}N(0,1).
    \end{align*}
    Conditional on $\calD$, since $w_a,\sigma_{n,a}$ are fixed and $\hat\delta_l\perp\hat\delta_s$ , we have that $(\frac{\hat\delta_l-\delta_{\rm MH,l}}{\sigma_{n,l}},\frac{\hat\delta_s-\delta_{\rm MH,s}}{\sigma_{n,s}})$ converges in distribution to a bivariate normal distribution with mean $0$ and covariance matrix equal to the $2\times2$ identity matrix. Then we complete the proof by the continuous mapping theorem and the bounded convergence theorem.
}
    
\subsection{Proof for Theorem \ref{theo: ate}}
\label{sec: proof thm2}
We decompose $\hat\delta-\delta_{\rm ATE}$ into the sum of two terms $U$ and $V$, where
\begin{align}\label{eq: decomposition U V}
\begin{split}
    &U=\hat{\delta}-\delta_{\mh},\\
    &V=\delta_{\mh} -\delta_{\rm ATE}.\\
    \end{split}
\end{align}
(a) {\bf Asymptotic distribution under large-stratum asymptotics.} We establish the asymptotic distribution of $\hat\delta-\delta_{\rm ATE}$ under large-stratum asymptotics. The asymptotic distribution of $U$ has been derived in the proof of Theorem \ref{theo:1} in \ref{sec: proof thm1}.
Now consider $V$. Note that if $\delta_1=\cdots=\delta_K=\delta_{\rm ATE}$, then $V=0$. In the following, we allow for varying risk differences. We rewrite $V$ as
\begin{align*}
    \sqrt{n}V &= \sqrt{n}(\delta_{\mh} - \delta_{\rm ATE}) 
    \\&= \sqrt{n} (\frac{\sum_k n_{.1k}n_{.0k}n_{..k}^{-1}\delta_k}{\sum_k n_{.1k}n_{.0k}n_{..k}^{-1}} -\delta_{\rm ATE}) \\
    &= \sqrt{n} (\frac{n^{-1}\sum_k n_{.1k}n_{.0k}n_{..k}^{-1}\delta_k}{n^{-1}\sum_k n_{.1k}n_{.0k}n_{..k}^{-1}} -\delta_{\rm ATE}) \\
    &=  \frac{\sqrt{n}\sum_k(\delta_k-\delta_{\rm ATE}) n_{.1k}n_{.0k}n_{..k}^{-1}n^{-1}}{\sum_k n_{.1k}n_{.0k}n_{..k}^{-1}n^{-1}} := 
    \frac{T_1}{T_2}.
\end{align*}
It's obvious that $T_2 \xrightarrow{p}\pi_1\pi_0$. Then, we show that asymptotically $T_1$ follows a normal distribution. 

Conditioning on $\mathcal{Z}=\{Z_1,\cdots,Z_n\}$, we calculate the conditional expectation, variance, and second moment of $n_{.1k}$ as follows:
\begin{align*}
    E(\frac{n_{.1k}}{n_{..k}}\mid\mathcal{Z})=\frac{\sum_{i=1}^nI(Z_i=z_k)E(A_i\mid\mathcal{Z})}{n_{..k}}=\frac{\sum_{i=1}^nI(Z_i=z_k)P(A_i=1)}{n_{..k}}=\pi_1,
\end{align*}
\begin{align*}
    \var(\frac{n_{.1k}}{n_{..k}}\mid\mathcal{Z})=\frac{\sum_{i=1}^nI(Z_i=z_k)\var(A_i)}{n_{..k}^2}=\frac{\pi_1\pi_0}{n_{..k}},
\end{align*} 
and $E(n_{.1k}^2\mid\mathcal{Z})=n_{..k}\pi_1\pi_0+n_{..k}^2\pi_1^2$. Therefore, it follows that 
\begin{align*}
E(T_1\mid\mathcal{Z})&=\sum_{k=1}^K(\delta_k-\delta_{\rm ATE})\frac{\sqrt{n}}{n}E(\frac{n_{.1k}n_{.0k}}{n_{..k}}\mid\mathcal{Z}
    )\\
    &=\sum_{k=1}^K(\delta_k-\delta_{\rm ATE})\frac{\sqrt{n}}{n}E\{\frac{n_{.1k}(n_{..k}-n_{.1k})}{n_{..k}}\mid\mathcal{Z}
    \}\\
    &=\sum_{k=1}^K(\delta_k-\delta_{\rm ATE})\frac{\sqrt{n}}{n}E(n_{.1k}-\frac{n_{.1k}^2}{n_{..k}}\mid\mathcal{Z}
    )\\
    &=\sum_{k=1}^K(\delta_k-\delta_{\rm ATE})\frac{\sqrt{n}}{n}(n_{..k}-1)\pi_1\pi_0
\end{align*}
and
\begin{align*}
    T_1-E(T_1\mid\mathcal{Z})=\sum_{k=1}^K(\delta_k-\delta_{\rm ATE})\sqrt{n_{..k}}\frac{\sqrt{n_{..k}}}{\sqrt{n}}(\frac{n_{.1k}n_{.0k}}{n_{..k}^2}-\frac{n_{..k}-1}{n_{..k}}\pi_1\pi_0).
\end{align*}
Using the fact that $n_{.1k}/n_{..k}$ is an average of independent terms conditioning on $\mathcal{Z}$, CLT gives
\begin{align*}
    \sqrt{n_{..k}}(\frac{n_{.1k}}{n_{..k}}-\pi_1)\mid\mathcal{Z}\xrightarrow{d}N(0,\pi_1\pi_0).
\end{align*}
From the Delta method with $x\mapsto x(1-x)$ and Slutsky's lemma, we have that
\begin{align*}
    \sqrt{n_{..k}}\frac{\sqrt{n_{..k}}}{\sqrt{n}}(\frac{n_{.1k}n_{.0k}}{n_{..k}^2}-\pi_1\pi_0)\mid\mathcal{Z}\xrightarrow{d}N(0,(\pi_1-\pi_0)^2\pi_1\pi_0\rho_k).
\end{align*}
Notice that $n_{.1k}$'s are mutually independent conditional on $\mathcal{Z}$. Since $K<\infty$, it turns out that conditional on $\mathcal{Z}$,
\begin{align*}
    &\ T_1-E(T_1\mid\mathcal{Z})\\
    &=\sum_{k=1}^K(\delta_k-\delta_{\rm ATE})\sqrt{n_{..k}}\frac{\sqrt{n_{..k}}}{\sqrt{n}}(\frac{n_{.1k}n_{.0k}}{n_{..k}^2}-\pi_1\pi_0)+\frac{\pi_1\pi_0}{\sqrt{n}}\sum_{k=1}^K(\delta_k-\delta_{\rm ATE})\\
    &=\sum_{k=1}^K(\delta_k-\delta_{\rm ATE})\sqrt{n_{..k}}\frac{\sqrt{n_{..k}}}{\sqrt{n}}(\frac{n_{.1k}n_{.0k}}{n_{..k}^2}-\pi_1\pi_0)+o(1)\\
    &\xrightarrow{d}N(0,(\pi_1-\pi_0)^2\pi_1\pi_0\sum_{k=1}^K(\delta_k-\delta_{\rm ATE})^2\rho_k),
\end{align*}
where we have used that $n^{-1/2}\sum_{k=1}^K(\delta_k-\delta_{\rm ATE})=o(1)$. By the bounded convergence theorem, it holds unconditionally. 

Then, we derive the asymptotic distribution of $E(T_1\mid \mathcal{Z})-E(T_1)$. Recall that $E(n_{..k})=n\rho_k$. Some calculations show that
\begin{align*}
    &\ E(T_1\mid\mathcal{Z})-E(T_1)\\
    &=\pi_1\pi_0\frac{\sqrt{n}}{n}\sum_{k=1}^K(\delta_k-\delta_{\rm ATE})(n_{..k}-n\rho_k)\\
    &=\pi_1\pi_0\frac{\sqrt{n}}{n}\sum_{k=1}^K(\delta_k-\delta_{\rm ATE})n_{..k}\\
    &=\pi_1\pi_0\frac{\sqrt{n}}{n}\sum_{k=1}^K(\delta_k-\delta_{\rm ATE})\sum_{i=1}^nI(Z_i=z_k)\\
    &=\pi_1\pi_0\frac{\sqrt{n}}{n}\sum_{i=1}^n\sum_{k=1}^K\{E(Y_i^{(1)}-Y_i^{(0)}\mid Z_i=z_k)-\delta_{\rm ATE}\}I(Z_i=z_k)\\
    &=\pi_1\pi_0\frac{\sqrt{n}}{n}\sum_{i=1}^n\{E(Y_i^{(1)}-Y_i^{(0)}\mid Z_i)-\delta_{\rm ATE}\}.
\end{align*}
By CLT, we have that $
     E(T_1\mid\mathcal{Z})-E(T_1)\xrightarrow{d}N(0,\pi_1^2\pi_0^2\var\{E(Y^{(1)}-Y^{(0)}\mid \mathcal{Z})\})$, 
where $\var\{E(Y^{(1)}-Y^{(0)}\mid \mathcal{Z})\}=\sum_k\rho_k(\delta_k^2-\delta_{\rm ATE}^2)$ by the definition of $\delta_k$. 

Let $\phi_u$ and $\phi_v$ be the random variables, the distributions of which are the same as the asymptotic distributions of $T_1-E(T_1\mid\mathcal{Z})$ and $E(T_1\mid\mathcal{Z})-E(T_1)$ respectively. Notice that $T_1-E(T_1\mid\mathcal{Z})$ and $E(T_1\mid\mathcal{Z})-E(T_1)$ are asymptotically independent because

\begin{equation}\label{eq: asymp indep}
\begin{aligned}
&P\bigl[\;T_1 - E(T_1\mid\mathcal{Z}) \le u,\;E(T_1\mid\mathcal{Z}) - E(T_1)\le v\;\bigr]\\
&=E\Bigl[\;I\bigl\{\;T_1 - E(T_1\mid\mathcal{Z}) \le u,\;E(T_1\mid\mathcal{Z}) - E(T_1)\le v\;\bigr\}\;\Bigr]\\
&=E\Bigl[\;I\bigl\{\;E(T_1\mid\mathcal{Z}) - E(T_1)\le v\;\bigr\}\;
    E\bigl\{\;I\bigl(T_1 - E(T_1\mid\mathcal{Z}) \le u\bigr)\;\bigm|\;\mathcal{Z}\bigr\}\;\Bigr]\\
&=E\Bigl[\;I\bigl\{\;E(T_1\mid\mathcal{Z}) - E(T_1)\le v\;\bigr\}\;
    \Bigl(\;E\bigl\{I\bigl(T_1 - E(T_1\mid\mathcal{Z}) \le u\bigr)\mid\mathcal{Z}\bigr\}
           -E\{\,I(\phi_u \le u)\}\Bigr)\;\Bigr]\\
&\quad +P(\phi_u\le u)\;P\bigl[\;E(T_1\mid\mathcal{Z}) - E(T_1)\le v\;\bigr]\\
&\longrightarrow P(\phi_u\le u)\;P(\phi_v\le v).
\end{aligned}
\end{equation}
where last line follows from the bounded convergence theorem.

Combining the results for $T_1-E(T_1\mid\mathcal{Z})$ and $E(T_1\mid\mathcal{Z})-E(T_1)$ and that these two terms are asymptotically independent (i.e. their asymptotic joint distribution is the product of their asymptotic marginal distributions),
we have that
\begin{align*}
    \frac{T_1-E(T_1)}{\sqrt{\sum_{k=1}^K\rho_k\left\{(\delta_k-\delta_{\rm ATE})^2(\pi_1-\pi_0)^2\pi_1\pi_0+\pi_1^2\pi_0^2(\delta_k^2-\delta_{\rm ATE}^2)\right\}}}\xrightarrow{d}N(0,1).
\end{align*}
Since $E(T_1)=\frac{\pi_1\pi_0}{\sqrt{n}}\sum_{k=1}^K(\delta_k-\delta_{\rm ATE})(n\rho_k-1)=o(1)$ and $T_2\xrightarrow{p}\pi_1\pi_0$, we derive the asymptotic distribution of $V$ as follows:
\begin{align*}
    \frac{V}{\sqrt{\pi_1^{-1}\pi_0^{-1}\sum_{k=1}^K\rho_k\left\{(\delta_k-\delta_{\rm ATE})^2(\pi_1-\pi_0)^2+\pi_1\pi_0(\delta_k^2-\delta_{\rm ATE}^2)\right\}}}\xrightarrow{d}N(0,1).
\end{align*}
Finally, since $U$ and $V$ are asymptotically independent following from the similar derivations in \eqref{eq: asymp indep}, we conclude that
\begin{align*}
    \frac{\hat\delta-\delta_{\rm ATE}}{\sqrt{\sigma^2_{n}+\nu^2_{n,1}}}\xrightarrow{d}N(0,1),
\end{align*}
where $ \sigma_{n}^2$ is defined in \eqref{eq:cmh var} and $\nu^2_{n,1}=n^{-1}\pi_1^{-1}\pi_0^{-1}\sum_{k=1}^K\rho_k\left\{(\delta_k-\delta_{\rm ATE})^2(\pi_1-\pi_0)^2+\pi_1\pi_0(\delta_k^2-\delta_{\rm ATE}^2)\right\}$.

(b) {\bf Asymptotic distribution under sparse-stratum asymptotics.} We use the same decomposition \eqref{eq: decomposition U V} for $\hat\delta-\delta_{\rm ATE}$. From the results in the proof of Theorem \ref{theo:1} in \ref{sec: proof thm1}, we have that
\begin{align*}
    \frac{U}{\sigma_{n}}\xrightarrow{d}N(0,1).
\end{align*}

To derive the asymptotic distribution of $V=T_1/T_2$ where $T_1=\sqrt{n}\sum_k(\delta_k-\delta_{\rm ATE}) n_{.1k}n_{.0k}n_{..k}^{-1}n^{-1}$ and $T_2=\sum_k n_{.1k}n_{.0k}n_{..k}^{-1}n^{-1}$, we follow the same steps to deal with $T_1$ by considering $T_1-E(T_1)=T_1-E(T_1\mid \mathcal{Z})+E(T_1\mid \mathcal{Z})-E(T_1)$. Because $T_1-E(T_1\mid \mathcal{Z})$ is the sum of independent terms conditional on $\mathcal{Z}$, by Lindeberg's CLT, it turns out that
\begin{align}
\label{eq: ATE cid var w_k}
    \frac{\sum_{k=1}^K(\delta_k-\delta_{\rm ATE})\left(\frac{w_{k}}{n}-\frac{n_{..k}-1}{n}\pi_1\pi_0\right)}{\sqrt{\sum_{k=1}^K(\delta_k-\delta_{\rm ATE})^2n^{-2}\var(w_{k}\mid \mathcal{Z})}}\mid\mathcal{Z}\xrightarrow{d}N(0,1).
\end{align}
Then, we focus on calculating $\var(w_{k}\mid\mathcal{Z})$. For notational simplicity, we denote $n_{..k}$ with $n_k$ in the following derivations. By the definition of $w_k$, it follows that
\begin{align}
    \label{eq: var w_k}\var(w_{k}\mid\mathcal{Z})&=E(w_{k}^2\mid\mathcal{Z})-E(w_{k}\mid\mathcal{Z})^2\nonumber\\
    &=\frac{1}{n_{k}^2}E\{n_{.1k}^2(n_{k}-n_{.1k})^2\mid\mathcal{Z}\}-(n_{k}-1)^2\pi_1^2\pi_0^2.
\end{align}
Some calculations show that
\begin{align}
&E(n_{.1k}^2\mid\mathcal{Z})=n_k\pi_1\pi_0+n_k^2\pi_1^2=n_k\pi_1+n_k(n_k-1)\pi_1^2, \nonumber\\
&E(n_{.1k}^3\mid\mathcal{Z})=n_k(n_k-1)(n_k-2)\pi_1^3 + 3n_k(n_k-1)\pi_1^2 + n_k\pi_1, \mbox{ and}\label{eq: w_k moments}\\ 
&E(n_{.1k}^4\mid\mathcal{Z})=n_k(n_k-1)(n_k-2)(n_k-3)\pi_1^4 + 6n_k(n_k-1)(n_k-2)\pi_1^3+7n_k(n_k-1)\pi_1^2+n_k\pi_1. \nonumber
\end{align}
Plugging the moments into \eqref{eq: var w_k}, we obtain the following result:
\begin{align*}
&\ E\{n_{.1k}^2(n_{k}-n_{.1k})^2\mid\mathcal{Z}\}\\
&=E(n_k^2n_{.1k}^2-2n_kn_{.1k}^3+n_{.1k}^4\mid\mathcal{Z})\\
&= n_k(n_k-1)(n_k-2)(n_k-3)\pi_1^4 + 6n_k(n_k-1)(n_k-2)\pi_1^3+7n_k(n_k-1)\pi_1^2+n_k\pi_1\\
&\ -2n_k^2(n_k-1)(n_k-2)\pi_1^3 - 6n_k^2(n_k-1)\pi_1^2 -2 n_k^2\pi_1 + n_k^3\pi_1+n_k^3(n_k-1)\pi_1^2\\
&=n_k\left\{(n_k-1)(n_k-2)(n_k-3)\pi_1^4+(6-2n_k)(n_k-1)(n_k-2)\pi_1^3+(n_k^2-6n_k+7)(n_k-1)\pi_1^2+(n_k-1)^2\pi_1\right\}.
\end{align*}
Notice that
\begin{align*}
    &\ (n_k-1)(n_k-2)(n_k-3)\pi_1^4+(6-2n_k)(n_k-1)(n_k-2)\pi_1^3\\
    &=(n_k-1)(n_k-2)\left\{(n_k-3)\pi_1^3(1-\pi_0)+(6-2n_k)\pi_1^3\right\}\\
    &=-(n_k-1)(n_k-2)(n_k-3)\pi_1^3(1+\pi_0)\\
    &=-(n_k-1)(n_k-2)(n_k-3)\pi_1^2(1-\pi_0^2)
\end{align*}
and
\begin{align*}
   &\ -(n_k-1)(n_k-2)(n_k-3)\pi_1^2(1-\pi_0^2)+(n_k^2-6n_k+7)(n_k-1)\pi_1^2\\
   &=(n_k-1)\left\{-(n_k-2)(n_k-3)+n_k^2-6n_k+7\right\}\pi_1^2+(n_k-1)(n_k-2)(n_k-3)\pi_1^2\pi_0^2\\
   &=-(n_k-1)^2\pi_1^2+(n_k-1)(n_k-2)(n_k-3)\pi_1^2\pi_0^2.
\end{align*}
Hence, we can simplify the expression of $E\{n_{.1k}^2(n_{k}-n_{.1k})^2\mid\mathcal{Z}\}$ as follows:
\begin{equation}\label{eq: tmp2}
\begin{aligned}
    &\ E\{n_{.1k}^2(n_{k}-n_{.1k})^2\mid\mathcal{Z}\}\\
    &=n_k\left\{-(n_k-1)^2\pi_1^2+(n_k-1)(n_k-2)(n_k-3)\pi_1^2\pi_0^2+(n_k-1)^2\pi_1\right\}\\
    &=\pi_1\pi_0n_k(n_k-1)\left\{n_k-1+(n_k-2)(n_k-3)\pi_1\pi_0\right\}.
\end{aligned}
\end{equation}
Then, substituting the simplified expression into \eqref{eq: var w_k} gives that 
\begin{align}
   \var(w_{k}\mid\mathcal{Z})&=\frac{1}{n_{k}^2}E\{Pn_{.1k}^2(n_{k}-n_{.1k})^2\mid\mathcal{Z}\}-(n_{k}-1)^2\pi_1^2\pi_0^2\nonumber\\
   &=\pi_1\pi_0\frac{n_k-1}{n_k}\left\{n_k-1+(n_k-2)(n_k-3)\pi_1\pi_0\right\}-(n_{k}-1)^2\pi_1^2\pi_0^2\label{eq: var w_k simplified}\\ 
   &=\pi_1\pi_0\frac{n_k-1}{n_k}\left\{n_k-1-(4n_k-6)\pi_1\pi_0\nonumber\right\}.
\end{align}
Plugging the value of $\var(w_k\mid\mathcal{Z})$ into \eqref{eq: ATE cid var w_k} leads to the following conditional convergence in distribution of $T_1-E(T_1\mid\mathcal{Z})$, that is
\begin{align*}
    \frac{\sum_{k=1}^K(\delta_k-\delta_{\rm ATE})\left(\frac{w_{k}}{n}-\frac{n_{k}-1}{n}\pi_1\pi_0\right)}{\sqrt{\sum_{k=1}^K(\delta_k-\delta_{\rm ATE})^2n^{-2}\pi_1\pi_0\frac{n_k-1}{n_k}\left\{n_k-1-(4n_k-6)\pi_1\pi_0\right\}}}\mid\mathcal{Z}\xrightarrow{d}N(0,1),
\end{align*}
and further by the bounded convergence theorem, it follows that
\begin{align*}
    \frac{\sum_{k=1}^K(\delta_k-\delta_{\rm ATE})\left(\frac{w_{k}}{n}-\frac{n_{k}-1}{n}\pi_1\pi_0\right)}{\sqrt{n^{-1}\sum_{k=1}^K(\delta_k-\delta_{\rm ATE})^2\pi_1\pi_0\frac{n_k-1}{n_k}\frac{n_k-1-(4n_k-6)\pi_1\pi_0}{n}}}\xrightarrow{d}N(0,1).
\end{align*}

The derivation for the asymptotic distribution of $E(T_1\mid\mathcal{Z})-E(T_1)$ follows the same steps under large-stratum asymptotics, where we can interchange $\sum_{i=1}^n$ and $\sum_{k=1}^K$ because $|\sum_{k=1}^K(\delta_k-\delta_{\rm ATE})I(Z_i=z_k)|\leq \sup_k |\delta_k-\delta_{\rm ATE}|<\infty$. Then, combining the results of $T_1-E(T_1\mid\mathcal{Z})$ and $E(T_1\mid\mathcal{Z})-E(T_1)$ and using the fact that they are asymptotically independent, we have that
\begin{align}\label{eq: ATE 1}
    \frac{n^{-1/2}\{T_1-E(T_1)\}}{\sqrt{n^{-1}\sum_{k=1}^K(\delta_k-\delta_{\rm ATE})^2\pi_1\pi_0\frac{n_k-1}{n_k}\frac{n_k-1-(4n_k-6)\pi_1\pi_0}{n} + n^{-1}\pi_1^2\pi_0^2\sum_{k}\rho_k(\delta_k^2-\delta_{\rm ATE}^2)}}\xrightarrow{d}N(0,1).
\end{align}
Since $n^{-1/2}E(T_1)=\sum_k(\delta_k-\delta_{\rm ATE})(\rho_k-1/n)\pi_1\pi_0=-\pi_1\pi_0\sum_{k}(\delta_k-\delta_{\rm ATE})/n=o(1/\sqrt{n})$ and the denominator in \eqref{eq: ATE 1} is $O_p(1)$, it results in that
\begin{align*}
    \frac{n^{-1/2}T_1}{\sqrt{n^{-1}\sum_{k=1}^K(\delta_k-\delta_{\rm ATE})^2\pi_1\pi_0\frac{n_k-1}{n_k}\frac{n_k-1-(4n_k-6)\pi_1\pi_0}{n} + n^{-1}\pi_1^2\pi_0^2\sum_{k}\rho_k(\delta_k^2-\delta_{\rm ATE}^2)}}\xrightarrow{d}N(0,1).
\end{align*}

Finally, we deal with $T_2=n^{-1}\sum_{k=1}^Kw_k$. By \eqref{eq: var w_k simplified}, we can see that
\begin{align*}
    \frac{1}{K^2}\sum_{k=1}^K\var(w_k\mid\mathcal{Z})&=\frac{1}{K^2}\sum_{k=1}^K\pi_1\pi_0\frac{n_k-1}{n_k}\left\{n_k-1+(n_k-2)(n_k-3)\pi_1\pi_0\right\}\leq \frac{1}{K^2}\sum_{k=1}^Kn_k=\frac{n}{K^2}=o(1).
\end{align*}
Since $w_k$'s are mutually independent conditional on $\mathcal{Z}$, by \eqref{eq: w_k moments} and Lemma \ref{lemma1}, we have that conditional on $\mathcal{Z}$, $T_2=(1-K/n)\pi_1\pi_0+o_p(1)$, which means $T_2$ converges to some constant in probability. Combining the results of $T_1$ and $T_2$ and using the Slutsky's lemma, we obtain the asymptotic distribution of $V$. 

From the results of $U$ and $V$, following the same steps in the proof under large-stratum asymptotics, we conclude that
\begin{align*}
    \frac{(\hat\delta-\delta_{\rm ATE})}{\sqrt{\sigma_{n}^2 + \nu_{n}^2}}\xrightarrow{d}N(0,1),
\end{align*}
where $\sigma_n^2$ is defined in \eqref{eq:cmh var} and 
\begin{align*}
     \nu_{n}^2 = n^{-1}(\sum_{k=1}^K\frac{w_{k}}{n})^{-2}\left\{\sum_{k=1}^K(\delta_k-\delta_{\rm ATE})^2\pi_1\pi_0\frac{n_k-1}{n_k}\frac{n_k-1-(4n_k-6)\pi_1\pi_0}{n}+\pi_1^2\pi_0^2\sum_{k=1}^K\rho_k(\delta_k^2-\delta_{\rm ATE}^2)\right\}.
\end{align*}

{\bf Proof of $T_1-E(T_1\mid\mathcal{Z})$ satisfying Lindeberg's condition.} For the completeness, we show that the Lindeberg's condition is satisfied by $T_1-E(T_1\mid\mathcal{Z})$. Since $n_{..k}\leq c$, $C<1$, and $\max_j(\delta_j-\delta_{\rm ATE})^2/\sum_{k=1}^K(\delta_k-\delta_{\rm ATE})^2=o(1)$, it follows that conditional on $\mathcal{Z}$,
\begin{align*}
    &\max_j \frac{(\delta_j-\delta_{\rm ATE})^2\frac{n_j-1}{n_j}\{n_j-1-(4n_j-6)\pi_1\pi_0\}}{\sum_{k=1}^K(\delta_k-\delta_{\rm ATE})^2\frac{n_k-1}{n_k}\{n_k-1-(4n_k-6)\pi_1\pi_0\}}\\
    &\leq \max_j \frac{(\delta_j-\delta_{\rm ATE})^2n_j^2/n_j}{\sum_{k=1}^K(\delta_k-\delta_{\rm ATE})^2\frac{n_k-1}{2n_k}}\\
    &\leq \frac{c}{4}\max_j\frac{(\delta_j-\delta_{\rm ATE})^2}{\sum_{k=1}^K(\delta_k-\delta_{\rm ATE})^2}\\
    &=o(1).
\end{align*}
 The rest of the proof follows the same steps in the proof for Theorem \ref{theo:1} in \ref{sec: proof thm1}.

{\red
(c) {\bf Asymptotic distribution under mixed asymptotics.} The proof follows analogous steps to those in part (c) of Section \ref{sec: proof thm1}.
}

 (d) {\bf $\nu_n^2$ is a unified variance.} We will show that $\nu_n^2-\nu_{n,1}^2$ converges to $0$ in probability under the large-stratum asymptotics. Under the large-stratum asymptotics, we have that $n^{-1}\sum_{k=1}^Kw_{k}\xrightarrow{p}\pi_1\pi_0$. Therefore, plugging that $n^{-1}\sum_{k=1}^Kw_k=\pi_1\pi_0+o_p(1)$ into $\nu_n^2$ gives that 
\begin{align*}
    \nu_{n}^2&=\frac{1}{n}\pi_1^{-1}\pi_0^{-1}\left\{\sum_{k=1}^K\rho_k(\delta_k-\delta_{\rm ATE})^2(1-4{\pi_1\pi_0})+\pi_1\pi_0\sum_{k=1}^K\rho_k(\delta_k^2-\delta_{\rm ATE}^2)\right\}+o_p(1)\\
    &=\frac{1}{n}\pi_1^{-1}\pi_0^{-1}\left\{\sum_{k=1}^K\rho_k(\delta_k-\delta_{\rm ATE})^2((\pi_1+\pi_0)^2-4{\pi_1\pi_0})+\pi_1\pi_0\sum_{k=1}^K\rho_k(\delta_k^2-\delta_{\rm ATE}^2)\right\}+o_p(1)\\
    &=\frac{1}{n}\pi_1^{-1}\pi_0^{-1}\sum_{k=1}^K\rho_k\left\{(\delta_k-\delta_{\rm ATE})^2(\pi_1-\pi_0)^2+\pi_1\pi_0(\delta_k^2-\delta_{\rm ATE}^2)\right\}+o_p(1)\\
    &=\nu_{n, 1}^2+o_p(1),
\end{align*}
which implies that $\nu_n^2$ is a unified variance under both large- and sparse-stratum asymptotics. This observation also indicates that $\hat\nu^2$ is a unified consistent variance estimator as long as $\hat\nu^2$ is consistent of $\nu_n^2$, which is proved in \ref{sec: mGR ATE}.


\subsection{Proof for Theorem \ref{gen_th}}\label{sec: proof thm3}
(a) {\bf Asymptotic distribution under sparse-stratum asymptotics.}
Decompose $(\hat{\delta}_{\rm PS}-\delta_{\rm ATE})$ as follows: $\hat{\delta}_{\rm PS}-\delta_{\rm ATE} = U+V$, where
\begin{align*}
    &U =\sum_k \frac{n_{..k}}{n}\{(\Bar{Y}_{1k} - \Bar{Y}_{0k})I(n_{.1k}n_{.0k}\neq 0) -\delta_k\}:= \sum_k U_k,\\
    &V=\sum_k \frac{n_{..k}}{n}\delta_k -\delta_{\rm ATE}.
\end{align*}
Following the steps in the proof of Theorem \ref{theo: ate} in \ref{sec: proof thm1}, we have that 
\begin{align*}
    &E(U_k\mid \calD)=-\frac{n_{..k}}{n}\delta_kI(n_{.1k}n_{.0k}=0),\\
    &\var(\sqrt{n}U_k\mid\calD)=I(n_{.1k}n_{.0k}\neq 0)\frac{n_{..k}}{n}\left\{\frac{\var(Y^{(1)}\mid Z=z_k)}{n_{.1k}/n_{..k}}+\frac{\var(Y^{(0)}\mid Z=z_k)}{n_{.0k}/n_{..k}}\right\}.
\end{align*}
By Lindeberg's CLT, we obtain the conditional convergence of $U$, that is
\begin{align}\label{eq: U PS}
    \frac{\sqrt{n}\left\{U+\sum_{k=1}^K\frac{n_{..k}}{n}\delta_kI(n_{.1k}n_{.0k}= 0)\right\}}{\sqrt{\sum_{k=1}^KI(n_{.1k}n_{.0k}\neq 0)\frac{n_{..k}}{n}\left\{\frac{\var(Y^{(1)}\mid Z=z_k)}{n_{.1k}/n_{..k}}+\frac{\var(Y^{(0)}\mid Z=z_k)}{n_{.0k}/n_{..k}}\right\}}}\mid\calD\xrightarrow{d}N(0,1).
\end{align}
From the bounded convergence theorem, it holds unconditionally.
Notice that the denominator is on the order of $1$ with probability approaching 1, which results from that
\begin{align*}
    &\sum_{k=1}^KI(n_{.1k}n_{.0k}\neq 0)\frac{n_{..k}}{n}\left\{\frac{\var(Y^{(1)}\mid Z=z_k)}{n_{.1k}/n_{..k}}+\frac{\var(Y^{(0)}\mid Z=z_k)}{n_{.0k}/n_{..k}}\right\}\\
    &\geq 2\frac{K-K_1}{n}\min_k\left\{\var(Y^{(1)}\mid Z=z_k)+\var(Y^{(0)}\mid Z=z_k)\right\}\\
    &\geq2(\frac{K-K_1}{cK})\min_k\left\{\var(Y^{(1)}\mid Z=z_k)+\var(Y^{(0)}\mid Z=z_k)\right\}
\end{align*}
and
\begin{align*}
    &\sum_{k=1}^KI(n_{.1k}n_{.0k}\neq 0)\frac{n_{..k}}{n}\left\{\frac{\var(Y^{(1)}\mid Z=z_k)}{n_{.1k}/n_{..k}}+\frac{\var(Y^{(0)}\mid Z=z_k)}{n_{.0k}/n_{..k}}\right\}\\
    &\leq \frac{c^2(K-K_1)}{n}\max_k\left\{\var(Y^{(1)}\mid Z=z_k)+\var(Y^{(0)}\mid Z=z_k)\right\}.
\end{align*}
Therefore, the bias term in the numerator in \eqref{eq: U PS} goes to zero because $|\sum_{k=1}^K\frac{n_{..k}}{n}\delta_kI(n_{.1k}n_{.0k}= 0)|\leq \frac{cK_1}{n}\sup_k|\delta_k|=o_p(n^{-1/2})$ under the condition $K_1/\sqrt{K}\to 0$ almost surely. 

Next, we derive the asymptotic distribution of $V$. Recall that $V=\sum_{k=1}^K\frac{n_{..k}}{n}\delta_k-\delta_{\rm ATE}.$ We rewrite $V$ as follows:
\begin{align*}
    V&=\sum_{k=1}^K\frac{n_{..k}}{n}\delta_k-\delta_{\rm ATE}\\
    &=\frac{1}{n}\sum_{k=1}^K\sum_{i=1}^n\delta_kI(Z_i=z_k)-\delta_{\rm ATE}\\
    &=\frac{1}{n}\sum_{i=1}^n\sum_{k=1}^KE(Y_i^{(1)}-Y_i^{(0)}\mid Z_i=z_k)I(Z_i=z_k)-\delta_{\rm ATE}\\
    &=\frac{1}{n}\sum_{i=1}^nE(Y_i^{(1)}-Y_i^{(0)}\mid Z_i)-\delta_{\rm ATE},
\end{align*}
where we can interchange $\sum_{k=1}^K$ and $\sum_{i=1}^n$ because $|\sum_{k=1}^K\delta_k I(Z_i=z_k)|\leq \sup_k|\delta_k|<\infty.$ Since $Z_i$'s are i.i.d and $E\{E(Y_i^{(1)}-Y_i^{(0)}\mid Z_i)\}=\delta_{\rm ATE}$, by CLT, we have that
\begin{align*}
    \frac{\sqrt{n}V}{\sqrt{\var\{E(Y^{(1)}-Y^{(0)}\mid\mathcal{Z})\}}}\xrightarrow{d}N(0,1).
\end{align*}

Combining the results regarding $U$ and $V$ and using that $U$ and $V$ are asymptotically independent, we conclude that
\begin{align*}
    \frac{\sqrt{n}(\hat\delta_{\rm PS}-\delta_{\rm ATE})}{\sqrt{\sum_{k=1}^KI(n_{.1k}n_{.0k}\neq 0)\frac{n_{..k}}{n}\left\{\frac{\var(Y^{(1)}\mid Z=z_k)}{n_{.1k}/n_{..k}}+\frac{\var(Y^{(0)}\mid Z=z_k)}{n_{.0k}/n_{..k}}\right\} + \var\{E(Y^{(1)}-Y^{(0)}\mid Z)\}}}\xrightarrow{d}N(0,1).
\end{align*}

{\bf Proof of $\{U_k\}$ satisfying Lindeberg's condition.} For the completeness, we show that $\{U_k\}_{k=1}^K$ satisfies Lindeberg's condition. We can see that conditional on $\calD$,
\begin{align*}
    &\max_j\frac{I(n_{.1j}n_{.0j}\neq 0)\frac{n_{..j}}{n}\left\{\frac{\var(Y^{(1)}\mid Z=z_j)}{n_{.1j}/n_{..j}}+\frac{\var(Y^{(0)}\mid Z=z_j)}{n_{.0j}/n_{..j}}\right\}}{\sum_{k=1}^KI(n_{.1k}n_{.0k}\neq 0)\frac{n_{..k}}{n}\left\{\frac{\var(Y^{(1)}\mid Z=z_k)}{n_{.1k}/n_{..k}}+\frac{\var(Y^{(0)}\mid Z=z_k)}{n_{.0k}/n_{..k}}\right\}}\\
    &\leq\frac{\max_j\frac{c}{n}\left\{\frac{\var(Y^{(1)}\mid Z=z_j)}{1/c}+\frac{\var(Y^{(0)}\mid Z=z_j)}{1/c}\right\}}{\sum_{k=1}^KI(n_{.1k}n_{.0k}\neq 0)\frac{1}{n}\left\{\frac{\var(Y^{(1)}\mid Z=z_k)}{1}+\frac{\var(Y^{(0)}\mid Z=z_k)}{1}\right\}}\\
    &\leq \frac{c^2}{2(K-K_1)}\frac{\max_k\{\var(Y^{(1)}\mid Z=z_k)+\var(Y^{(0)}\mid Z=z_k)\}}{\min_k\{\var(Y^{(1)}\mid Z=z_k)+\var(Y^{(0)}\mid Z=z_k)\}}=o(1),
\end{align*}
where we have used that $n_{..k}\leq c$.
The rest of the proof follows the same derivation in the proof for Theorem \ref{theo:1} in \ref{sec: proof thm1}.

(b) {\bf Asymptotic distribution under mixed asymptotics.} The proof follows analogous steps to those in part (c) of Section \ref{sec: proof thm1}.

(c) {\bf $\sigma_{n,\rm PS}^2$ is a unified variance.} We will show that $\sigma_{n,\rm PS}^2=E\{\frac{\var(Y^{(1)}\mid Z)}{\pi_1}+\frac{\var(Y^{(0)}\mid Z)}{\pi_0}\}+o_p(1)$ under the large-stratum asymptotics. Under the large-stratum asymptotics, we have that $n_{.1k}n_{.0k}\neq 0$ almost surely, $n_{..k}/n\xrightarrow{p}\rho_k$, and $n_{.ak}/n_{..k}\xrightarrow{p}\pi_a$. Therefore, it follows that
\begin{align*}
    &\ \sum_{k=1}^KI(n_{.1k}n_{.0k}\neq 0)\frac{n_{..k}}{n}\left\{\frac{\var(Y^{(1)}\mid Z=z_k)}{n_{.1k}/n_{..k}}+\frac{\var(Y^{(0)}\mid Z=z_k)}{n_{.0k}/n_{..k}}\right\}\\
    &= \sum_{k=1}^K\rho_k\left\{\frac{\var(Y^{(1)}\mid Z=z_k)}{\pi_1}+\frac{\var(Y^{(0)}\mid Z=z_k)}{\pi_0}\right\}+o_p(1)\\
    &=E\{\frac{\var(Y^{(1)}\mid Z)}{\pi_1}+\frac{\var(Y^{(0)}\mid Z)}{\pi_0}\}+o_p(1),
\end{align*}
which implies that $\sigma_{n,\rm PS}^2$ is a unified variance under both large- and sparse-stratum asymptotics. This observation also indicates that $\hat\sigma_{\rm PS}^2$ is a unified consistent variance estimator as long as $\hat\sigma_{\rm PS}^2$ is consistent of $\sigma_{n,\rm PS}^2$, which is proved in \ref{sec: PS consistent}.

\subsection{Consistency of the GR variance estimator}\label{sec: consistency GR}
Recall that under both large- and sparse-stratum asymptotics, it holds that
\begin{align*}
    \frac{\hat\delta-\delta_{\mh}}{\sigma_n}\xrightarrow{d}N(0,1), \mbox{ where }\sigma_n^2=\frac{1}{(\sum_{k=1}^Kw_k)^2}\sum_{k=1}^K\frac{n_{.1k}n_{.0k}^2p_{1k}(1-p_{1k})+n_{.1k}^2n_{.0k}p_{0k}(1-p_{0k})}{n_{..k}^2}.
\end{align*}
Also, recall that the GR variance estimator $\hat\sigma_{\gr}^2$ uses $n_{11k}/n_{.1k}$ and $n_{10k}/n_{.0k}$ to estimate $p_{1k}$ and $p_{0k}$ respectively. 

(a)
{\bf Consistency under large-stratum asymptotics.} Under Assumption \ref{as1_assumption}, since $K<\infty$, $\hat p_{ak}=p_{ak}+o_p(1)$, $n_{.ak}/n_{..k}=\pi_a+o_p(1)$, and $n_{..k}/n=\rho_k+o_p(1)$, we have that $n\hat\sigma_{\gr}^2-n\sigma_n^2=o_p(1)$, i.e., the GR variance estimator is consistent.

(b) {\bf Inconsistency under sparse-stratum asymptotics.}
However, the GR variance estimator is biased under Assumption \ref{sparse_assumption}. We first calculate the expectation of $\hat p_{ak}(1-\hat p_{ak})$ conditional on $\calD$ as follows:
\begin{align*}
    E\{\hat p_{ak}(1-\hat p_{ak})\mid\calD\}&=\frac{1}{n_{.ak}^2}E(n_{.ak}n_{1ak}-n_{1ak}^2\mid\calD)=\frac{n_{.ak}-1}{n_{.ak}}p_{ak}(1-p_{ak}).
\end{align*}
Plugging it into  $E(\hat\sigma_\gr^2\mid\calD)$ gives
\begin{align*}
    E(\hat\sigma_{\gr}^2\mid\calD)=\frac{1}{(\sum_{k=1}^Kw_k)^2}\sum_{k=1}^K\frac{n_{.1k}n_{.0k}^2\frac{n_{.1k}-1}{n_{.1k}}p_{1k}(1-p_{1k})
+n_{.1k}^2n_{.0k}\frac{n_{.0k}-1}{n_{.0k}}p_{0k}(1-p_{0k})}{n_{..k}^2},
\end{align*}
the difference of which between $\sigma_n^2$ is that
\begin{align*}
    \sigma_n^2-E(\hat\sigma_{\gr}^2\mid\calD)=\frac{1}{(\sum_{k=1}^Kw_k)^2}\sum_{k=1}^K\frac{n_{.0k}^2p_{1k}(1-p_{1k})
+n_{.1k}^2p_{0k}(1-p_{0k})}{n_{..k}^2}.
\end{align*}
By plugging $E(n_{.ak}^2\mid\mathcal{Z})=n_{..k}\pi_1\pi_0 + n_{..k}^2\pi_a^2$ as shown in \eqref{eq: w_k moments} into the difference, we obtain that
\begin{align*}
    \sum_{k=1}^K\frac{n_{.0k}^2p_{1k}(1-p_{1k})
+n_{.1k}^2p_{0k}(1-p_{0k})}{n_{..k}^2}&=\sum_{k=1}^K\frac{(\pi_1\pi_0+n_{..k}\pi_0^2)p_{1k}(1-p_{1k})+(\pi_1\pi_0+n_{..k}\pi_1^2)p_{0k}(1-p_{0k})}{n_{..k}}\\
&\geq \sum_{k=1}^K \frac{2\pi_1\pi_0\epsilon(1-\epsilon)}{c}+(\pi_1^2+\pi_0^2)\epsilon(1-\epsilon)\\
&=K\epsilon(1-\epsilon)\left\{\frac{2\pi_1\pi_0}{c}+(\pi_1^2+\pi_0^2)\right\}.
\end{align*}
Recall that $n^{-1}\sum_{k=1}^Kw_k$ is positive and on the order of 1 with probability approaching 1, and $K/n\geq 1/c$. Therefore, we can see that $E\{n(\sigma_n^2-\hat\sigma_{\gr}^2)\mid\mathcal{Z}\}\geq c'\left\{\frac{2\pi_1\pi_0\epsilon(1-\epsilon)}{c}+(\pi_1^2+\pi_0^2)\epsilon(1-\epsilon)\right\}$ with probability approaching 1, where $c'$ is a positive constant. Since the right-hand side is constant and positive, taking expectation on both sides shows that $\hat\sigma_\gr^2$ is biased and underestimates the true variance under the sparse-stratum asymptotics.

\subsection{Consistency of the $\text{mGR}_{\mh}$ variance estimator}\label{sec: mGR MH}
Under the large-stratum asymptotics, $(\frac{n_{.ak}}{n_{.ak}-1})^{I(n_{.ak}>1)}=1+o_p(1)$. So, $\hat\sigma^2$ is consistent following the same steps in \ref{sec: consistency GR}. We show it is also consistent under the sparse-stratum asymptotics as follows. Using that $E(n_{1ak}n_{0ak}\mid \mathcal{D})=n_{.ak}(n_{.ak}-1)p_{ak}(1-p_{ak})$ gives
\begin{align*}
    &\quad E(\hat\sigma^2\mid\calD)\\
    &=\frac{1}{(\sum_{k=1}^Kw_k)^2}\sum_{k=1}^Kw_k^2\left\{\frac{n_{.1k}-1}{n_{.1k}^2}(\frac{n_{.1k}}{n_{.1k}-1})^{I(n_{.1k}>1)}p_{1k}(1-p_{1k})+\frac{n_{.0k}-1}{n_{.0k}^2}(\frac{n_{.0k}}{n_{.0k}-1})^{I(n_{.0k}>1)}p_{0k}(1-p_{0k})\right\}.
\end{align*}
Therefore, the difference of $\sigma_n^2$ between $ E(\hat\sigma^2\mid\calD)$ is 
\begin{align*}
    \sigma^2_n-E(\hat\sigma^2\mid\calD)&=\frac{1}{(\sum_{k=1}^Kw_k)^2}\sum_{k=1}^K\frac{n_{.0k}^2p_{1k}(1-p_{1k})I(n_{.1k}=1)
+n_{.1k}^2p_{0k}(1-p_{0k})I(n_{.0k}=1)}{n_{..k}^2}\\
&\leq \frac{1}{(\sum_{k=1}^Kw_k)^2}\sum_{k=1}^K\frac{I(n_{.1k}=1)
+I(n_{.0k}=1)}{4}\\
&=o_p(\frac{1}{n}),
\end{align*}
 where the last equality follows from that $\sum_{k=1}^K \{I(n_{.1k}=1) +I(n_{.0k}=1)  \}/K = o_p(1)$. 
 
 To use Lemma \ref{lemma1}, we focus on the variance of $$X_k:=\frac{1}{n}w_{k}^2 \left\{  \frac{n_{11k}n_{01k}}{n_{.1k}^3} \left( \frac{n_{.1k}}{n_{.1k}-1}\right)^{ I(n_{.1k}>1)}+  \frac{n_{10k}n_{00k}}{n_{.0k}^3} \left( \frac{n_{.0k}}{n_{.0k}-1}\right)^{ I(n_{.0k}>1)} \right\}$$ conditional on $\calD$. We can see that
\begin{align*}
    \var(X_k\mid\calD)&\leq E(X_k^2\mid \calD)\\
    &=\frac{1}{n^2}E\left[w_{k}^4 \left\{  \frac{n_{11k}n_{01k}}{n_{.1k}^3} \left( \frac{n_{.1k}}{n_{.1k}-1}\right)^{ I(n_{.1k}>1)}+  \frac{n_{10k}n_{00k}}{n_{.0k}^3} \left( \frac{n_{.0k}}{n_{.0k}-1}\right)^{ I(n_{.0k}>1)} \right\}^2\mid \calD\right]\\
    &\leq \frac{4}{n^2}E\left\{w_{k}^4 \left(  \frac{n_{11k}n_{01k}}{n_{.1k}^3} +  \frac{n_{10k}n_{00k}}{n_{.0k}^3}  \right)^2\mid \calD\right\}\\
    &\leq\frac{4}{n^2}E\left\{\frac{n_{.1k}^4n_{.0k}^4}{n_{..k}^4}\left(\frac{1}{n_{.1k}}+\frac{1}{n_{.0k}}\right)^2\mid\calD\right\}\\
    &\leq \frac{16n_{..k}^2}{n^2}\\
    &\leq \frac{16c^2}{n^2}.
\end{align*}
Therefore, $\sum_{k=1}^K\var(X_k\mid\calD)=16c^2K/n^2=o(1)$. By Lemma \ref{lemma1}, we conclude that conditional on $\calD$, $n(\sigma_n
^2-\hat\sigma^2)=o_p(1)$. By the bounded convergence theorem, it holds unconditionally, which means that $\hat\sigma^2$ is consistent under the sparse-stratum asymptotics.

\subsection{Consistency of the $\text{mGR}_{\rm ATE}$ variance estimator} \label{sec: mGR ATE}

We have already shown that $\hat\sigma^2$ is a consistent estimator of $\sigma_n^2$ under both asymptotics in \ref{sec: mGR MH}. To show $\text{mGR}_{\rm ATE}$ is consistent, it suffices to show $\hat\nu^2$ is a consistent estimator of $\nu_n^2$. Under the large-stratum asymptotics, the proof of consistency is the same as the proof of $\hat\sigma^2$ in \ref{sec: mGR MH}. Next, we show the consistency under the sparse-stratum asympototic. Recall that
\begin{equation*}
    n\hat\nu^2 = \frac{ \sum_{k=1}^K I(n_{.1k}n_{.0k}\neq 0) \left\{(\widehat{\delta_k^2}-2\hat\delta_k\hat\delta+\hat\delta^2) \hat\pi_1\hat\pi_0 \frac{n_{..k}-1}{n_{..k}}\frac{n_{..k}-1-(4n_{..k}-6)\hat\pi_1\hat\pi_0}{n}+\hat\pi_1^2\hat\pi_0^2\hat\rho_k(\widehat{\delta_k^2}-\hat\delta^2)\right\}}{(n^{-1} \sum_{k=1}^K w_{k})^2}.
\end{equation*}

First, we construct $n\Tilde{\nu}^2$ by replacing $\hat\pi_1$ and $\hat\pi_0$ with $\pi_1$ and $\pi_0$ in $n\hat\nu^2$ respectively. Then, it turns out that
\begin{align*}
    |n\hat\nu^2-n\Tilde{\nu}^2|&=|\frac{ \sum_{k=1}^K I(n_{.1k}n_{.0k}\neq 0) \left\{(\widehat{\delta_k^2}-2\hat\delta_k\hat\delta+\hat\delta^2) \frac{n_{..k}-1}{n_{..k}}\frac{n_{..k}-1-(4n_{..k}-6)\hat\pi_1\hat\pi_0}{n}o_p(1)+\hat\rho_k(\widehat{\delta_k^2}-\hat\delta^2)o_p(1)\right\}}{(n^{-1} \sum_{k=1}^K w_{k})^2}|\\
    &\leq \frac{ \sum_{k=1}^K 4 \frac{n_{..k}}{n}o_p(1)+2\hat\rho_ko_p(1)}{(n^{-1} \sum_{k=1}^K w_{k})^2}\\
    &=\frac{6}{(n^{-1} \sum_{k=1}^K w_{k})^2}o_p(1)=o_p(1).
\end{align*}
Similarly, we can replace $\hat\delta$ with $\delta_{\rm ATE}$ with an asymptotically negligible term remaining. Therefore, we have that
\begin{align*}
    n\hat\nu^2 = \frac{ \sum_{k=1}^K \left\{(\widehat{\delta_k^2}-2\hat\delta_k\delta_{\rm ATE}+\delta_{\rm ATE}^2) \pi_1\pi_0 \frac{n_{..k}-1}{n_{..k}}\frac{n_{..k}-1-(4n_{..k}-6)\pi_1\pi_0}{n}+\pi_1^2\pi_0^2\hat\rho_k(\widehat{\delta_k^2}-\delta_{\rm ATE}^2)\right\}}{(n^{-1} \sum_{k=1}^K w_{k})^2}+o_p(1).
\end{align*}
Then, since $\hat\delta_k$ is an unbiased estimator of $\delta_k$ conditional on $\calD$ and $\widehat{\delta_k^2}$ is also an unbiased estimator of $\delta_k^2$ in strata with $n_{.1k},n_{.0k}>1$ conditional on $\calD$ resulting from that
\begin{align*}
    E(\hat p_{ak}^2-S_{ak}^2/n_{.ak}\mid\calD) = \frac{p_{ak}(1-p_{ak})}{n_{.ak}}+p_{ak}^2-\frac{p_{ak}(1-p_{ak})}{n_{.ak}}=p_{ak}^2,
\end{align*}
we conclude that $\hat\nu^2$ is consistent, following the same proof of $\hat\sigma^2$ in \ref{sec: mGR MH}.
\subsection{Consistency of the PS variance estimator}\label{sec: PS consistent}
We show that the PS variance estimator is consistent of $n\sigma_{n,\rm PS}^2+n\nu_{n,\rm PS}^2$ under both large- and sparse-asymptotics.

(a) {\bf Consistency under large-stratum asymptotics.} Under the large-stratum regime, $n_{.1k},n_{.0k}>1$ almost surely. Therefore, since $S_{.ak}^2\xrightarrow{p}\var(Y^{(a)}\mid Z=z_k)$ and $n_{.ak}/n_{..k}\xrightarrow{p}\pi_a$, we have that $n\hat\sigma_{\rm PS}^2\xrightarrow{p}E\{\pi_1^{-1}\var(Y^{(1)}\mid Z) + \pi_0^{-1}\var(Y^{(0)}\mid Z)\}$ and $n\hat\nu^2_{\rm PS}\xrightarrow{p}\var\{E(Y^{(1)}-Y^{(0)}\mid Z)\}$. Therefore, the PS variance estimator is consistent.

(b) {\bf Consistency under sparse-stratum asymptotics.} Notice that $E(S_{ak}^2\mid\calD)=\var(Y^{(a)}\mid Z=z_k)$ and $E(\Bar{Y}_{ak}^2-S_{ak}^2/n_{.ak}\mid\calD)=p_{ak}^2$. Therefore, $n\hat\sigma_{\rm PS}^2$ is consistent of $n\sigma_{n,\rm PS}^2$ and $n\hat\nu_{\rm PS}^2$ is consistent of $n\nu_{n,\rm PS}^2$ given that $\sum_{k=1}^K \{I(n_{.1k}=1) +I(n_{.0k}=1)  \}/K = o_p(1)$, following the same steps in \ref{sec: mGR MH}.

\subsection{Asymptotic variance of the {\red MH test statistic}}
Recall that {\red the MH test statistic} can be written as
\begin{align*}
\frac{\big\{ n^{-1/2}\sum_{k=1}^K n_{11k} - n_{.1k}n_{1.k}/n_{..k}\big\}^2}{n^{-1}\sum_{k=1}^K \frac{ n_{1.k} n_{0.k}n_{.1k} n_{.0k} }{(n_{..k}- 1)n_{..k}^2}} .
\end{align*}

We focus on the convergence of the denominator under different regimes and show that the asymptotic variance of $n^{-1/2}\sum_{k=1}^Kn_{11k}-n_{.1k}n_{1.k}/n_{..k}$ is $\frac{1}{n}\sum_{k=1}^K\frac{n_{.1k}n_{.0k}}{n_{..k}}p_{k}(1-p_k)$ under both large- and sparse asymptotics. Note that under the null hypothesis $H_0:\delta_k=0$ for all $k$, we have that $E(Y^{(1)}\mid Z=z_k)=E(Y^{(0)}\mid Z=z_k)$, which implies that $E(Y^{(1)}\mid Z=z_k)=E(Y^{(0)}\mid Z=z_k)=E(Y\mid Z=z_k):=p_k$. It follows that $\pi_1p_{1k}+\pi_0p_{0k}=(\pi_1+\pi_0)p_{k}=p_k.$ 

{\bf Under the large-stratum asymptotics}. Under the large-stratum asymptotics, we have that $n_{1.k}/n_{..k}\xrightarrow{p}p_k$ and $n_{0.k}/n_{..k}\xrightarrow{p}1-p_k$. Therefore, it follows that
\begin{align*}
    \sum_{k=1}^K\frac{n_{1.k}n_{0.k}n_{.1k}n_{.0k}}{(n_{..k}-1)n_{..k}^2n}\xrightarrow{p}\sum_{k=1}^Kp_k(1-p_k)\pi_1\pi_0\rho_k=\pi_1\pi_0E\{\var(Y\mid Z)\},
\end{align*}
where we have used $\var(Y\mid Z=z_k)=p_k(1-p_k)$.

{\bf Under sparse-stratum asymptotics.} Denote $\frac{n_{1.k}n_{0.k}n_{.1k}n_{.0k}}{(n_{..k}-1)n_{..k}^2}$ by $X_k$. Conditional on $\calD$, we have that
\begin{align*}
    E(X_k\mid\calD)&=\frac{n_{.1k}n_{.0k}}{(n_{..k}-1)n_{..k}^2}E\{n_{1.k}(n_{..k}-n_{1.k})\mid\calD\}\\
    &=\frac{n_{.1k}n_{.0k}}{(n_{..k}-1)n_{..k}^2}\left\{n_{..k}^2p_k-n_{..k}p_k(1-p_k)-n_{..k}^2p_k^2\right\}\\
    &=\frac{n_{.1k}n_{.0k}}{n_{..k}}p_{k}(1-p_k).
\end{align*}

Recall that in \ref{sec: proof thm2} we have the equation \eqref{eq: tmp2}:
\begin{align*}
    E\{n_{.1k}^2(n_{k}-n_{.1k})^2\mid\mathcal{Z}\}
    =\pi_1\pi_0n_{..k}(n_{..k}-1)\left\{n_{..k}-1+(n_{..k}-2)(n_{..k}-3)\pi_1\pi_0\right\}.
\end{align*}
Following the same derivation of \eqref{eq: tmp2} gives that
\begin{align*}
E(n_{1.k}^2n_{0.k}^2\mid\calD)&=p_k(1-p_k)n_{..k}(n_{..k}-1)\left\{n_{..k}-1+(n_{..k}-2)(n_{..k}-3)p_k(1-p_k)\right\}.
\end{align*}
Therefore, it follows that
\begin{align*}
\quad\var(X_k\mid \calD)&\leq E\{\frac{n_{1.k}^2n_{0.k}^2n_{.1k}^2n_{.0k}^2}{(n_{..k}-1)^2n_{..k}^4}\mid \calD\}\\ 
    &=\frac{p_k(1-p_k)\left\{n_{..k}-1+(n_{..k}-2)(n_{..k}-3)p_k(1-p_k)\right\}n_{.1k}^2n_{.0k}^2}{(n_{..k}-1)n_{..k}^3}\\
    &\leq \frac{\left\{n_{..k}-1+(n_{..k}-1)^2\right\}n_{..k}^2n_{..k}^2}{(n_{..k}-1)n_{..k}^3}\\
    &\leq n_{..k}^2\\
    &\leq c^2.
\end{align*}
Therefore, we have that $K^{-2}\sum_{k=1}^K\var(X_k\mid\calD)=K^{-1}c^2=o(1)$.  Since $K/n=O(1)$, by Lemma \ref{lemma1}, it turns out that conditional on $\calD$,
\begin{align*}
    \sum_{k=1}^K\frac{X_k-E(X_k\mid\calD)}{n}=\frac{K}{n}\sum_{k=1}^K\frac{X_k-E(X_k\mid\calD)}{K}\xrightarrow{p}0.
\end{align*}
By the bounded convergence theorem, it also holds unconditionally, which implies that
\begin{align*}
    \sum_{k=1}^K\frac{n_{1.k}n_{0.k}n_{.1k}n_{.0k}}{(n_{..k}-1)n_{..k}^2n}&=\frac{1}{n}\sum_{k=1}^K\frac{n_{.1k}n_{.0k}}{n_{..k}}p_{k}(1-p_k)+o_p(1).
\end{align*}

 Notice that under the large-stratum asymptotics, $\frac{1}{n}\sum_{k=1}^K\frac{n_{.1k}n_{.0k}}{n_{..k}}p_{k}(1-p_k)\xrightarrow{p}\pi_1\pi_0E\{\var(Y\mid Z)\}$. Therefore, we conclude that the asymptotic variance of $n^{-1/2}\sum_{k=1}^Kn_{11k}-n_{.1k}n_{1.k}/n_{..k}$ is $\frac{1}{n}\sum_{k=1}^K\frac{n_{.1k}n_{.0k}}{n_{..k}}p_{k}(1-p_k)$ under both large- and sparse-stratum asymptotics.

\bibliographystyle{plainnat}
\bibliography{export}